\documentclass[%
     aip,
     amsmath,amssymb,
     reprint,%
]{revtex4-1}

\usepackage{graphicx}% Include figure files
\usepackage{dcolumn}% Align table columns on decimal point
\usepackage{bm}% bold math
\usepackage[utf8]{inputenc}
\usepackage[T1]{fontenc}
\usepackage{mathptmx}
\usepackage{etoolbox}

\makeatletter
\def\@email#1#2{%
 \endgroup
 \patchcmd{\titleblock@produce}
  {\frontmatter@RRAPformat}
  {\frontmatter@RRAPformat{\produce@RRAP{*#1\href{mailto:#2}{#2}}}\frontmatter@RRAPformat}
  {}{}
}%
\makeatother

\usepackage[margin=2cm]{geometry}
\allowdisplaybreaks

\usepackage{subfig}

\usepackage[hyperindex,breaklinks]{hyperref}
\usepackage[dvipsnames]{xcolor}
\usepackage{amsmath}
\usepackage{amsfonts,amssymb}

\usepackage{tikz}
\usetikzlibrary{quantikz}

\usepackage{nicematrix}
\usepackage{arydshln}
\usepackage{soul}  % to use the strikeout text

\usepackage{pgfplots}
\pgfplotsset{compat=1.17}

\usepackage{mathbbol}
\usepackage{xifthen}

\usepackage[titletoc,title]{appendix}

\usepackage{hhline}
\usepackage{multirow}

\colorlet{ygray}{gray!20}

\newcommand{\aeq}{\begin{equation}}
\newcommand{\eeq}{\end{equation}}
\newcommand{\aeqn}{\begin{eqnarray}}
\newcommand{\eeqn}{\end{eqnarray}}
\newcommand{\aeqns}{\begin{eqnarray*}}
\newcommand{\eeqns}{\end{eqnarray*}}

\newcommand{\ybs}[1]{\boldsymbol{#1}}

\newcommand{\yi}{\mathrm{i}}

\newcommand{\oO}{\mathcal{O}}
\newcommand*\diff{\mathop{}\!\mathrm{d}}
\newcommand{\itimes}[2]{\langle #1|#2 \rangle}

\newcommand{\ylb}{\left(}
\newcommand{\yrb}{\right)}

\newcommand{\yggr}[5]{\gategroup[wires=#1, steps=#2, style={#4, dashed, inner sep= #5 pt}, label style={yshift=0.05 cm}]{#3}}

\newcommand{\yEx}[1][]{\ifthenelse{\isempty{#1}}{\tilde{E}_x}{\tilde{E}_{x,#1}}}
\newcommand{\yEy}[1][]{\ifthenelse{\isempty{#1}}{\tilde{E}_y}{\tilde{E}_{y,#1}}}
\newcommand{\yBz}[1][]{\ifthenelse{\isempty{#1}}{\tilde{B}_z}{\tilde{B}_{z,#1}}}

\usepgfplotslibrary{fillbetween}
\begin{document}
\preprint{AIP/123-QED}

%% -------------------------------------------------------
%% --- TITLE
%% -------------------------------------------------------
% If necessary make line breaks with \\
\title[]{Simulation of linear non-Hermitian boundary-value problems with quantum singular value transformation}
\author{I. Novikau}
\email{inovikau@pppl.gov}
\affiliation{Princeton Plasma Physics Laboratory, Princeton, New Jersey 08543, USA}
\author{I. Y. Dodin}
\affiliation{Princeton Plasma Physics Laboratory, Princeton, New Jersey 08543, USA}
\affiliation{Department of Astrophysical Sciences, Princeton University, Princeton, New Jersey 08544, USA}
\author{E. A. Startsev}
\affiliation{Princeton Plasma Physics Laboratory, Princeton, New Jersey 08543, USA}

\date{\today}

%% -------------------------------------------------------
%% --- Abstract ---
%% -------------------------------------------------------
\begin{abstract}
We propose a quantum algorithm for simulating dissipative waves in inhomogeneous linear media as a boundary-value problem. 
Using the so-called quantum singular value transformation (QSVT), we construct a quantum circuit that models the propagation of  electromagnetic waves in a one-dimensional system with outgoing boundary conditions.
The corresponding measurement procedure is also discussed.
% A quantum circuit is presented that models one-dimensional electromagnetic waves using the so-called quantum singular value transformation (QSVT), and the corresponding measurement procedure is also discussed. 
Limitations of the QSVT algorithm are identified in connection with the large condition numbers that the dispersion matrices exhibit at weak dissipation. 

% Modeling of radio-frequency wave dynamics in boundary-value problems have significant practical applications in plasma diagnostics and plasma heating. 
% However, due to the substantial discrepancy between typical wave lengths and sizes of simulated domains, even linear computations become challenging for classical computers.
% In this paper, we investigate applicability of the state-of-art quantum singular value transform (QSVT) method for computation of dissipative electromagnetic waves in simplified one-dimensional systems. 
% The main interest is focused on the direct encoding of the wave non-Hermitian system into an oracle and on the circuit measurements to extract useful classical information such as the wave energy and wave number spectra.
% In particular, we discuss an algorithm to calculate wave absorption power by using a separate oracle to encode the environment conductivity.
% It is demonstrated that even noiseless emulations of the QSVT circuits are accompanies by a set of challenges mainly related to typically large condition numbers of classical wave matrices.
\end{abstract}
\maketitle

\section{Introduction}

First-principle (`full-wave') modeling of linear waves in inhomogeneous linear media is important for various applications, for example, plasma heating in fusion research.\cite{Fisch87, Pinsker01, Prater04}
However, it can be computationally expensive, especially for waves with wavelengths that are orders magnitude smaller than the characteristic inhomogeneity scales. Such simulations can be facilitated by quantum computing (QC). 

Quantum algorithms have been proposed for initial-value problems that involve purely Hermitian interactions, such as the propagation of electromagnetic (EM) waves in cold magnetized plasma\cite{Novikau22} and also Landau damping of kinetic plasma waves.\cite{Engel19}
However, practical applications are usually concerned with dissipative waves and are set up as boundary-value problems, also called antenna problems.
The corresponding codes (Refs.~\onlinecite{Lancellotti06, Shiraiwa17, Svidzinski18}, to name a few) are often used to model EM waves in fusion plasmas.\cite{Hillairet21, Zhang22, Guttenfelder22}
After discretization, such problems can be represented as linear vector equations of the form
\begin{equation}\label{eq:axb}
    A\psi = b,
\end{equation}
where $A$ of size $N\times N$ is generally a non-Hermitian invertible matrix, $b$ is a given vector, and the vector $\psi$ represents the field(s) of interest.\cite{Dodin20}

The first quantum method to solve Eq.~\eqref{eq:axb}, the so-called HHL algorithm, was presented in Ref.~\onlinecite{HHL09}.
This method was developed further in Refs.~\onlinecite{Ambainis12, Childs17}, where its scaling with the condition number $\kappa$ of the matrix $A$ and the absolute error has been improved.
A thorough analysis of the HHL algorithm in application to EM classical-wave problems was given in Ref.~\onlinecite{Scherer17}.
% , where the most demanding subroutine was found to be the Hamiltonian simulation.
It was demonstrated there that the time necessary for a single run of the corresponding HHL circuit to achieve quantum advantage is comparable with the age of the Universe (even without taking into account the costs of encoding the matrix $A$ into a quantum circuit).
Hence, a concern has emerged to what extent quantum algorithms are actually applicable to boundary-value wave problems.

% and obtained further development in Refs.~\onlinecite{Ambainis12, Childs17}, where the main efforts were put to achieve the optimal scaling with respect to the matrix condition number $\kappa$ and the resulting absolute error.
% Thorough analysis of the HHL modeling of electromagnetic classical problems was given in Ref.~\onlinecite{Scherer17}, where the Hamiltonian simulation was revealed as the most resource-demanding subroutine within the HHL.
% This algorithm was also used in Ref.~\onlinecite{Cao13} to solve the Poisson equation discretized on a regular grid with Dirichlet boundary conditions.
% Quantum lattice simulations of Maxwell equations were demonstrated in Refs.~\onlinecite{Vahala20}.

Here, we address this matter by developing a different approach to solving Eq.~\eqref{eq:axb} on a quantum computer, namely, by using so-called quantum signal processing (QSP).
Although originally developed\cite{Low17, Low19} for Hermitian matrices, the QSP has been recently extended to general matrices using the quantum singular value transformation (QSVT).\cite{Gilyen19, Martyn21} 
% Here, we explore how to solve~\eqref{eq:axb} using a different, state-of-the-art quantum algorithm, namely, quantum signal processing (QSP)\cite{Low17, Low19} originally developed to work with Hermitian operators.
% Recently, it was extended to the so-called quantum singular value transform (QSVT) algorithm\cite{Gilyen19, Martyn21} to deal with arbitrary non-Hermitian matrices.
The QSVT provides a near-optimal dependence of the query complexity (the number of calls to the subcircuit encoding $A$) on both the condition number and the error, so it is considered as a promising algorithm for solving linear equations. 
We apply the QSVT to a boundary-value wave problem for the first time. 

We consider an EM wave propagating in an inhomogeneous one-dimensional dielectric medium with a source and outgoing boundary conditions. 
In the first part of our work, we construct a quantum circuit for the corresponding matrix $A$ using the QSVT, emulate quantum simulations on a classical computer, and benchmark our results against those of conventional classical simulations. 
The second part of the work is concerned with extracting classical information from quantum circuits using measurements. 
In general, this step is computationally expensive, so including it is necessary when assessing the efficiency of quantum simulations~\cite{Montanaro16}. 
We discuss how to perform relevant measurements in the wave problem and estimate the necessary resources. 
Our finding is that even with the measurement costs included, the overall quantum simulations of dissipative waves based on the QSVT scale favorably compared to classical simulations in multi-dimensional systems. 
We expect the gain to be particularly efficient in kinetic plasma problems, where the wave modeling is done in phase space with six or even more dimensions.

Our paper is organized as follows. 
In Sec.~\ref{sec:qsvt}, we briefly describe the QSVT algorithm and the corresponding circuit.
The QSVT scaling and its comparison with classical methods are discussed in Sec.~\ref{sec:qsvt-scaling}.
In Sec.~\ref{sec:theoretical-model}, we outline our one-dimensional wave system and describe its discretization.
The encoding of the system into a quantum circuit is discussed in detail in Sec.~\ref{sec:encoding}.
The classical modeling of the system and its quantum simulation on a digital emulator of quantum circuits are compared in Sec.~\ref{sec:comparison}.
There, we explore circuits for basic measurements of the wave-number spectrum and the wave energy.
Besides, an algorithm is proposed to measure wave absorption power, where the QSVT is used for both EM field computations and emulation of the electrical conductivity.
The efficiency of the QSVT algorithm for simulations of classical waves and the related challenges that remain are discussed in Sec.~\ref{sec:conclusions}.
Auxiliary information and technical details are also discussed in appendices~\ref{app:basics} and~\ref{app:ae-gaussian}.

% Additional information concerning basic quantum gates and the quantum arithmetic operators is given in Appendix~\ref{app:basics}.
% QSVT implementation of a Gaussian used in Sec.~\ref{sec:JE} to emulate electric conductivity is presented in Appendix~\ref{app:ae-gaussian}.

\section{Quantum Singular Value Transformation}\label{sec:qsvt}

% ---------------------------------------------------------
% --- QSVT circuit ---
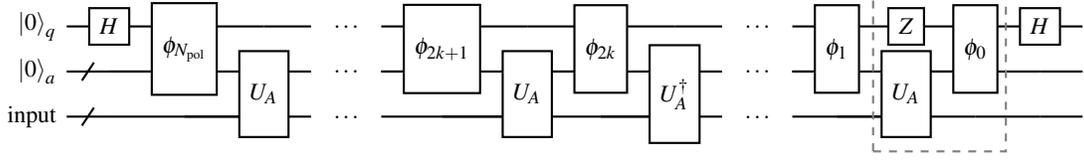
\begin{figure*}[t!]
\centering
\begin{quantikz}[transparent, row sep={0.6cm,between origins},column sep={0.3cm}]
\lstick{$\ket{0}_q$}  &\gate{H}   &\gate[2]{\phi_{N_{\rm pol}}}&\qw           &\qw &\dots & &\gate[2]{\phi_{2k+1}}&\qw          &\gate[2]{\phi_{2k}}&\qw                    &\qw &\dots &              
    &\gate[2]{\phi_1} &\gate{Z}\yggr{3}{2}{}{gray}{0}&\gate[2]{\phi_0}&\gate{H}&\qw\\
\lstick{$\ket{0}_a$}  &\qwbundle{}&\qw                         &\gate[2]{U_A} &\qw &\dots & &\qw                  &\gate[2]{U_A}&\qw                &\gate[2]{U^\dagger_A}  &\qw &\dots &   
    &\qw              &\gate[2]{U_A}                 &\qw             &\qw     &\qw\\
\lstick{${\rm input}$}&\qwbundle{}&\qw                         &\qw           &\qw &\dots & &\qw                  &\qw          &\qw                &\qw                    &\qw &\dots &   
    &\qw              &\qw                           &\qw             &\qw     &\qw 
\end{quantikz}
\caption{\label{circ:qsvt-odd} 
A schematic of the QSVT circuit encoding a matrix real polynomial of order $N_{\rm pol}$, where $N_{\rm pol}$ is odd, by using $N_{\rm pol} + 1$ angles.
For encoding a polynomial with even $N_{\rm pol}$, the gates in the dashed box should be removed. 
The gates denoted as $\phi_k$ represent the controlled rotations $\exp(\yi\phi_k Z_\Pi)$ shown in Fig.~\ref{circ:controlled-rotation}.
The upper (lower) qubit is the most (least) significant one.
}
\end{figure*}

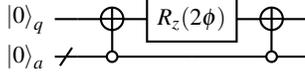
\begin{figure}[t!]
\centering
\begin{quantikz}[row sep={0.5cm,between origins},column sep={0.3cm}]
\lstick{$\ket{0}_q$} &\qw         &\targ{}   &\gate{R_z(2\phi)}&\targ{}   &\qw\\
\lstick{$\ket{0}_a$} &\qwbundle{} &\octrl{-1}&\qw              &\octrl{-1}&\qw
\end{quantikz}
\caption{\label{circ:controlled-rotation} 
The circuit of the controlled rotation $\exp(\yi\phi Z_\Pi)$.
Here, $R_z(2\phi) \equiv \exp(-\yi\phi Z)$, as in Eq.~\eqref{eq:rz}.
The empty circles, called here zero-control nodes, activate the $X$ gates if the qubit $a$ is in the zero state.
}
\end{figure}
% ---------------------------------------------------------

% ----------------------------------------------------------
% --- General idea ---
% ----------------------------------------------------------
\subsection{General idea}
One of the ways to solve Eq.~\eqref{eq:axb} is by calculating the inverse of the matrix $A$.
The polynomial approximation $P_{\rm inv}(A)$ of this function with a Hermitian $A$ can be found on a quantum computer by using the QSP.\cite{Low17, Low19}
% A given function $f(A)$ can be approximated with a matrix polynomial, $f(A) \approx P_f(A)$.
% To construct $P_f(A)$ of a Hermitian $A$ on a quantum compute, one can use the so-called Quantum Signal Processing (QSP)\cite{Low17, Low19}.
To compute $P_{\rm inv}(A)$ of a non-Hermitian $A$, one can dilate the matrix to make it the Hermitian one and then use the QSP.
Another possibility, which does not require the dilation,  is to apply the QSVT.\cite{Gilyen19, Martyn21}
The idea of this approach is based on a generalization of the classical singular value transformation
\begin{equation}\label{eq:svd}
A = U_L S U_R^\dagger,
\end{equation}
where $S = {\rm diag}(s_1, \dots, s_N)$ is a diagonal $N\times N$ matrix with real nonnegative (or strictly positive, if $A$ is invertible) diagonal elements $s_i$ called singular values of $A$; also, $U_L$ and $U_R$ are complex unitary matrices.
The QSVT operates with a matrix polynomial 
% $P_{\rm qsvt}(A)$, whose action is defined by a given complex scalar polynomial $p(s)$ of the singular values:
\begin{equation}\label{eq:qsvt-definition}
    P_{\rm qsvt}(A) = U_L p(S) U_R^\dagger,
\end{equation}
% with $p(S) = {\rm diag}(p(s_1), \dots, p(s_N))$, and $P_{\rm qsvt}(A)$ is calculated by a QSVT quantum circuit.
where $p(S) = {\rm diag}(p(s_1), \dots, p(s_N))$, and $p(s_i)$ is a complex polynomial of a scalar $s_i$. 
In particular, let us consider the case when $P_f$ is a polynomial approximation of some given function $f$. 
If $A$ is Hermitian, then $U_L = U_R$ and, clearly, $P_{\rm qsvt}(A) = P_f(A)$. 
If $A$ is not Hermitian, then $P_{\rm qsvt}(A)$ generally does not coincide with $P_f(A)$, but it still can be used to approximate $P_f(A)$ if $f$ is the inverse function. This is seen as follows.

% In the case when $p(s)$ is a polynomial approximation of an arbitrary function $f(s)$, the polynomial $P_{\rm qsvt}(A)$ defined by Eq.~\eqref{eq:qsvt-definition} generally does not coincide with the polynomial approximation $P_f(A)$ of the matrix function $f(A)$ if $A$ is a non-Hermitian matrix, and $P_{\rm qsvt}(A) = P_f(A)$ if $A$ is Hermitian.
% Yet, the QSVT can be used to approximate the inverse function of a non-Hermitian $A$.
% Indeed, the QSVT allows one to find the Moore--Penrose pseudoinverse $A^+$ (which must not be confused with the Hermitian adjoint $A^\dagger$).
First, note that the QSVT allows one to find the Moore--Penrose pseudoinverse $A^+$ (which must not be confused with the Hermitian adjoint $A^\dagger$).
From the definition of $A^+$, and assuming that $A^\dag A$ is invertible, one has $A^+ = (A^\dagger A)^{-1} A^\dagger$, so, by Eq.~\eqref{eq:svd}, one finds
\begin{equation}\label{eq:Aplus}
A^+ = U_R S^{-1} U_L^\dagger.
\end{equation}
% Now, by introducing a symbol of polynomial approximation denoted $\simeq$, 
Now, let us consider $P_{\rm qsvt}$ specifically with $p(s) \simeq s^{-1}$, where the symbol $\simeq$ denotes a polynomial approximation. 
By Eq.~\eqref{eq:qsvt-definition}, one has $P_{\rm qsvt}(A^\dagger) = U_R S^{-1} U_L^\dagger$, and by comparing this with Eq.~\eqref{eq:Aplus}, one finds that $A^+ = P_{\rm qsvt}(A^\dag)$.
If the matrix $A$ is invertible, then $A^+ = A^{-1}$ and therefore one has
\begin{gather}
A^{-1} = P_{\rm qsvt}(A^\dag).
\end{gather}
This shows that $A^{-1}$ can be approximated with a QSVT polynomial constructed for $p(s) \simeq s^{-1}$.

\subsection{Block encoding}
Because a quantum circuit can implement only unitary operations, to decompose a matrix polynomial of a given non-unitary matrix $A$, one first needs to encode it as a subblock of an auxiliary unitary $U_A$.
This procedure is called block encoding and involves introducing additional ancilla qubits. 
Specifically, $U_A$ acts as $A$ when the ancillae are in the zero state, $\ket0_a$, so $U_A$ has a form
\begin{equation}\label{eq:UA}
U_A = 
\begin{pmatrix}
A & \cdot \\
\cdot & \cdot
\end{pmatrix}.
\end{equation}
For this, $A$ must be normalized such that $\varsigma||A||_{\rm max} \leq 1$, where
\begin{equation}
    ||A||_{\rm max} = \max_k\sum_j\sqrt{|A_{kj}|^2},
\end{equation}
and $\varsigma$ is related to the matrix sparsity as detailed in Sec.~\ref{sec:enc-nonH}. 
(We define the sparsity as the number of nonzero matrix elements in a row maximized over all rows.)
Otherwise, $A$ should be renormalized as follows:
\begin{equation}
    A \to A/\ylb||A||_{\rm max}\varsigma\yrb.\label{eq:gen-A-norm}
\end{equation}

% ----------------------------------------------------------
% --- Polynomials via matrix rotations ---
% ----------------------------------------------------------
\subsection{Polynomials via matrix rotations}
The sets of right and left singular vectors $u_r$ and $u_l$, which are columns of the matrices $U_L$ and $U_R$, correspondingly, form two orthogonal sets.
Any other vector can be represented as a linear superposition of the vectors from one of these sets.
Therefore, to construct $U_A$, it is sufficient to define the action of this operator on the vectors $u_r$ and $u_l$.
% we need to consider the action of the block-encoding oracle $U_A$ only on the vectors $u_r$ and $u_l$.
Assuming that $u_r$ and $u_l$ are associated with a given singular value $s$, one obtains from Eq.~\eqref{eq:svd} that
\begin{subequations}
\begin{eqnarray}
&&U_A\ket0_a\ket{u_r} = s\ket{0}_a\ket{u_l} + \sqrt{1 - s^2}\ket{\perp_l},\label{eq:ur-space}\\
&&U_A^\dagger\ket0_a\ket{u_l} = s\ket{0}_a\ket{u_r} + \sqrt{1 - s^2}\ket{\perp_r},
\end{eqnarray}
\end{subequations}
where $\ket{\perp_l}$ and $\ket{\perp_r}$ are vectors orthogonal to the subspace that corresponds to the ancillae $a$ being in the zero state.
% By multiplying the first and second equations by $U_A^\dagger$ and $U_A$, respectively, one obtains
Using that $U_A^\dag U_A = 1$, one obtains
\begin{subequations}
\begin{eqnarray}
&&U_A\ket{\perp_r} = \sqrt{1 - s^2}\ket{0}_a\ket{u_l} - s\ket{\perp_l},\\
&&U_A^\dagger\ket{\perp_l} = \sqrt{1 - s^2}\ket{0}_a\ket{u_r} - s\ket{\perp_r}.\label{eq:perp-l-space}
\end{eqnarray}
\end{subequations}
The above equations indicate that $U_A$ maps the Hilbert space spanned by $\ket{0}_a\ket{u_r}$ and $\ket{\perp_r}$ to the space spanned by $\ket{0}_a\ket{u_l}$ and $\ket{\perp_l}$. 
Likewise, $U_A^\dagger$ maps $\ket{0}_a\ket{u_l}$ and $\ket{\perp_l}$ back to $\ket{0}_a\ket{u_r}$ and $\ket{\perp_r}$.
Note that all these spaces remain invariant under the action of the projector $\Pi$:
% , whose matrix representation in these spaces is
\begin{equation}
\Pi = 
\begin{pmatrix}
    1 & 0 \\
    0 & 0
\end{pmatrix}.
\end{equation}
This property also extends to any function of $\Pi$.
In particular, the reflector $Z_\Pi = 2\Pi - 1$ maps $\ket{0}_a\ket{u_{r,l}}$ to $\ket{0}_a\ket{u_{r,l}}$, and $\ket{\perp_{r,l}}$ to $- \ket{\perp_{r,l}}$.
By using the reflector, one can compose an elementary block of the QSVT circuit: 
\begin{equation}
    W = U_A^\dagger e^{\yi\phi_x Z_\Pi}  U_A e^{\yi\phi_y Z_\Pi},
\end{equation}
where $\phi_x$ and $\phi_y$ are real scalars that can be understood as rotation angles.
A sequence of copies of this operator creates a complex polynomial $P_{\rm qsvt}(A)$ of definite parity.
An odd polynomial of $A$ with a degree not exceeding $N_{\rm pol}$ can be calculated as\cite{Gilyen19}
\begin{equation}\label{eq:qsvt-odd}
\begin{split}
    P_{\rm qsvt}^{\rm odd}(A) = \bra{0}_{q,a}\ylb 
        e^{\yi\phi_0 Z_\Pi} U_A e^{\yi\phi_1 Z_\Pi} \prod^{(N_{\rm pol}-1)/2}_{k=1} G_k
    \yrb\ket{0}_{q,a},
\end{split}
\end{equation}
where
\begin{equation}
    G_k = U_A^\dagger e^{\yi\phi_{2k} Z_\Pi}  U_A e^{\yi\phi_{2k+1} Z_\Pi}.
\end{equation}
The polynomial $P_{\rm qsvt}$ is defined up to a global phase and requires $N_{\rm pol}+1$ classically precalculated angles $\phi_i$ as explained in Sec.~\ref{sec:qsvt-angles}. 
In the spaces defined by Eqs.~\eqref{eq:ur-space}-\eqref{eq:perp-l-space} and spanned by $\ket{0}_a\ket{u_r}$ and $\ket{\perp_r}$, as well as $\ket{0}_a\ket{u_l}$ and $\ket{\perp_l}$, the matrix polynomial becomes the scalar polynomial $p(s)$ of a real argument $s$.
We are interested only in real polynomials, ${\rm Re}\ p(s)$, whose circuit representation is shown in Fig.~\ref{circ:qsvt-odd}.
(Real polynomials have only real coefficients, but maps complex domains into complex images.)
The ancilla register $a$ is used to construct the block-encoding oracle $U_A$, and the ancilla qubit $q$ is used to construct the controlled rotations $\exp(\yi\phi_{k} Z_\Pi)$ (Fig.~\ref{circ:controlled-rotation}).
% \begin{subequations}\label{sys:controlled-rotation}
% \begin{eqnarray}
% e^{\yi\phi Z_\Pi}\ket{0}_q\ket{0}_a &=& e^{\yi\phi}\ket{0}_q\ket{0}_a,\\
% e^{\yi\phi Z_\Pi}\ket{0}_q\ket{\perp}_a &=& e^{-\yi\phi}\ket{0}_q\ket{\perp}_a,
% \end{eqnarray}
% \end{subequations}
% where the operator controlled by the register $a$ acts on the qubit $q$.
The QSVT can be used to solve Eq.~\eqref{eq:axb} in that it computes $\ket{\psi_{\rm out}} = \ket{0}_{q,a}\ket{\psi_x} + \ket{\dots}$ for given $\ket{b}$ as
\begin{equation}\label{eq:qsvt-rescaling}
    \ket{\psi_x} = \frac{e^{\yi\phi_{\rm glob}}}{\beta_{\rm sc}\kappa_{\rm qsvt}}A^{-1}\ket{b}.
\end{equation}
The QSVT circuit returns $\ket{\psi_x}$ in the input register when the ancilla registers $a$ and $q$ are output in the zero state.
The rescaling by the condition number $\kappa_{\rm qsvt}$ and the additional factor $\beta_{\rm sc}$ ensures that $||\psi_x|| \leq 1$, and $\phi_{\rm glob}$ is an arbitrary global angle.
Equation~\eqref{eq:qsvt-rescaling} indicates that measurements of the state $\ket{\psi_x}$ have the success probability $\oO(1/\kappa_{\rm qsvt}^2)$.

Note that we distinguish the actual condition number $\kappa$ of the matrix $A$ and the condition number $\kappa_{\rm qsvt}$ that is used as a parameter in the calculation of the QSVT angles as explained below. 
Throughout this paper, $\kappa$ is defined as the ratio of the maximum and minimum singular values.
The QSVT algorithm properly approximates $A^{-1}$ if $\kappa_{\rm qsvt} \gtrsim \kappa$.

For more details about the QSVT, see, for example, the recent overview in Ref.~\onlinecite{Lin22}.

% ----------------------------------------------------------
% --- Computation of the QSVT angles ---
% ----------------------------------------------------------
\subsection{Computation of QSVT angles}\label{sec:qsvt-angles}
To compute the angles $\phi_k$ necessary for the construction of the QSVT circuit, one compares the polynomials $P_f(s)$ and $p(s)$.
The polynomial $P_f(s)$ can be taken as a sum of Chebyschev polynomials:
\begin{equation}\label{eq:series-Tk}
    P_f(s) = \sum_{k=0}^{N_c} c_k T_k(s).
\end{equation}
The choice is justified in that the Chebyshev approximation is close to the minimax polynomial,\cite{William07}
which is the best possible polynomial approximation but requires the application of the Remez algorithm\cite{Remez34} that can be computationally expensive.
% heavy for computation and requires the application of the Remez algorithm.\cite{Remez34}
% The choice is justified by the commonly known fact\cite{William07} that the Chebyschev approximation is close to the minimax polynomial, which is the best possible polynomial approximation but heavy for computation and requires the application of the Remez algorithm.\cite{Remez34}
The coefficients $c_k$, however, can be computed without involving the Remez algorithm, specifically, using the Fourier series
\begin{equation}\label{eq:fourier-ck}
    c_k \approx \frac{2 - \delta_{k0}}{2N_q} (-1)^k \sum_{j = 0}^{2N_q - 1} f\ylb-\cos(j\pi/N_q)\yrb e^{\yi \frac{kj\pi}{N_q}},
\end{equation}
where the number of points $N_q$ should be not less than $N_c$.
In our work, we used both algorithms, specifically, the numerical implementation of the Remez algorithm as presented in Ref.~\onlinecite{Dong21code} and also our GPU-parallelized version of the Fourier approach.\cite{QSP-code}

The polynomial $p(s)$ can also be represented as a linear combination of Chebyschev polynomials with coefficients depending on $\phi_k$.
The general algorithm to compute $\phi_k$ involves comparing the polynomial coefficients in $p(s)$ and $P_f(s)$.
Usually, this requires arbitrary-precision arithmetic.\cite{Haah20, Haah20code}
In this work, however, $\phi_k$ are computed by minimizing the difference between $p(s)$ and $P_f(s)$ as proposed in Refs.~\onlinecite{Dong21, Dong21code}. 
This algorithm works with the standard double-precision arithmetic.

% In this work, however, $\phi_k$ are computed by minimizing the function
% \begin{equation}\label{eq:L-min}
%     L(\phi_0, \phi_1, \dots) \sim \sum_j \bigg|\text{Re}[P_{\rm qsvt}(x_j, \phi_0, \phi_1, \dots)] - P_f(x_j)\bigg|,
% \end{equation}
% where $x_j$ are the known Chebyschev nodes.
% This method was proposed in Refs.~\onlinecite{Dong21, Dong21code} and works with the standard double-precision arithmetic.

% The matrix polynomial $P_f(A)$ acting on an arbitrary vector $\ket{u}$ can be reduced to the polynomial of a real scalar $s$ because $\ket{u}$ can be represented as a sum, for example, of the right singular vectors:
% \begin{equation}
%     P_f(A) \sum_k u_k \ket{u_{r,k}} = \sum_k u_k P_f(s_k) \ket{u_{l,k}},
% \end{equation}
% where $s_k$ are the singular values corresponding to the right and left vectors $\ket{u_{r,k}}$ and $\ket{u_{l,k}}$.
% Hence, further we consider $f(s)$ and $P_f(s)$ instead of $f(A)$ and $P_f(A)$, correspondingly.

To compute the polynomial for the inverse function $s^{-1}$, one considers the interval $s\in[-1,-1/\kappa_{\rm qsvt}]\cup[1/\kappa_{\rm qsvt},1]$ and approximates the original function by an auxiliary one that coincides with $s^{-1}$ at the chosen interval and is analytic within $[-1/\kappa_{\rm qsvt}, 1/\kappa_{\rm qsvt}]$.
For instance, one can take~\cite{Ying22}
\begin{equation}
    f(s) = \frac{1 - e^{-(5s\kappa_{\rm qsvt})^2}}{s}.
\end{equation}
Then, one can use an odd polynomial to approximate $f(s)$ 
with some absolute approximation error $\epsilon_{\rm qsvt}$:
\begin{equation}
    |f(s) - P_f(s)| \leq \epsilon_{\rm qsvt}.
\end{equation}
Our calculations show that the polynomial approximation of $s^{-1}$ using the Remez algorithm works efficiently for small condition numbers $(\kappa_{\rm qsvt} \lesssim 200)$ and, generally, results in a smaller number of terms in $P_f(s)$ than the Fourier approach.
For larger $\kappa_{\rm qsvt}$, we use the Fourier approach since the Remez algorithm fails to deliver the coefficients of higher-order polynomials within a reasonable time.
(Further optimization of the Remez algorithm, including its parallelization, might be possible but is not considered in this work.)
In both cases, though, the number of terms in $P_f(x)$ grows linearly with $\kappa_{\rm qsvt}$ and logarithmically with $\epsilon_{\rm qsvt}^{-1}$.

% ----------------------------------------------------------
% --- Scaling ---
% ----------------------------------------------------------
\section{Scaling}\label{sec:qsvt-scaling}

\subsection{General scaling of the QSVT}

% *****************************
% *** Scan: query complexity ***
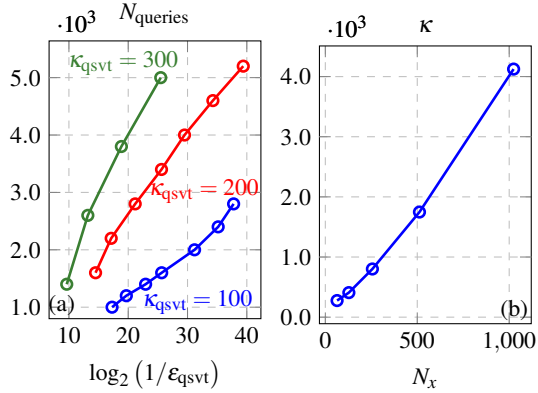
\begin{figure}[!t]
\centering
\begin{tikzpicture}
\begin{axis}[
 	title={$N_{\rm queries}$},
	xlabel={$\log_2\ylb1/\epsilon_{\rm qsvt}\yrb$},
	ylabel={},
	legend pos=north west,
    ymin = 800,
	extra x ticks={},
	extra y ticks={1000, 3000, 5000},
	extra y tick style = {scaled y ticks= base 10:-3},
	xmajorgrids=true, ymajorgrids=true, grid style=dashed,
	height=0.3\textwidth, 
 	width =0.25\textwidth,
    title style={yshift=-4},
	ylabel style={yshift=-0.5ex},
    yticklabel = {
        \pgfmathprintnumber[
            % fixed,
            precision=1,
            zerofill,
        ]{\tick}
    },
    scaled y ticks= base 10:-3,
	legend pos = north west,
	legend style={
	    fill=white, fill opacity=0.9, draw opacity=1,text opacity=1,
	    font=\small,  %\tiny, \small
	},
]

\addplot [blue, solid, mark=o, line width=1.0pt] table [y=Y, x=X]{./data/scan_angles_eps_k100.dat};
% \addlegendentry{$\kappa = 100$};
\addplot [red, solid, mark=o, line width=1.0pt] table [y=Y, x=X]{./data/scan_angles_eps_k200.dat};
% \addlegendentry{$\kappa = 200$};
\addplot [OliveGreen, solid, mark=o, line width=1.0pt] table [y=Y, x=X]{./data/scan_angles_eps_k300.dat};
% \addlegendentry{$\kappa = 300$};

% \node[blue]       at (rel axis cs: 0.75, 0.10) {$\ybs{\kappa_{\rm qsvt} = 100}$};
% \node[red]        at (rel axis cs: 0.73, 0.48) {$\ybs{\kappa_{\rm qsvt} =$}\\$200$};
% \node[OliveGreen] at (rel axis cs: 0.35, 0.92) {$\ybs{\kappa_{\rm qsvt} = 300}$};
\node[blue]       at (rel axis cs: 0.70, 0.06) {$\kappa_{\rm qsvt} = 100$};
\node[red]        at (rel axis cs: 0.73, 0.46) {$\kappa_{\rm qsvt} = 200$};
\node[OliveGreen] at (rel axis cs: 0.35, 0.92) {$\kappa_{\rm qsvt} = 300$};

\node[black] at (rel axis cs: 0.06, 0.05) {(a)};
\end{axis}
\end{tikzpicture}
% ---------------------------------------------------------------------------------------
% ---------------------------------------------------------------------------------------
\begin{tikzpicture}
\begin{axis}[
 	title={$\kappa$},
	xlabel={$N_x$},
	ylabel={},
	legend pos=north west,
	extra x ticks={},
	extra y ticks={1000, 3000},
	extra y tick style = {scaled y ticks= base 10:-3},
	xmajorgrids=true, ymajorgrids=true, grid style=dashed,
	height=0.3\textwidth, 
 	width =0.25\textwidth,
    title style={yshift=-4},
	ylabel style={yshift=-0.5ex},
    yticklabel = {
        \pgfmathprintnumber[
            fixed,
            precision=1,
            zerofill,
        ]{\tick}
    },
    scaled y ticks= base 10:-3,
	legend pos = north west,
	legend style={
	    fill=white, fill opacity=0.9, draw opacity=1,text opacity=1,
	    font=\small,  %\tiny, \small
	},
]

\addplot [blue, solid, mark=o, line width=1.0pt] table [y=Y, x=X]{./data/scan_two_diels_kappa_Nx.dat};
% \addlegendentry{$\kappa = |s_{\rm max}|/|s_{\rm min, \neq 0}|$};

% \node[black] at (rel axis cs: 0.24, 0.92) {$\ybs{\kappa = |s_{\rm max}|/|s_{\rm min, \neq 0}|}$};

\node[black] at (rel axis cs: 0.92, 0.05) {(b)};
\end{axis}
\end{tikzpicture}
% ---------------------------------------------------------------------------------------
% ---------------------------------------------------------------------------------------
\caption{
\label{fig:scan-angles-kappa-eps} 
(a): Query complexity, $N_{\rm queries}$, versus the QSVT approximation error, 
$\epsilon_{\rm qsvt} \in [10^{-3},\dots  10^{-12}]$, for various $\kappa_{\rm qsvt}$. 
(b): $\kappa$ versus the number of spatial points, $N_x$, for the system described in Sec.~\ref{sec:discr}.
}
\end{figure}
% *****************************

Here, we consider the scaling of the QSVT algorithm and compare it with classical methods for the matrix inversion.
Because it is difficult to thoroughly analyze how preconditioning scales with the system size,\cite{Montanaro16} we consider only a conservative scaling of the problem without preconditioning.
However, even in this case, a polynomial speedup of the quantum method is possible for high-dimensional problems.

According to Ref.~\onlinecite{Martyn21}, the query complexity of the QSVT algorithm scales as $\oO(\kappa\ln(\kappa/\epsilon_{\rm qsvt}))$, as also  validated by numerical simulations (Fig.~\ref{fig:scan-angles-kappa-eps}).
Here, we assume that the QSVT angles are computed using the parameter $\kappa_{\rm qsvt}$ close to the matrix condition number, $\kappa_{\rm qsvt} \approx \kappa$.
During each query, the QSVT circuit addresses the block-encoding oracle, which scales as $\oO(\varsigma \ln(N))$, where $N$ is the size of the encoded matrix $A$, and $\varsigma$ is the matrix sparsity.
According to Eq.~\eqref{eq:qsvt-rescaling}, the probability of the zero-ancilla state after inverting a given matrix using the QSVT is $\oO(1/\kappa^2)$.
Therefore, subsequent amplitude estimation, which includes amplitude amplification as a subroutine, requires $\oO(\kappa)$ repetitions of the original QSVT procedure to achieve the probability $\geq 0.5$.
Thus, the algorithm complexity is
\begin{equation}
    \oO\ylb \kappa^2\varsigma\ln(N)\ln\ylb\frac{\kappa}{\epsilon_{\rm qsvt}} \yrb \yrb,
\end{equation}
assuming that the initialization of the starting quantum state is trivial, which is true in our case as explained in Sec.~\ref{sec:init}.

% -------------------------------------------------------------------------------------
% --- Scaling of the algorithm for modeling classical waves ---
% -------------------------------------------------------------------------------------
\subsection{Scaling of the algorithm for modeling classical waves}

In this paper, we are focusing on modeling wave dynamics and consider measurements relevant to this problem.
Usually, in such problems, the spatial distribution of the corresponding fields is discretized by a finite-difference scheme or a finite-element method (FEM) over a grid with $N_x^D$ points (assuming the resolution is the same along all axes), where $D$ is the number of spatial dimensions, and $N_x$ is the number of points in the spatial grid along a single axis.
This distribution is then encoded into $N_x^D N_{\rm fields}$ complex amplitudes of the state-vector elements (Sec.~\ref{sec:encoding}), where $N_{\rm fields}$ is the number of simulated variables (we assume $N_{\rm fields} = 1$).

If one measures, for example, the wave-number spectrum in a spatial domain with $N_x$ points by applying the quantum Fourier transform, one first needs to obtain the state projection where this spatial domain is encoded.
The probability of this state equals the probability to extract $N_x$ elements from the set with $N_x^D$ elements (assuming that the field amplitudes are comparable at all $N_x^D$ spatial points), so it scales as $1/N_x^{D - 1}$.
To increase this probability, the amplitude-amplification technique should be used, which involves $\oO(\sqrt{N_x^{D-1}})$ repetitions of the whole QSVT circuit.
Hence, the whole algorithm scales as
\begin{equation}\label{eq:qsvt-scaling-1}
    \oO\ylb \sqrt{N_x^{D-1}}\kappa^2 \varsigma\ln(N_x^{D})\ln\ylb\frac{\kappa}{\epsilon_{\rm qsvt}}\yrb\yrb.
\end{equation}

To eliminate the factor $\sqrt{N_x^{D-1}}$ from the above equation, one can dilate the system by adding $\eta_{\rm copies} \geq 1$ copies, denoted as  $F_j$, of the field amplitudes $F_{j_0}$ at each position $j_0$ within the spatial domain of interest to the original state vector.
Correspondingly, it is necessary to supplement the original system of equations (as the one presented in Eqs.~\eqref{eq:ampere} and~\eqref{eq:faraday}) by $\eta_{\rm copies}$ copies of the equation $F_j - F_{j_0} = 0$ for $N_x$ positions $j_0$.
By adding $N_{\rm copies} = \eta_{\rm copies} N_x$ additional equations, one increases the desired probability up to $(1 + \eta_{\rm copies})N_x^{1-D}$.
If $\eta_{\rm copies}\sim\oO(N_x^{D-1})$ (in other words, if one doubles the size of the original matrix $A$, $N_x^D \to 2N_x^D$), then the probability becomes $\oO(1)$ and the scaling~\eqref{eq:qsvt-scaling-1} turns into
\begin{equation}\label{eq:qsvt-scaling-2}
    \oO\ylb \kappa^2 \varsigma\ln(N_x^{D})\ln\ylb\frac{\kappa}{\epsilon_{\rm qsvt}}\yrb\yrb.
\end{equation}
However, by adding $\eta_{\rm copies}N_x$ equations we also increase the number of operations in the block-encoding oracle as
\begin{equation}
\begin{split}
    \oO&\ylb\varsigma\ln\ylb N_x^{D}\yrb\yrb \to \oO\ylb\varsigma\ln\ylb 2 N_x^{D}\yrb\yrb. 
\end{split}
\end{equation}
Furthermore, the condition number of the dilated system increases as well.
According to Theorem 3.1 in Ref.~\onlinecite{Bank89}, the condition number of FEM matrices scales as $\oO(N_{\rm full}^{2/D})$ for $N_{\rm full} = N_x^{D}$.
For instance, the condition number of our one-dimensional system described in Sec.~\ref{sec:discr} scales even better, as $\oO(N_x)$, according to the numerical results shown in Fig.~\ref{fig:scan-angles-kappa-eps}.
Thus, by doubling the matrix size, we change the condition number scaling as $\oO(N_x^2)\to\oO(2^{2/D}N_x^2)$.
As a result, one obtains the following scaling of the QSVT for the dilated matrix:
\begin{equation}\label{eq:scaling-qsvt}
    \oO\ylb 2^{4/D} N_x^4 \varsigma\ln\ylb 2 N_x^{D}\yrb \left[\frac{2}{D}\ln2 + \ln\ylb\frac{N_x^2}{\epsilon_{\rm qsvt}}\yrb\right]\yrb.
\end{equation}

% -------------------------------------------------------------------------------------
% --- Comparison with classical methods ---
% -------------------------------------------------------------------------------------
\subsection{Comparison with classical iterative methods}
The QSVT can be compared with the best-known conjugate-gradient-based classical iterative methods for the inversion of sparse matrices.
These methods\cite{Saad03}, such as BiCGSTAB\cite{Vorst92}, GMRES\cite{Saad86} and TFQMR\cite{Freund93}, which work with non-symmetric matrices, generally require preconditioning to converge.
Here, we assume that they scale at least as the conjugate gradient method (which works only with symmetric matrices).
As shown in Ref.~\onlinecite{Shewchuk94}, the number of iterations in this algorithm scales as $\oO(\kappa \ln(1/\epsilon_{\rm qsvt}))$, and each iteration, where the sparse matrix-vector multiplication is the main operation, scales as $\oO(\varsigma N_x^{D})$.
Thus, the scaling of the classical iterative method is
\begin{equation}\label{eq:gradient-method-scaling}
    \oO\ylb\varsigma N_x^{D}\kappa \ln(1/\epsilon_{\rm qsvt})\yrb.
\end{equation}
Here, we take the resulting absolute error in the classically calculated signal to be $\epsilon_{\rm qsvt}$.
Taking into account the dependence of the condition number on the matrix size, $\kappa\sim\oO(N_x^2)$, the classical scaling becomes
\begin{equation}\label{eq:scaling-classical}
\oO\ylb\varsigma N_x^{D + 2} \ln(1/\epsilon_{\rm qsvt})\yrb.
\end{equation}
By comparing Eqs.~\eqref{eq:scaling-qsvt} and~\eqref{eq:scaling-classical}, one can see that the QSVT becomes more efficient than the classical methods in high-dimensional problems, namely, those with $D \geq 3$. 
Examples of such (linear) problems include modeling of the wave propagation in plasma that is described hydrodynamically in three spatial dimensions or kinetically in six phase-space dimensions. 

The QSVT scaling may be improved by reducing the condition number by means of preconditioning like that described in  Ref.~\onlinecite{Clader13}.
We also stress that the QSVT algorithm currently has a bottleneck in that there are no optimized methods for calculating the rotation angles $\phi_k$ (used in Eq.~\eqref{eq:qsvt-odd}) for large condition numbers. 
The latter problem is solved for our one-dimensional system, where $\kappa\sim 500$, by using the GPU-parallelized Fourier approach described in Sec.~\ref{sec:qsvt-angles}.
For more complex problems, one may have to consider the method for computing $\phi_k$ that was proposed in Ref.~\onlinecite{Ying22}.

\section{Theoretical model}\label{sec:theoretical-model}

% -----------------------------------------------------------------------------------
% --- Problem specification ---
% -----------------------------------------------------------------------------------
\subsection{Problem specification}

We consider EM waves in a linear medium with a dielectric permittivity $\epsilon$.
For simplicity, we assume the magnetic permittivity equal to unity as is the case, for example, in classical plasmas. 
Then, Maxwell's equations governing the waves can be written as
\begin{subequations}\label{sys:maxwell-gen}
\begin{eqnarray}
&&\epsilon\partial_t \ybs{E} = \ybs{\nabla}\times\ybs{B},\\
&&\partial_t \ybs{B} = - \ybs{\nabla}\times\ybs{E},
\end{eqnarray}
\end{subequations}
assuming units such that the speed of light equals unity.
% assuming the units, where the light speed $c$ equals $1$.
We consider the simplest model that allows investigating various measurement techniques of the wave spectrum and wave energy.
Specifically, $\epsilon$ will be assumed a piecewise-constant function of the spatial coordinates, and we will be interested in modeling the wave propagation across discontinuities of this function. 
In the absence of surface current and charge densities, the following boundary conditions are satisfied on each discontinuity of $\varepsilon$:
\begin{subequations}
\begin{eqnarray}
&&\ybs{E}_{I,t} - \ybs{E}_{II,t} = 0,\label{eq:int-Et}\\
&&\ybs{B}_{I,t} - \ybs{B}_{II,t} = 0,\label{eq:int-Bt}\\
&&\epsilon_I\ybs{E}_{I,n} - \epsilon_{II}\ybs{E}_{II,n} = 0,\\
&&\ybs{B}_{I,n} - \ybs{B}_{II,n} = 0, 
\end{eqnarray}
\end{subequations}
where $\ybs{E}_t$ and $\ybs{B}_t$ are the fields tangent to the interface, $\ybs{E}_n$ and $\ybs{B}_n$ are the fields normal to the interface.

% This is a simplest model of an inhomogeneous medium, where one can investigate the measurement techniques of the wave spectrum and wave energy.
% We will be interested in modeling waves propagating across abrupt variations of $\epsilon$. 
% On a boundary separating regions I and II with different $\epsilon$ and in the absence of surface current and charge densities, the following boundary conditions should be applied:
% \begin{subequations}
% \begin{eqnarray}
% &&\ybs{E_{I,t}} - \ybs{E_{II,t}} = 0,\label{eq:int-Et}\\
% &&\ybs{B_{I,t}} - \ybs{B_{II,t}} = 0,\label{eq:int-Bt}\\
% &&\epsilon_I\ybs{E_{I,n}} - \epsilon_{II}\ybs{E_{II,n}} = 0,\\
% &&\ybs{B_{I,n}} - \ybs{B_{II,n}} = 0, 
% \end{eqnarray}
% \end{subequations}
% where $\ybs{E_t}$ and $\ybs{B_t}$ are the fields tangent to the surface, $\ybs{E_n}$ and $\ybs{B_n}$ are the fields normal to the interface.

% \begin{subequations}
% \begin{eqnarray}
% &&\ybs{E_{I,t}} - \ybs{E_{II,t}} = 0,\label{eq:int-Et}\\
% &&\ybs{B_{I,t}} - \ybs{B_{II,t}} = \ybs{J_{\rm surf}}\times\ybs{r_n},\label{eq:int-Bt}\\
% &&\epsilon_I\ybs{E_{I,n}} - \epsilon_{II}\ybs{E_{II,n}} = \rho_{\rm surf},\\
% &&\ybs{B_{I,n}} - \ybs{B_{II,n}} = 0, 
% \end{eqnarray}
% \end{subequations}
% where $\ybs{r_n}$ is the unit vector normal to the boundary; $\ybs{E_t}$ and $\ybs{B_t}$ are the fields tangent to the surface, $\ybs{E_n}$ and $\ybs{B_n}$ are the fields normal to the interface; $\ybs{J_{\rm surf}}$ and $\rho_{\rm surf}$ are the surface current and charge densities, correspondingly, assumed to be $0$.

We consider a one-dimensional problem where waves propagate along the $x$ axis. 
The magnetic and electric fields are polarized along the $z$ and $y$ axes, respectively.
Hence, Eqs.~\eqref{sys:maxwell-gen} can be expressed in a simple form:
\begin{subequations}\label{sys:EyBz}
\begin{eqnarray}
&&\epsilon\partial_t E = - \partial_x B,\\
&&\partial_t B = - \partial_x E,
\end{eqnarray}
\end{subequations}
where we omit the subindices $y$ and $z$ in $E_y$ and $B_z$.
Also, we introduce the spatial coordinate $r_x$ that changes from $0$ to $1$ and is defined as 
\begin{equation}\label{eq:coord-rx}
    r_x = x / \max(|x|).
\end{equation}
We assume a discontinuity of $\epsilon$ at $r_{x, \rm int} = 1/2$, so that we have $N_{\rm layers} = 2$ dielectric layers of different permittivities:
\begin{equation}\label{eq:epsilon-rint}
\epsilon
    = \left\{
    \begin{array}{cc}
        \epsilon_0, & r_x < r_{x,\rm int},\\
        \epsilon_1, & r_x > r_{x,\rm int}.
    \end{array}
    \right.
\end{equation}

A monochromatic source $Q=Q_0\exp(\yi\omega t)$ of the frequency $\omega$ and amplitude $Q_0$ is placed at $r_x = 1$. 
The corresponding boundary-value problem, where we take $E,B\sim\exp(\yi\omega t)$, results in the following equations:
\begin{subequations}\label{sys:model}
\begin{eqnarray}
&&\yi\omega\epsilon_L E(x) + \partial_x B(x) = 0,\label{eq:model-E}\\
&&\yi\omega B(x) + \partial_x E(x) = 0,\label{eq:model-B}
\end{eqnarray}
\end{subequations}
with outgoing boundary conditions at $r_x = 0$ and $r_x = 1$:
\begin{subequations}\label{eq:model-boundaries}
\begin{eqnarray}
&&(\yi\omega - \partial_x) E\big|_{r_x = 0} = 0,\label{eq:left-cond}\\
&&(\yi\omega + \partial_x) B\big|_{r_x = 1} = Q_0,\label{eq:right-cond}
\end{eqnarray}
\end{subequations}
and
\begin{subequations}
\begin{eqnarray}
&&E(r_{x,\rm int})\big|_{{\rm layer}\ 0} = E(r_{x,\rm int})\big|_{{\rm layer}\ 1},\\
&&B(r_{x,\rm int})\big|_{{\rm layer}\ 0} = B(r_{x,\rm int})\big|_{{\rm layer}\ 1},
\end{eqnarray}
\end{subequations}
with $L = [0,1]$.
% results in the following equations:
% \begin{subequations}\label{sys:model}
% \begin{eqnarray}
% &&\yi\omega\epsilon_L E(x) + \partial_x B(x) = 0,\\
% &&\yi\omega B(x) + \partial_x E(x) = 0,\\
% &&E(r_{x,\rm int})\big|_{{\rm layer}\ 0} = E(r_{x,\rm int})\big|_{{\rm layer}\ 1},\\
% &&B(r_{x,\rm int})\big|_{{\rm layer}\ 0} = B(r_{x,\rm int})\big|_{{\rm layer}\ 1},\\
% &&(\yi\omega - \partial_x) E\big|_{r_x = 0} = 0,\label{eq:left-cond}\\
% &&(\yi\omega + \partial_x) B\big|_{r_x = 1} = Q_0,\label{eq:right-cond}
% \end{eqnarray}
% \end{subequations}
% with $L = [0,1]$.
(Here, the square brackets $[i_{\rm left}, i_{\rm right}]$ indicate a set of all integers from $i_{\rm left}$ to $i_{\rm right}$, including $i_{\rm left}$ and $i_{\rm right}$.
To denote the same set but excluding $i_{\rm right}$, we use the parenthesis as in $[i_{\rm left}, i_{\rm right})$.
The same notation will also be used to indicate open, $(x_{\rm left}, x_{\rm right})$, and closed, $[x_{\rm left}, x_{\rm right}]$, continuous intervals with some rational numbers $x_{\rm left}$ and $x_{\rm right}$.)
% The spatial coordinate $x$ is normalized to $k_{x,0}^{-1}$ where $k_{x,0} = \omega/c$, the frequency is normalized to $\omega$, and we keep the same notations for the normalized variables.
% (Although $\omega = 1$ in the normalized units, we keep $\omega$ explicitly in Eqs.~\eqref{sys:model} to track its influence on the block-encoding of the modeled system.)
% According to Eq.~\eqref{eq:right-cond}, on the right boundary, the right-propagating wave escape the simulated domain, and the source $Q$ oscillating with the constant amplitude $Q_0$ excites a left-propagating wave moving inside the domain.
% Thus, the system becomes non-Hermitian because of the constant energy pumping by the source $Q$ and due to the dissipation of the wave energy through the open boundaries.
Note that the system is not conservative in that the wave energy is lost through radiation on the left and right boundaries and replenished by the source $Q$. 
In the left half of the spatial domain, where $r_x < r_{x,\rm int}$, the EM field is a left-propagating wave. 
% In the right half, both left- and right-propagating waves are present due to the reflection at the interface between the dielectric layers.
In the right half, the reflection at the interface between the dielectric layers results in the interference of left- and right-propagating waves and in the formation of a standing wave.
% \ycb{The interference of these waves results in a standing wave, and the magnitudes of the wave real and imaginary components depend on the position of the discontinuity and the permittivities in different layers.}
% The oscillation magnitudes of the real and imaginary components of the resulting standing wave depend on the These waves interfere with each other}
% \ycb{These waves interfere with each other and the oscillation amplitudes Depending on the position of the discontinuity of $\epsilon$ and the values of the permittivities in different layers, the waves the Due to the interference of these waves, }
% \ycr{
% The electric fields of the incident and reflected waves are related by the expression
% \begin{equation}\label{eq:Eref-Einc}
%     E_{\rm ref} = E_{\rm inc} \frac{\sqrt{\epsilon_0} -\sqrt{\epsilon_1}}{\sqrt{\epsilon_0} + \sqrt{\epsilon_1}}.
% \end{equation}
% If $\epsilon_0 < \epsilon_1$, there is a phase shift of $\pi$ between $E_{\rm ref}$ and $E_{\rm inc}$ resulting in the destructive interference.
% Comment: to compare the theory with simulations in Figs.~\eqref{fig:comp-CL-QC-case2}, it is also necessary to take into account the position of the $\epsilon_0-\epsilon_1$ interface and the initial phase (the wave phase on the right edge).
% Since we already have comparison between classical and quantum simulations, i would skip the comparison with the theory.
% }
Also, the wave propagating within the layer with the permittivity $\epsilon$ has a wave number equal to 
\begin{equation}\label{eq:kx-epsilon}
    k_x = \sqrt{\epsilon}\omega.
\end{equation}

Below, we show how to develop a QSVT algorithm for this boundary-value problem and perform quantum measurements in the corresponding circuit to infer information about the spatial spectrum and the wave energy.
% We also show that our quantum scheme produces results that are in agreements with Eqs.~\eqref{eq:Eref-Einc} and~\eqref{eq:kx-epsilon}.

% We will see how the quantum scheme to be developed reproduces these results via a direct calculation.
% Specifically, in this model, a wave excited by $Q$ at $r_x=1$ propagates to the left and falls at the interface between the dielectric layers.
% The electric fields of the incident and reflected waves are related by the expression
% \begin{equation}
%     E_{\rm ref} = E_{\rm inc} \frac{\sqrt{\epsilon_0} -\sqrt{\epsilon_1}}{\sqrt{\epsilon_0} + \sqrt{\epsilon_1}}.
% \end{equation}
% If $\epsilon_0 < \epsilon_1$, there is a phase shift of $\pi$ between $E_{\rm ref}$ and $E_{\rm inc}$ resulting in the destructive interference.
% Apart from that, a wave propagating within the layer with the permittivity $\epsilon$ has a wave number equal to 
% \begin{equation}\label{eq:kx-epsilon}
%     k_x = \sqrt{\epsilon}\omega.
% \end{equation}
% We will see how the quantum scheme to be developed reproduces these results via a direct calculation.

% *****************************
% *** x-grid figure ***
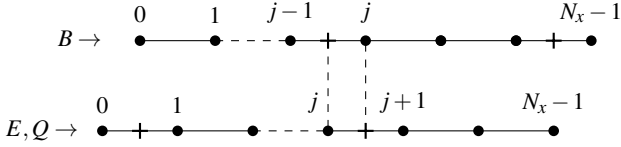
\begin{figure}[!t]
\centering
\begin{tikzpicture}[
        dot/.style = {circle, fill=black, inner sep=0pt, minimum size=4pt},
        every label/.append style = {inner sep=4pt,},
        thick
    ]
        \newcommand{\yshf}{-5.4}
        \newcommand{\yshs}{-5.9}
        \newcommand{\yhh}{1.0}
        \newcommand{\ygap}{1.0}
        \newcommand{\yvsh}{1.2}
        
        % --- magnetic field grid ---
        % \node[label={$B,Q\rightarrow$}] at (\yshf-2*\yhh,\yvsh-0.4) {};
        \node[label={$B\rightarrow$}] at (\yshf-0.8*\yhh,\yvsh-0.35) {};
        
        \draw[thin] (\yshf,\yvsh) -- (\yshf+\yhh,\yvsh);
        \draw[thin,dashed] (\yshf+\yhh,\yvsh) -- (\yshf+\yhh+\ygap,\yvsh);
        \draw[thin] (\yshf+\yhh+\ygap,\yvsh) -- (\yshf+5*\yhh+\ygap,\yvsh);
        
        \node[dot,label={$0$}] at (\yshf,\yvsh) {};
        \node[dot,label={$1$}] at (\yshf+\yhh,\yvsh) {};
        \node[dot,label={$j-1$}] at (\yshf+\yhh+\ygap,\yvsh) {};
        \node[cross] at (\yshf+1.5*\yhh+\ygap,\yvsh) {};
        \draw[thin,dashed] (\yshf+1.5*\yhh+\ygap,\yvsh) -- (\yshs+2*\yhh+\ygap,0); % vertical line
        \node[dot,label={$j$}] at (\yshf+2*\yhh+\ygap,\yvsh) {};
        \node[dot] at (\yshf+3*\yhh+\ygap,\yvsh) {};
        \node[dot] at (\yshf+4*\yhh+\ygap,\yvsh) {};
        \node[cross] at (\yshf+4.5*\yhh+\ygap,\yvsh) {};
        \node[dot,label={$N_x - 1$}] at (\yshf+5*\yhh+\ygap,\yvsh) {};
        
        % \node[dot,label={[shift={(-0.2, 0.0)}] below:$s=-1+h$}] at (\yshf,\yvsh) {};
        \node[dot] at (\yshf+3*\yhh+\ygap,\yvsh) {};
        \node[dot] at (\yshf+4*\yhh+\ygap,\yvsh) {};
        % \node[dot,label={below:$1+h$}] at (\yshf+5*\yhh+\ygap,\yvsh) {};
        
        % --- electric field grid ---
        % \node[label={$E\rightarrow$}] at (\yshs-2*\yhh,-0.4) {};
         \node[label={$E,Q\rightarrow$}] at (\yshs-0.8*\yhh,-0.4) {};
        
        \draw[thin] (\yshs,0) -- (\yshs+2*\yhh,0);
        \draw[thin,dashed] (\yshs+2*\yhh,0) -- (\yshs+2*\yhh+\ygap,0);
        \draw[thin] (\yshs+2*\yhh+\ygap,0) -- (\yshs+5*\yhh+\ygap,0);
        
        \node[dot,label={$0$}] at (\yshs,0) {};
        \node[cross] at (\yshs+0.5*\yhh,0) {};
        \node[dot,label={$1$}] at (\yshs+\yhh,0) {};
        \node[dot] at (\yshs+2*\yhh,0) {};
        \node[dot,label={[shift={(-0.2, 0.0)}] $j$}] at (\yshs+2*\yhh+\ygap,0) {};
        \node[cross] at (\yshs+2.5*\yhh+\ygap,0) {};
        \draw[thin,dashed] (\yshs+2.5*\yhh+\ygap,0) -- (\yshf+2*\yhh+\ygap,\yvsh); % vertical line
        \node[dot,label={$j+1$}] at (\yshs+3*\yhh+\ygap,0) {};
        \node[dot] at (\yshs+4*\yhh+\ygap,0) {};
        \node[dot,label={$N_x - 1$}] at (\yshs+5*\yhh+\ygap,0) {};
        
        % \node[dot,label={[shift={(-0.2, 0.0)}] below:$s=-1$}] at (\yshs,0) {};
        % \node[dot,label={below:$x_{q_1}$}] at (\yshs+2*\yhh+\ygap,0) {};
        % \node[dot,label={below:$x_{q_2}$}] at (\yshs+3*\yhh+\ygap,0) {};
        % \node[dot,label={below:$1$}] at (\yshs+5*\yhh+\ygap,0) {};

\end{tikzpicture}
\caption{\label{fig:x-grid} 
The spatial domain is discretized with two staggered grids. 
Each grid has $N_x$ points, and the spatial step $h$ is defined in Eq.~\eqref{eq:sigma-h}.
The magnetic field is defined on the upper grid shifted by $h$ to the right from the lower grid, on which the electric field is defined.
The source $Q$ is placed at the right boundary of the lower grid.
The outgoing boundary conditions~\eqref{eq:model-boundaries} on $E$ and $B$ are calculated in the middle of the leftmost cell of the lower grid and of the rightmost cell of the upper grid, respectively.}
\end{figure}
% *****************************

% -----------------------------------------------------------------------------------
% --- Discretization ---
% -----------------------------------------------------------------------------------
\subsection{Discretization}\label{sec:discr}

We discretize Eqs.~\eqref{sys:model} using the central finite difference scheme as shown in Fig.~\ref{fig:x-grid}.
Separate staggered grids, each having $N_x = 2^{n_x}$ points, are used for $E$ and $B$, and each dielectric layer contains $M_x = 2^{n_x - 1}$ spatial points.
At the bulk spatial points, the discretized Eq.~\eqref{eq:model-E} is
\begin{equation}\label{eq:ampere}
\yi\omega\epsilon_L E_j + \sigma (B_j - B_{j-1}) = 0,
\end{equation}
where $j = (M_x L + k)$ with $k = [1,M_x)$ for $L=0$ and $k = [0,M_x)$ for $L = 1$.
Also, $\sigma = (2h)^{-1}$, where $2h$ is the spatial cell size:
\begin{equation}\label{eq:sigma-h}    
2h = x_{j+1}-x_j,
\end{equation}
for all $j$.
% \begin{subequations}\label{sys:discr-Amper}
% \begin{eqnarray}
% &&\yi\omega\epsilon_L E_j + \sigma (B_j - B_{j-1}) = 0,\label{eq:ampere}\\ 
% &&j = (M_x L + k)= [1, N_x),\\
% &&\sigma = (2h)^{-1}, \quad 2h = (x_{j+1}-x_j),\label{eq:sigma-h}
% \end{eqnarray}
% \end{subequations}
% where $L = [0,1]$ and $k = [0,M_x)$.
The discretized Eq.~\eqref{eq:model-B} is
\begin{equation}
\yi\omega B_j + \sigma (E_{j+1} - E_j) = 0, \quad j = [0,N_x-2].\label{eq:faraday}
\end{equation}
The EM fields satisfy the outgoing boundary conditions~\eqref{eq:model-boundaries} expressed as 
\begin{subequations}\label{sys:boundary}
\begin{eqnarray}
&&\eta_{+} E_0 + \eta_{-} E_1 = 0,\\
&&\eta_{-} B_{N_x - 2} + \eta_{+} B_{N_x-1} = Q_0,\\
&&\eta_{-} = \yi\omega - h^{-1},\\
&&\eta_{+} = \yi\omega + h^{-1}.
\end{eqnarray}
\end{subequations}
Equations~\eqref{eq:ampere} and~\eqref{eq:faraday} with the boundary conditions~\eqref{sys:boundary} can be represented as a set of linear equations~\eqref{eq:axb} where the vector $\psi$ stores $N_{\rm vars} = 2$ variables, $E$ and $B$:
\begin{equation}
    \psi_{d N_x + j} = 
        \left\{ \begin{aligned}
        E_j&,\quad d = 0,\\
        B_j&,\quad d = 1,
        \end{aligned} \right.
\end{equation}
with $j = [0,N_x)$.
The right-hand-side vector $b$ in Eq.~\eqref{eq:axb} has dimension $N_{\rm vars} N_x$ and encodes the source with the amplitude $Q_0 = 1$:
\begin{equation}\label{eq:b}
b_{d N_x + j} = 
    \left\{ \begin{aligned}
        1&,\quad (d N_x + j) = 2 N_x - 1,\\
        0&,\quad \text{otherwise},
    \end{aligned} \right. 
\end{equation}
with $j = [0,N_x)$ and $d = [0,1]$.
The $N_{\rm vars}N_x\times N_{\rm vars}N_x$ matrix $A$ is represented as a sum of two matrices:
\begin{equation}\label{eq:A}
    A = A^{\rm bulk} + A^{\rm edge},
\end{equation}
with
\begin{equation}\label{eq:A-bulk}
A^{\rm bulk}_{kj} = 
    \left\{ \begin{aligned}
        (-1)^l \sigma,    \quad &j = k + N_x - l,\ k = [1, N_x),&\\
        (-1)^{l+1} \sigma,\quad &j = k - N_x + l,&\\
                                &k = [N_x, 2N_x),&\\
        \yi\omega,\quad &j=k,\ \ \ k = [N_x,2N_x - 2],&\\
        \yi\omega\epsilon_L,\quad &j=k,\ \ \ k\neq0,&\\
                                  &k=[M_x L, M_x(L+1)-1],&\\
        0,\quad &\text{otherwise},&
    \end{aligned} \right.
\end{equation}
and 
\begin{equation}\label{eq:A-edge}
A^{\rm edge}_{kj} = 
    \left\{ \begin{aligned}
        \eta_{+},\quad &j = k,\ \ \ k = 0 \text{ and } 2N_x-1,&\\
        \eta_{-},\quad &j = k + 1 = 1,& \\
                       &j = k - 1 = 2 N_x - 2,&\\
        0,\quad &\text{otherwise},&
    \end{aligned} \right.
\end{equation}
where $L = [0, 1]$, $l = [0, 1]$, and $k$ is the row index.
The matrix $A^{\rm bulk}$ contains information about the system evolution in the bulk spatial points.
The matrix $A^{\rm edge}$ describes the boundary conditions.

% -----------------------------------------------------------------------------------
% --- Encoding of the system into the quantum circuit ---
% -----------------------------------------------------------------------------------
\section{Encoding of the classical system into a quantum circuit}\label{sec:encoding}

Various formulations suitable for the QC of Maxwell's equations were proposed in Refs.~\onlinecite{Clader13, Vahala20, Ram21, Vahala21, Koukoutsis22}, 
which considered analytical descriptions and the algorithm complexity but not actual circuits. 
Below, we show for the first time how to construct quantum circuits for this problem explicitly.
% where authors stopped on either the analytical description of the problem or on the discussion of the query complexity of their algorithms without considering actual circuit representation of the system matrix.
% In this work, however, we describe quantum circuits for all quantum oracles necessary for the quantum modeling of the system introduced in Sec.~\ref{sec:discr}.

% -----------------------------------------------------------------------------------
% --- Input registers ---
% -----------------------------------------------------------------------------------
\subsection{Input registers}\label{sec:input-registers}
To map the problem on a quantum circuit, we introduce two input registers.
The first one, $r_d$, consists of one qubit and encodes the variable index: $\ket{0}_{r_d}$ for the electric field and $\ket{1}_{r_d}$ for the magnetic field. 
The second register, $r_j$, contains $n_x = \log_2 N_x$ qubits and encodes the coordinate index on the spatial grid.
The state amplitude corresponds to the field magnitude at a given spatial point.
For instance, the amplitude of the state $\ket{1}_{r_d}\ket{j}_{r_j}$ stores the magnitude of the magnetic field at $x_j$, where $x_j$ is taken on the upper spatial grid in Fig.~\ref{fig:x-grid}.
All indices are numbered from $0$.

% -----------------------------------------------------------------------------------
% --- Initialization ---
% -----------------------------------------------------------------------------------
\subsection{Initialization}\label{sec:init}
To solve Eq.~\eqref{eq:axb} with the matrix~\eqref{eq:A} and the source~\eqref{eq:b} using the QSVT, one needs to block-encode the matrix $A$ into the unitary $U_A$ (Eq.~\eqref{eq:UA}) and encode the vector $b$ as the initial state, denoted as $\ket{b}$.
At the beginning, all qubits are initialized in the zero state.
Using the registers $r_j$ and $r_d$, the right-hand-side vector can be encoded as
\begin{equation}\label{eq:b-enc}
\ket{b} = \sum_{k=0}^{N_x - 1} \ylb
\alpha_k^{(E)} \ket{0}_{r_d} \ket{k}_{r_j}
+
\alpha_k^{(B)} \ket{1}_{r_d} \ket{k}_{r_j}\yrb.
\end{equation}
The coefficients $\alpha_k^{(E, B)}$ are determined from Eq.~\eqref{eq:b}, whence $\alpha_k^{(E)} = 0$ for all $k$ and $\alpha_k^{(B)} = \delta_{k,N_x - 1}$. 
Because the bit string encoding the number $N_x-1$ consists only of units, the state $\ket{b}$ can be initialized by applying the $X$ gate to each qubit in the register $r_j$. Also, the $X$ gate is applied to the qubit $r_d$.
Importantly, the depth of the corresponding initialization circuit, henceforth denoted `INIT', does not depend on $N_x$.

% All elements of the vector $b$ are equal to zero except the element at $j=2N_x-1$, which equals $1$.
% In terms of the introduced registers $r_j$ and $r_d$, this is equivalent to the state $\ket{1}_{r_d}\ket{N_x-1}_{r_j}$.
% So the initialization can be done simply by applying the Pauli $X$ gate to each qubit in the register $r_j$ and to the qubit $r_d$.
% Throughout the paper, the initialization circuit is denoted as `INIT'.
% It should be emphasized that the depth of the initialization circuit does not depend on $N_x$ that is a huge benefit of the model.

% -----------------------------------------------------------------------------------
% --- Encoding of a non-Hermitian matrix ---
% -----------------------------------------------------------------------------------
\subsection{Block encoding of a non-Hermitian matrix}\label{sec:enc-nonH}
To block-encode the matrix $A$, one should normalize it first according to Eq.~\eqref{eq:gen-A-norm}.
Here, we normalize $A$ in the following way:
\begin{equation}\label{eq:norm-of-A}
    A \to A/\ylb d_H^2 ||A||_{\rm max}\yrb.
\end{equation}
The origin of the factor $d_H = 2$, where $d_H^2$ is close to the actual sparsity of $A$, is explained in Sec.~\ref{sec:oracle-structure}.
% later by the action of Hadamard gates used for encoding the correct number of nonzero elements at a given matrix row.

To block-encode a non-Hermitian matrix, one usually extends it first to a Hermitian one:\cite{HHL09, Clader13, Jin22}
\begin{equation}\label{eq:extension}
    A_{\rm ext} = 
    \begin{pmatrix}
    0 & A \\
    A^\dagger & 0
    \end{pmatrix}.
\end{equation}
After that, one can use the standard state-preparation technique\cite{Berry12, Novikau22} to block-encode $A_{\rm ext}$, where the oracle $U_A$ is represented through unitary operators $O_F$, $O_H$, $O_M$: 
\begin{equation}\label{eq:state-prep-stand}
    U_A = O_F^\dagger O_M O_H O_F.
\end{equation}
The circuit implementing $U_A$ operates with `input' qubit registers (in our case, registers $r_j$ and $r_d$) that contain initial and final (output) data, and also ancilla registers that are initialized in the zero states and used for intermediate computations. 
The operator $O_F$ reads a row index from the input registers, calculates the column indices of all nonzero elements of $A_{\rm ext}$ at the given row and writes the computed indices into the ancillae.
The operator $O_H$ computes the values of these nonzero elements and encodes them into the amplitudes of the ancilla-qubits' states. 
The operator $O_M$ transfers the computed column indices to the input registers so that $U_A$ could return the indices as an output.
After that, $O_F^\dagger$ uncomputes (sets back to zero) the used ancilla registers for the state encoding the matrix nonzero elements.

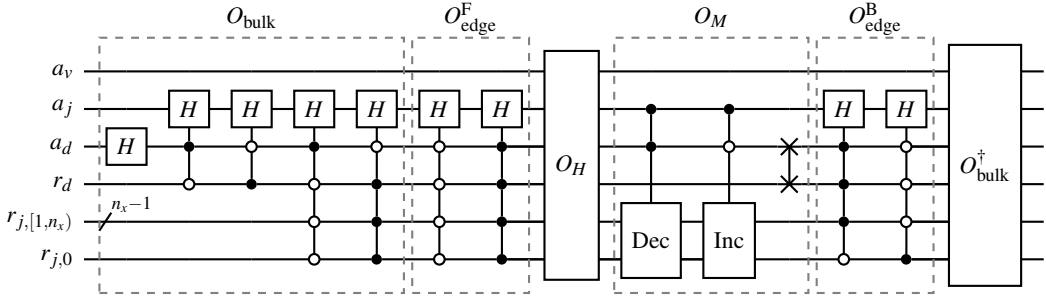
\begin{figure*}[t!]
\centering
\begin{quantikz}[row sep={0.5cm,between origins},column sep={0.3cm}]
\lstick{$a_v$}          &\qw\yggr{6}{5}{$O_{\rm bulk}$}{gray}{-0.5}&\qw&\qw&\qw&\qw
                            &\qw\yggr{6}{2}{$O^{\rm F}_{\rm edge}$}{gray}{-0.5}&\qw
                            &\gate[6]{O_H}
                            &\qw\yggr{6}{3}{$O_M$}{gray}{-0.5}&\qw&\qw
                            &\qw\yggr{6}{2}{$O^{\rm B}_{\rm edge}$}{gray}{-0.5}&\qw
                            &\gate[6]{O^\dagger_{\rm bulk}}&\qw\\
\lstick{$a_j$}          &\qw             &\gate{H}  &\gate{H}  &\gate{H}  &\gate{H}  &\gate{H}  &\gate{H} &\qw&\ctrl{1}           &\ctrl{1}           &\qw     &\gate{H}  &\gate{H}  &\qw&\qw\\
\lstick{$a_d$}          &\gate{H}        &\ctrl{-1} &\octrl{-1}&\ctrl{-1} &\octrl{-1}&\octrl{-1}&\ctrl{-1}&\qw&\ctrl{2}           &\octrl{2}          &\swap{1}&\ctrl{-1} &\octrl{-1}&\qw&\qw\\
\lstick{$r_d$}            &\qw             &\octrl{-1}&\ctrl{-1} &\octrl{-1}&\ctrl{-1} &\octrl{-1}&\ctrl{-1}&\qw&\qw                &\qw                &\targX{}&\ctrl{-1} &\octrl{-1}&\qw&\qw\\
\lstick{$r_{j,[1,n_x)}$}&\qwbundle{n_x-1}&\qw       &\qw       &\octrl{-1}&\ctrl{-1} &\octrl{-1}&\ctrl{-1}&\qw&\gate[2]{{\rm Dec}}&\gate[2]{{\rm Inc}}&\qw     &\ctrl{-1} &\octrl{-1}&\qw&\qw\\
\lstick{$r_{j,0}$}          &\qw             &\qw       &\qw       &\octrl{-1}&\ctrl{-1} &\octrl{-1}&\ctrl{-1}&\qw&\qw                &\qw                &\qw     &\octrl{-1}&\ctrl{-1} &\qw&\qw
\end{quantikz}
\caption{\label{circ:UA} 
The circuit implementing the oracle $U_A$ described in Eq.~\eqref{eq:BE-mod-state-prep}.
The circuit for the operator $O_H$ is shown in Fig.~\ref{circ:OH}.
The decrementor (Dec) and incrementor (Inc) are described in Eqs.~\eqref{eq:dec-inc}, and their circuits can be found in Ref.~\onlinecite{Novikau22}.
The ancillae $a_d$, $a_j$ and $a_v$ are initialized in the zero state and described in Sec.~\ref{sec:ancillae}.
}
\end{figure*}

In our implementation, we amend the above technique.
In order to shorten the circuit for $U_A$, we avoid extending $A$ to a Hermitian matrix.
Instead, we split $O_F$ into two oracles $O_{\rm bulk}$ and $O^{\rm F}_{\rm edge}$, and the adjoint $O_F^\dagger$ is replaced with the product of $O^\dagger_{\rm bulk}$ and $O^{\rm B}_{\rm edge}$. 
The index F in $O^{\rm F}_{\rm edge}$ stands for `forward', and the index 'B' in $O^{\rm B}_{\rm edge}$ stands for 'backward'.
The pair of oracles $O^{\rm F}_{\rm edge}$ and $O^{\rm B}_{\rm edge}$ encodes the locations of the elements $\eta_{\pm}$ in Eq.~\eqref{eq:A-edge}.
The oracles $O_{\rm bulk}$ and $O^\dagger_{\rm bulk}$ encode the positions of the matrix elements in Eq.~\eqref{eq:A-bulk}.
Hence, the decomposition~\eqref{eq:state-prep-stand} is replaced with
\begin{equation}\label{eq:BE-mod-state-prep}
    U_A = O^\dagger_{\rm bulk} O^{\rm B}_{\rm edge} O_M O_H O^{\rm F}_{\rm edge}O_{\rm bulk}.
\end{equation}
Note that for a Hermitian $A$, one has $O_{\rm edge}^{\rm B} = (O_{\rm edge}^{\rm F})^\dagger$, but generally this is not the case (Sec.~\ref{sec:oracle-structure}). 
% where $O_F$ is split into two oracles $O_{\rm bulk}$ and $O^{\rm F}_{\rm edge}$, where the last one is formally called `push-forward` oracle.
% The adjoint operator $O_F^\dagger$ is split into $O^\dagger_{\rm bulk}$ and $O^{\rm B}_{\rm edge}$, and the latter is called `pull-backward` oracle.
% The pair of oracles $O^{\rm F}_{\rm edge}$ and $O^{\rm B}_{\rm edge}$ encodes the location of the elements $\eta_{\pm}$ in Eq.~\eqref{eq:A-edge}.
% The oracles $O_{\rm bulk}$ and $O^\dagger_{\rm bulk}$ encode the positions of the matrix elements in Eq.~\eqref{eq:A-bulk}.
The decomposition~\eqref{eq:BE-mod-state-prep} eliminates the need for the supplemental ancilla used for the extension~\eqref{eq:extension}, and one can also avoid additional gates needed for the block-encoding of $A^\dagger$ in Eq.~\eqref{eq:extension}.
The circuit for $U_A$ is shown in Fig.~\ref{circ:UA}. 
To describe the action of each operator in the above product, we need to consider the ancilla qubits used in the intermediate computations.

% ******************************

% ******************************
% *** OH circuit ***
\begin{figure}[b!]
\centering
\begin{quantikz}[row sep={0.4cm,between origins},column sep={0.2cm}]
\lstick{$a_v$} &\qw           &\gate[5]{O_{H,\omega}}&\gate[5]{O_{H,\eta}}&\gate[5]{O_{H,\sigma}}&\gate{R_y(\theta_\pi)} &\gate{R_y(\theta_{-\pi})}&\gate{X}&\qw\\
\lstick{$a_j$} &\qw           &\qw                   &\qw                 &\qw                   &\qw                    &\qw                      &\qw     &\qw\\
\lstick{$a_d$} &\qw           &\qw                   &\qw                 &\qw                   &\ctrl{-2}              &\octrl{-2}               &\qw     &\qw\\
\lstick{$r_d$} &\qw           &\qw                   &\qw                 &\qw                   &\octrl{-1}             &\ctrl{-1}                &\qw     &\qw\\
\lstick{$r_j$} &\qwbundle{n_x}&\qw                   &\qw                 &\qw                   &\octrl{-1}             &\ctrl{-1}                &\qw     &\qw
\end{quantikz}\caption{\label{circ:OH} 
The circuit implementing the operator $O_H$.
The subcircuits are described in Fig.~\ref{circ:OH-parts}.
% The values $A_{kj}$ from Eq.~\eqref{eq:A} are encoded into the amplitudes of the states $\ket{1}_{a_v}\ket{j}_{a_j}\ket{k}_{r_j}$.
% To keep the final conditioning on zero ancillae (as discussed in Eq.~\eqref{eq:UA}), the Pauli $X$ gate is applied at the end of the circuit.
The rotation angles $\theta_{\pm\pi}$ are computed using Eqs.~\eqref{sys:angles-OH}.
}
\end{figure}
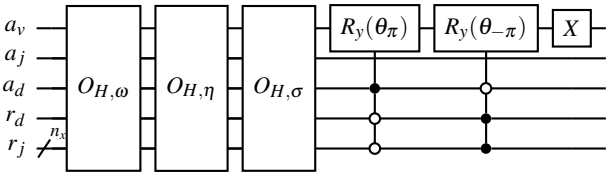
% ******************************

% ******************************
% *** OH parts circuit ***
\begin{figure*}[t!]
\centering
\begin{quantikz}[row sep={0.4cm,between origins},column sep={0.2cm}]
\lstick{$a_v$}          &\gate{R_x(\theta_{\omega,\epsilon_0})}\yggr{7}{7}{$O_{H,\omega}$}{gray}{-0.05} 
                            &\gate{R_x(\theta_{\omega,\epsilon_1})}&\gate{R_x(\theta_\omega)}
                            &\gate{R_x^\dagger(\theta_{\omega,\epsilon_0})}&\gate{R_x^\dagger(\theta_\omega)}
                            &\gate{R_x(\theta_{\omega,\epsilon_0,\rm e})}&\gate{R_x(\theta_{\omega,\rm e})}&\qw\\
\lstick{$a_j$}            &\qw         &\qw            &\qw               &\qw                       &\qw                       &\qw                 &\qw                 &\qw\\
\lstick{$a_d$}            &\octrl{-2}  &\octrl{-2}     &\ctrl{-2}         &\octrl{-2}                &\ctrl{-2}                 &\octrl{-2}          &\ctrl{-2}           &\qw\\
\lstick{$r_d$}            &\octrl{-1}  &\octrl{-1}     &\ctrl{-1}         &\octrl{-1}                &\ctrl{-1}                 &\octrl{-1}          &\ctrl{-1}           &\qw\\
\lstick{$r_{j,n_x-1}$}    &\octrl{-1}  &\ctrl{-1}      &\qw               &\octrl{-1}                &\ctrl{-1}                 &\octrl{-1}          &\ctrl{-1}           &\qw\\
\lstick{$r_{j,[1,n_x-2]}$}&\qwbundle{} &\qw            &\qw               &\octrl{-1}                &\ctrl{-1}                 &\octrl{-1}          &\ctrl{-1}           &\qw\\
\lstick{$r_{j,0}$}        &\qw         &\qw            &\qw               &\qw                       &\qw                       &\ctrl{-1}           &\octrl{-1}          &\qw
\end{quantikz}
\\
\begin{quantikz}[row sep={0.4cm,between origins},column sep={0.2cm}]
\lstick{$a_v$}          &\gate{R_c(\eta_{+})}\yggr{6}{4}{$O_{H,\eta}$}{gray}{-0.5} 
                                                 &\gate{R_c(\eta_-)}&\gate{R_c(\eta_+)}&\gate{R_c(\eta_-)}
                                                                                                          &\gate{R_y(\theta_\sigma)}\yggr{6}{6}{$O_{H,\sigma}$}{gray}{-0.5} 
                                                                                                                      &\gate{R_y(\theta_{-\sigma})}
                                                                                                                                 &\gate{R_y(\theta_{\sigma,\rm e})}  
                                                                                                                                                &\gate{R_y(\theta_\sigma)}
                                                                                                                                                             &\gate{R_y(\theta_{-\sigma})} 
                                                                                                                                                                        &\gate{R_y(\theta_{-\sigma,\rm e})} &\qw\\
\lstick{$a_j$}            &\octrl{-1}            &\ctrl{-1}         &\octrl{-1}        &\ctrl{-1}         &\octrl{-1} &\ctrl{-1} &\octrl{-1}     &\ctrl{-1}  &\octrl{-1}&\octrl{-1}                     &\qw\\
\lstick{$a_d$}            &\octrl{-1}            &\octrl{-1}        &\ctrl{-1}         &\ctrl{-1}         &\ctrl{-1}  &\ctrl{-1} &\ctrl{-1}      &\octrl{-1} &\octrl{-1}&\octrl{-1}                     &\qw\\
\lstick{$r_d$}            &\octrl{-1}            &\octrl{-1}        &\ctrl{-1}         &\ctrl{-1}         &\octrl{-1} &\octrl{-1}&\octrl{-1}     &\ctrl{-1}  &\ctrl{-1} &\ctrl{-1}                      &\qw\\
\lstick{$r_{j,[1,n_x-1]}$}&\octrl{-1}\qwbundle{} &\octrl{-1}        &\ctrl{-1}         &\ctrl{-1}         &\qw        &\qw       &\ctrl{-1}      &\qw        &\qw       &\octrl{-1}                     &\qw\\
\lstick{$r_{j,0}$}        &\octrl{-1}            &\octrl{-1}        &\ctrl{-1}         &\ctrl{-1}         &\qw        &\qw       &\ctrl{-1}      &\qw        &\qw       &\octrl{-1}                     &\qw
\end{quantikz}
\caption{\label{circ:OH-parts} 
The subcircuits of the circuit for $O_H$ shown in Fig.~\ref{circ:OH}. The operator $R_c$ is described in Eq.~\eqref{eq:Rc}.
The corresponding rotation angles are computed using Eqs.~\eqref{sys:angles-OH}.
}
\end{figure*}
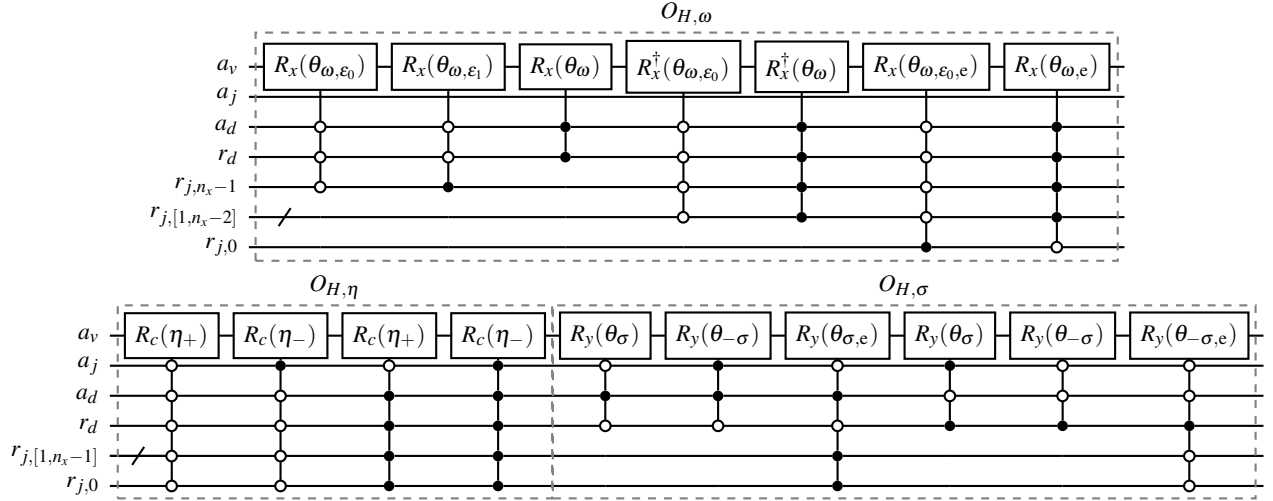
% ******************************

% -----------------------------------------------------------------------------------
% --- Ancilla qubits ---
% -----------------------------------------------------------------------------------
\subsection{Ancilla qubits}\label{sec:ancillae}

% Given the input row index $(lN_x + j)=[0,...2N_x)$ encoded to the input registers $r_d$ and $r_j$ as $\ket{l}_{r_d}\ket{k}_{r_j}$, the oracle $U_A$ returns the state $\sum_{k_c}A_{lN_x + j, k_c}\ket{0}_{r_a}\ket{k_c}_{r_d,r_j} + \ket{\perp}_{r_a}\ket{\dots}_{r_d,r_j}$, where $A_{lN_x + j, k_c}$ are the values of all nonzero matrix elements on the row $(lN_x + j)$, where each element is addressed by the column index $k_c$ saved in the registers $r_d$ and $r_j$ in the output state. 
% Here, $r_a$ denotes all ancilla qubits necessary for the construction of the oracle $U_A$. 
% Next, we consider what ancillae should be introduced in our case.

The matrix $A$ of size $2N_x\times 2N_x$ can be separated into four blocks of size $N_x\times N_x$ each.
We introduce the ancilla qubit $a_d$ to address the first $N_x$ columns of $A$ when this qubit is in the zero state and to the last $N_x$ columns when it is in the unit state. 
The input register $r_d$ described in Sec.~\ref{sec:input-registers} is used to address the first $N_x$ matrix rows when it is in the state $\ket{0}_{r_d}$ and the last $N_x$ rows when it is in the state $\ket{1}_{r_d}$.
Hence, any block can be addressed by using just the qubits $a_d$ and $r_d$.

Another ancilla, denoted $a_v$, is used to store the values of the matrix nonzero elements as explained in Sec.~\ref{sec:OH}. 

An additional qubit, $a_j$, encodes the relative positions of matrix elements with respect to the diagonal within each block.
The zero state of the qubit $a_j$ indicates a matrix element on the diagonal of a given block.
(The position of the block itself in the matrix $A$ is encoded by the qubits $a_d$ and $r_d$ as described above.)
Having $a_j$ in the unit state encodes different information depending on the state of $a_d$.
If $a_d$ is in the zero state, then $\ket{1}_{a_j}$ indicates the matrix element shifted by one cell to the right from the block diagonal.
If $a_d$ is in the unit state, then $\ket{1}_{a_j}$ indicates the element shifted by one cell to the left:
\begin{subequations}\label{sys:aj}
\begin{eqnarray}
&&\ket{0}_{a_j} \to i_c = i_r,\\
&&\ket{1}_{a_j} \to 
    \left\{\begin{aligned}  
        &i_c = i_r + 1,\quad\text{if $a_d$ is in state $\ket{0}_{a_d}$},&\\
        &i_c = i_r - 1,\quad\text{if $a_d$ is in state $\ket{1}_{a_d}$},&\\
    \end{aligned}\right.
\end{eqnarray}
\end{subequations}
where $i_c,i_r=[0,...N_x)$ are the column and row indices, respectively, of a matrix element within a given block.

The dependence on $a_d$ is due to the fact that in the upper and lower left blocks (corresponding to $a_d$'s being in the zero state), the matrix~\eqref{eq:A} has nonzero elements only on the block diagonals and in cells shifted by one cell to the right from the diagonals. 
In the upper and lower right blocks (corresponding to $a_d$'s being in the unit state), we work only with block-diagonal elements and elements shifted by one cell to the left.
% The meaning of the qubit $a_j$ explained above can be schematically represented in the following way:
% \begin{subequations}\label{sys:aj}
% \begin{eqnarray}
%     \text{if }\ket{\cdot}_{a_d} \text{ (any state)}:\quad &&\ket{0}_{a_j} \rightarrow i_c = i_r,\\
%     \text{if }\ket{0}_{a_d}:    \quad &&\ket{1}_{a_j} \rightarrow i_c = i_r + 1,\\
%     \text{if }\ket{1}_{a_d}:    \quad &&\ket{1}_{a_j} \rightarrow i_c = i_r - 1,
% \end{eqnarray}
% \end{subequations}

% -----------------------------------------------------------------------------------
% --- Description of the matrix structure ---
% -----------------------------------------------------------------------------------
\subsection{Matrix structure}\label{sec:oracle-structure}
% Let us consider separately the following operator:
% \begin{equation}\label{eq:OS}
%     O_S = O_{\rm sym}^\dagger O_{\rm edge}^{\rm inv}O_M O_{\rm edge}O_{\rm sym},
% \end{equation}
% which is a part of the whole block-encoding oracle $U_A$ (Eq.~\eqref{eq:BE-mod-state-prep}) and is responsible for the description of the matrix structure i.e. location of nonzero matrix elements without computing their values.
% The formal splitting of $U_A$ into the oracle $O_S$ and $O_H$ is possible due to the fact that $O_H$ only modifies the state amplitudes of the ancilla $a_v$ whilst $O_S$ does not act on $a_v$ at all.
% Although we have to preserve the order of operators in $U_A$ according to Eq.~\eqref{eq:BE-mod-state-prep} (we can not assume that $U_A = O_H O_S$), we still can consider the modification of the state amplitudes performed separately by $O_S$ and $O_H$.
% It is reasonable to consider $O_S$ isolated from $O_H$ since the action of $O_S$ explains the appearance of the factor $\varsigma$ in Eq.~\eqref{eq:gen-A-norm} or the factor $d_H^2$ in Eq.~\eqref{eq:norm-of-A}.
To explain the action of $U_A$ given by Eq.~\eqref{eq:BE-mod-state-prep}, let us first ignore the effect of $O_H$ and consider an auxiliary operator
\begin{equation}\label{eq:OS}
O_S = O_{\rm bulk}^\dagger O^{\rm B}_{\rm edge}O_M O^{\rm F}_{\rm edge} O_{\rm bulk},
\end{equation}
The purpose of introducing this auxiliary operator is to explain the appearance of the factor $d_H^2$ in Eq.~\eqref{eq:norm-of-A}, which is determined by the matrices entering~\eqref{eq:OS} but not $O_H$.

The operators $O_{\rm bulk}$ and $O^{\rm F}_{\rm edge}$ (Fig.~\ref{circ:UA}) encode the column indices of nonzero elements to a linear superposition of states of the ancillae $a_d$ and $a_j$ using a given row index from the registers $r_j$ and $r_d$.
For instance, for the matrix row with the index $k = 1$ (the second top matrix row encoded as $\ket{0}_{r_d}\ket{1}_{r_j}$), one has
\begin{align}\label{eq:example-OF}
O_{\rm bulk}&\ket{0}_{a_j}\ket{0}_{a_d}\ket{0}_{r_d}\ket{1}_{r_j} = (d_H^{-1/2}\ket{0}_{a_j}\ket{0}_{a_d} \\  
    &+ d_H^{-1}\ket{0}_{a_j}\ket{1}_{a_d} + d_H^{-1}\ket{1}_{a_j}\ket{1}_{a_d})\ket{0}_{r_d}\ket{1}_{r_j}.
\end{align}
% \begin{align}\label{eq:example-OF}
% &O_{\rm bulk}\ket{0}_{a_j}\ket{0}_{a_d}\ket{0}_{r_d}\ket{1}_{r_j}\\
%  = (&d_H^{-1/2}\ket{0}_{a_j}\ket{0}_{a_d} + d_H^{-1}\ket{0}_{a_j}\ket{1}_{a_d}\\ 
%  &+ d_H^{-1}\ket{1}_{a_j}\ket{1}_{a_d})\ket{0}_{r_d}\ket{1}_{r_j}.
% \end{align}
% \begin{align}\label{eq:example-OF}
% &O^{\rm F}_{\rm edge}O_{\rm bulk}\ket{0}_{a_j}\ket{0}_{a_d}\ket{0}_{r_d}\ket{1}_{r_j} = (d_H^{-1/2}\ket{0}_{a_j}\ket{0}_{a_d} \\  
%     &+ d_H^{-1}\ket{0}_{a_j}\ket{1}_{a_d} + d_H^{-1}\ket{1}_{a_j}\ket{1}_{a_d})\ket{0}_{r_d}\ket{1}_{r_j}.
% \end{align}
% \begin{equation}\label{eq:example-OF}
%     \begin{split}
%         &O_{\rm edge}O_{\rm sym}\ket{0}_{a_j}\ket{0}_{a_d}\ket{0}_{r_d}\ket{1}_{r_j} = (\\
%         &d_H^{-1/2}\ket{0}_{a_j}\ket{0}_{a_d} + d_H^{-1}\ket{0}_{a_j}\ket{1}_{a_d} + d_H^{-1}\ket{1}_{a_j}\ket{1}_{a_d}\\
%         &)\ket{0}_{r_d}\ket{1}_{r_j}.
%     \end{split}
% \end{equation}
Here, the first term in parenthesis encodes the position of the element $\yi\omega\epsilon_0$ on the main matrix diagonal in Eq.~\eqref{eq:A-bulk} in the row with $k = 1$.
The second and third terms encode the positions of the elements $(-1)^l\sigma$ in Eq.~\eqref{eq:A-bulk} at $k=1$.
Thus, the column indices are written to the states of the qubits $a_d$ and $a_j$.
The various orders of the factor $d_H$ introduced in Eq.~\eqref{eq:norm-of-A} appear here due to the action of Hadamard gates, because each Hadamard gate modifies the state probability by $2^{-1/2}$.

As mentioned in Sec.~\ref{sec:enc-nonH}, the oracle $U_A$ returns the state encoding matrix column indices.
To transfer the column indices saved in the ancillae $a_d$ and $a_j$ to the input registers $r_d$ and $r_j$, we use the operator $O_M$, where the decrementor and incrementor (`Dec' and `Inc' gates in Fig.~\ref{circ:UA}) map the relative positions (encoded into the ancilla $a_j$) to the absolute column indices and write them to the input register $r_j$.
These operators act on a given state $\ket{k}$ in the following way:
\begin{subequations}\label{eq:dec-inc}
\begin{eqnarray}
\text{Inc} \ket{k} &=& \ket{k+1},\\
\text{Dec} \ket{k} &=& \ket{k-1},
\end{eqnarray}
\end{subequations}
and they are controlled by the ancillae $a_j$ and $a_d$ (as shown in Fig.~\ref{circ:UA}) to correctly perform the mapping described in Eq.~\eqref{sys:aj}.
Apart from that, $O_M$ transfers the absolute column indices stored in the ancilla $a_d$ to the input qubit $r_d$ by swapping these qubits. 

% The operator $O_{\rm nsym}^{\rm inv}$ in Fig.~\ref{circ:UA} is used to correctly describe the location of nonzero matrix elements at the zeroth and the last matrix rows.
% If we apply $O_{\rm nsym}^\dagger$ instead $O_{\rm nsym}^{\rm inv}$ (as the standard-state preparation technique does),
% the elements $\eta_{-}$ in Eq.~\eqref{eq:A} do not appear during the block encoding.
% In other words, the operator $O_S$ used with $O_{\rm nsym}^\dagger$ does not return correct column indices of the elements $\eta_{-}$. 
% This happens because the standard technique can encode only Hermitian matrices.
% The elements $\eta_{-}$, however, are present in $A$ due to the outgoing boundary conditions and are responsible for the non-hermiticity of the matrix, so they can not be correctly encoded by $O_{\rm nsym}^\dagger$.
% Therefore, we use $O_{\rm nsym}^{\rm inv}$ instead of the simple Hermitian adjoint of $O_{\rm nsym}$.

The oracle $O^{\rm B}_{\rm edge}$ is close to $O^{F\dagger}_{\rm edge}$, but the control nodes of the Hadamard gates in $O^{\rm B}_{\rm edge}$ are adjusted in such a way that the operator $O_S$ returns the correct column indices of the element $\eta_{-}$ in Eq.~\eqref{eq:A-edge}.
The oracle $O^{\rm B}_{\rm edge}$ is found ad hoc specifically to encode the non-Hermitian matrix~\eqref{eq:A}.
% , and it has to be done differently if one needs to block-encoding a different matrix.

To describe the oracle $O_H$, note first that $O_S$ does not only encode the column indices into $r_d$ and $r_j$ but also modifies the amplitudes of the corresponding quantum states as demonstrated in Eq.~\eqref{eq:example-OF}.
Specifically, 
\begin{equation}\label{eq:dh-matrix}
\begin{split}
    &\bra{k_c}_{r_j}\bra{d_c}_{r_d}O_S\ket{d_r}_{r_d}\ket{k_r}_{r_j} = \\
    &\left\{ \begin{aligned}
        d_H^{-1},\quad &d_r = d_c,\ k_r = k_c,\ l = [2, 2N_x-3],&\\
        d_H^{-3/2},\quad &d_r = d_c,\ k_r = k_c,&\\
                        &l = [0,1] \cup [2N_x - 2, 2N_x-1],&\\
        d_H^{-3/2},\quad &d_r \neq d_c,\ k_r = k_c,\ l = [0]\cup[2N_x-1],&\\
        d_H^{-2},\quad   &d_r = d_c,\ |k_c - k_r| = 1,\ l = [0]\cup[2N_x-1],&\\
        d_H^{-2},\quad   &d_r \neq d_c,\ k_c = k_r,\ k_r = [1, N_x - 2],&\\
        d_H^{-2},\quad   &d_c = d_r + 1,\ k_c = k_r - 1,\ k_c = [0, N_x - 2],&\\
        d_H^{-2},\quad   &d_c = d_r - 1,\ k_c = k_r + 1,\ k_c = [1, N_x),&
    \end{aligned} \right.
\end{split}
\end{equation}
where $l = d_r N_x + k_r=[0, 2N_x)$, $d_r=[0,1]$, $k_r=[0,N_x)$,
and the ancillae $a_d$ and $a_j$ are assumed to be returned in the zero states.
% where $\ket{d_r}_{r_d}\ket{k_r}_{r_j}$ is the input state and $\bra{k_c}_{r_j}\bra{d_c}_{r_d}$ is the output state.
The oracle $O_H$ must take into account the values from Eq.~\eqref{eq:dh-matrix} and output the correct amplitudes $A_{kj}$ from Eqs.~\eqref{eq:A-bulk} and~\eqref{eq:A-edge}.

% -----------------------------------------------------------------------------------
% --- Operator O_H ---
% -----------------------------------------------------------------------------------
\subsection{Operator \texorpdfstring{$\boldsymbol{O_H}$}{$O_H$}}\label{sec:OH}
The operator $O_H$ encodes the values $A_{kj}$ into the state amplitudes of the qubit $a_v$.
This is done by applying the rotation gates described in Appendix~\ref{app:basics}.
The corresponding rotation angles are computed by taking into account the target values from Eq.~\eqref{eq:A}, denoted $v_{\rm des}$, and the multiplication factors from Eq.~\eqref{eq:dh-matrix}, denoted $c_d$, that appear due to the action of the operator $O_S$.
For instance, the value $v_{\rm des}$ can be encoded in the state amplitude by using the standard gate $R_x(\theta)$ acting on the zero state of the qubit $a_v$:
\begin{equation}\label{eq:Rx-example}
\begin{split}
    &R_x(\theta)\ylb c_d\ket{0}_{a_v}\ket{\dots} + \dots\yrb\\
    &= \ylb c_d\cos(\theta/2)\ket{0}_{a_v} - \yi c_d\sin(\theta/2)\ket{1}_{a_v}\yrb\ket{\dots}  + \dots,
\end{split}
\end{equation}
where $\ket{\dots}$ denotes states of other ancillae.
The target value $v_{\rm des}$ can be encoded to the amplitude of either the zero state $\ket{0}_{a_v}$ or the unit state $\ket{1}_{a_v}$.
Since some of the values from Eq.~\eqref{eq:A} are purely imaginary, we choose the unit state to store $v_{\rm des}$, so $v_{\rm des} = -\yi c_d\sin(\theta/2)$.
This requires the following angle:
\begin{equation}\label{eq:arsin}
    \theta = 2 \arcsin\ylb-\iota_v |v_{\rm des}|/c_d \yrb, 
\end{equation}
where $\iota_v = \pm 1$ is the sign of ${\rm Im}\ v_{\rm des}$.
The above transformation works only for purely real or imaginary values.
To encode a complex value, one can use the combined rotation $R_c(v)$ presented in Eq.~\eqref{eq:Rc}.

The circuit implementing $O_H$ is shown in Fig.~\ref{circ:OH}, and its individual blocks are detailed in Fig.~\ref{circ:OH-parts}.
The rotation gates of the oracle $O_H$ are controlled by the states of the ancillae $a_d$ and $a_j$ and the input registers $r_d$ and $r_j$ to compute the values of the matrix elements at the positions encoded in these states.
The calculation of the rotation angles for the oracle $O_H$ is summarized in Appendix~\ref{app:basics}.

% The last two $R_y$ gates zero the factors $d_H^{-3/2}$ at $d_c = d_r + 1, k_r = 0$ and at $d_c = d_r - 1, k_r = N_x - 1$ (Eq.~\eqref{eq:dh-matrix}), because the matrix $A$ has zero-value elements at these positions.

Assuming availability of gates that can be controlled by multiple qubits, the circuit depth of the oracle $U_A$ scales as $\oO(n_x)$ due to the scaling of the decrementor and incrementor operators~\eqref{eq:dec-inc}.
 The number of ancilla qubits is independent of $n_x$.
 The main reason for this is that the ancilla $a_j$ stores the relative positions of the matrix elements, but not the absolute column indices. 
 (In the latter case, the number of ancillae would be proportional to $n_x$.)
 If multicontrolled gates are not available, though, they would have to be transformed into elementary gates. 
 An $n$-controlled single-target gate can be transformed into $\oO(n^2)$ elementary gates using $\oO(n)$ ancillae.\cite{Barenco95}

\section{Comparison with classical simulations}\label{sec:comparison}

\subsection{Direct comparison without measurements}\label{sec:comparison-common}

% ********************************************
% *** Convergence with kappa: eps_0 = eps_1 ***
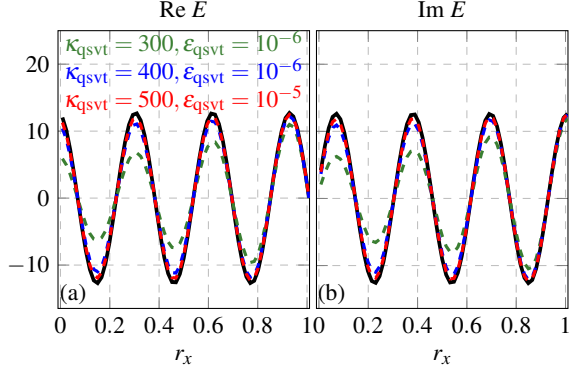
\begin{figure}
\centering
% ---------------------------------------------------------------------------------------
% -------------------- Fig a ------------------------------------------------------------
\begin{tikzpicture}
\begin{axis}[
 	title={$\text{Re}\ E$},
 	xlabel={$r_x$},
	ylabel={},
	legend pos=north west,
	xmin = -0.01,
	xmax = 1.01,
	ymax = 25.,
	extra x ticks={},
	extra y ticks={},
	xmajorgrids=true, ymajorgrids=true, grid style=dashed,
	height=0.3\textwidth, 
	width=0.28\textwidth,
    title style={yshift=-4},
	ylabel style={yshift=-0.5ex},
    % yticklabel = {
    %         \pgfmathprintnumber[
    %             fixed,
    %             precision=0,
    %             zerofill,
    %         ]{\tick}
    % },
    xticklabel style={
        text width=2.5em,
        align=center
    },
	legend pos = north west,
	legend style={
	    fill=white, fill opacity=0.9, draw opacity=1,text opacity=1,
	    font=\small,  %\tiny, \small
	},
]

\addplot [black,        solid, line width=1.4pt] table [y=Y, x=X]{./data/classical_k5_Ey_real.dat};
\addplot [OliveGreen,  dashed, line width=1.2pt] table [y=Y, x=X]{./data/conv_k5_k300_Ey_real.dat};
\addplot [blue,        dashed, line width=1.2pt] table [y=Y, x=X]{./data/conv_k5_k400_Ey_real.dat};
\addplot [red,         dashed, line width=1.2pt] table [y=Y, x=X]{./data/conv_k5_k500_Ey_real.dat};

\node[OliveGreen] at (rel axis cs: 0.5, 0.95) {$\kappa_{\rm qsvt} = 300, \epsilon_{\rm qsvt} = 10^{-6}$};
\node[blue]       at (rel axis cs: 0.5, 0.85) {$\kappa_{\rm qsvt} = 400, \epsilon_{\rm qsvt} = 10^{-6}$};
\node[red]        at (rel axis cs: 0.5, 0.75) {$\kappa_{\rm qsvt} = 500, \epsilon_{\rm qsvt} = 10^{-5}$};

\node[black] at (rel axis cs: 0.06, 0.05) {(a)};
\end{axis}
\end{tikzpicture}
\hspace{-1cm}
% ---------------------------------------------------------------------------------------
% --------------------- Fig b ------------------------------------------------------------
\begin{tikzpicture}
\begin{axis}[
  	title={$\text{Im}\ E$},
	xlabel={$r_x$},
	legend pos=north west,
	xmin = -0.01,
	xmax = 1.01,
    ymax = 25.,
 	yticklabels=\empty,
	extra x ticks={},
	extra y ticks={},
	xmajorgrids=true, ymajorgrids=true, grid style=dashed,
	height=0.3\textwidth, 
 	width=0.28\textwidth,
    title style={yshift=-4},
    xticklabel style={
        text width=2.5em,
        align=center
    },
	legend pos = north west,
	legend style={
	    fill=white, fill opacity=0.9, draw opacity=1,text opacity=1,
	    font=\small,  %\tiny, \small
	},
]

\addplot [black,        solid, line width=1.4pt] table [y=Y, x=X]{./data/classical_k5_Ey_imag.dat};
\addplot [OliveGreen,  dashed, line width=1.2pt] table [y=Y, x=X]{./data/conv_k5_k300_Ey_imag.dat};
\addplot [blue,        dashed, line width=1.2pt] table [y=Y, x=X]{./data/conv_k5_k400_Ey_imag.dat};
\addplot [red,         dashed, line width=1.2pt] table [y=Y, x=X]{./data/conv_k5_k500_Ey_imag.dat};

\node[black] at (rel axis cs: 0.06, 0.05) {(b)};
\end{axis}
\end{tikzpicture}
% ---------------------------------------------------------------------------------------
% ---------------------------------------------------------------------------------------
\caption{
\label{fig:comp-CL-QC-case1} 
Comparison of classical simulations (solid black lines) with quantum computations (colored dashed lines) done on an emulator\cite{QSP-code, Jones19} for $\epsilon_0 = \epsilon_1 = 1$ and $n_x = 6$: (a) ${\rm Re}\ E(x)$, (b) ${\rm Im}\ E(x)$.
The global phase of the signals from the QC is adjusted to that of the classical signal.
}
\end{figure}
% *********************************************

% ***************************************************
% *** Convergence with kappa: 4*eps_0 = eps_1 = 4 ***
\input{./figs/fig_conv_kappa_2.tex}
% ***************************************************

% *****************************************
% *** Scan: signal error ***
\begin{figure}
\centering
% ---------------------------------------------------------------------------------------
% ---------------------------------------------------------------------------------------
\begin{tikzpicture}
    \begin{axis}[
     	title={$\log_{10}\ylb\max_x|E^{\rm CL} - E^{\rm QC}|\yrb$},
     	xlabel={$\log_{10}(\epsilon_{\rm qsvt})$},
    	legend pos=north west,
    	extra x ticks={},
    	extra y ticks={},
    	xmajorgrids=true, ymajorgrids=true, grid style=dashed,
    	height=0.3\textwidth, 
     	width=0.50\textwidth,
        title style={yshift=-3},
    	ylabel style={yshift=-0.5ex},
        xticklabel style={text width=2.5em,  align=center},
    	legend pos = south east,
    	legend style={
    	    fill=white, fill opacity=0.9, draw opacity=1,text opacity=1,
    	    font=\small,  %\tiny, \small
    	},
    ]
    
    \addplot [blue,       dashed, mark=o, mark options={solid}, line width=1.2pt] table [y=Y, x=X]{./data/scan_signal_abs_err_n6_k500.dat};
    % \addlegendentry{$n_x = 6, \kappa_{\rm qsvt} = 500$};
    % \addplot [red,        dashed, mark=o, mark options={solid}, line width=1.2pt] table [y=Y, x=X]{./data/scan_signal_abs_err_n7_k500.dat};
    \addplot [red, dashed, mark=o, mark options={solid}, line width=1.2pt] table [y=Y, x=X]{./data/scan_signal_abs_err_n6_k600.dat};
    % \addlegendentry{$n_x = 6, \kappa_{\rm qsvt} = 600$};
    \addplot [OliveGreen,       dashed, mark=o, mark options={solid}, line width=1.2pt] table [y=Y, x=X]{./data/scan_signal_abs_err_n7_k600.dat};
    % \addlegendentry{$n_x = 7, \kappa_{\rm qsvt} = 600$};
    
    \node[blue]       at (rel axis cs: 0.3, 0.50) {$n_x = 6, \kappa_{\rm qsvt} = 500$};
    \node[red]        at (rel axis cs: 0.4, 0.10) {$n_x = 6, \kappa_{\rm qsvt} = 600$};
    \node[OliveGreen] at (rel axis cs: 0.6, 0.82) {$n_x = 7, \kappa_{\rm qsvt} = 600$};

    % \node[blue]       at (rel axis cs: 0.2, 0.64)  {$n_x = 6, \kappa_{\rm qsvt} = 500$};
    % \node[red]        at (rel axis cs: 0.8, 0.24)  {$n_x = 7, \kappa_{\rm qsvt} = 500$};
    % \node[OliveGreen] at (rel axis cs: 0.64, 0.84) {$n_x = 6, \kappa_{\rm qsvt} = 600$};
    % \node[gray]       at (rel axis cs: 0.8, 0.34)  {$n_x = 7, \kappa_{\rm qsvt} = 600$};
    
    % \node[black] at (rel axis cs: 0.04, 0.90) {(a)};
    \end{axis}
\end{tikzpicture}
% ---------------------------------------------------------------------------------------
% ---------------------------------------------------------------------------------------
\caption{\label{fig:scan-signal-error} 
The dependence of the maximum absolute difference between the electric fields calculated classically and by the QSVT on the absolute error $\epsilon_{\rm qsvt}$ for different values of $\kappa_{\rm qsvt}$ and $n_x$ for $\epsilon_0 = \epsilon_1 = 1$.
}
\end{figure}
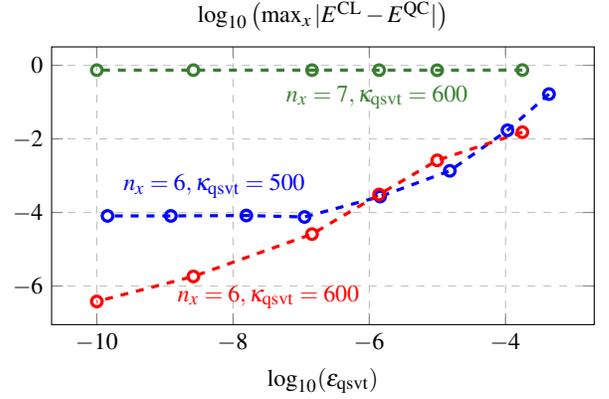
% *****************************************

We solve the model equations~\eqref{sys:model} by computing the inverse of the corresponding matrix~\eqref{eq:A} classically (by applying the standard Gauss--Jordan elimination) and by using the QSVT, whose quantum circuit is emulated using our open-source code.\cite{QSP-code}
Two sets of parameters are considered: $L_x k_{x,0} = 20.0$, $\epsilon_0 = \epsilon_1 = 1$, $n_x = 6$ and $L_x k_{x,0} = 28.8$, $4 \epsilon_0 = \epsilon_1 = 4$, $n_x = 7$, where $L_x$ is the spatial size of the system.
% The spatial size of the system is $L_x = 4.0$ cm.
% Two sets of parameters are considered: $k_{x,0} = 5.0$ cm$^{-1}$, $\epsilon_0 = \epsilon_1 = 1$, $n_x = 6$ and $k_{x,0} = 7.2$ cm$^{-1}$, $4 \epsilon_0 = \epsilon_1 = 4$, $n_x = 7$.
% The QSVT simulation returns the following state:
% \begin{equation}\label{eq:out-qsvt-state}
%     \psi_{\rm out, qsvt} = \frac{e^{\yi\phi_{\rm glob}}}{\beta_{\rm sc}\kappa} \left[E(x), B(x)\right],
% \end{equation}
% where the rescaling from Eq.~\eqref{eq:qsvt-rescaling} is taken into account, and $\phi_{\rm glob}$ is an unknown global phase.
The comparison between the classical and QSVT simulations is shown in Figs.~\ref{fig:comp-CL-QC-case1}-\ref{fig:comp-CL-QC-case2}, where the QSVT angles are computed using the indicated parameter $\kappa_{\rm qsvt}$ and the approximation error $\epsilon_{\rm qsvt}$.

The first case, with $\epsilon_0 = \epsilon_1 = 1$ (Fig.~\ref{fig:comp-CL-QC-case1}), corresponds to the propagation of a vacuum EM wave to the left from the source, which is placed at the right boundary.
The matrix $A$ corresponding to this case has a condition number $\kappa \approx 150$, which is estimated as the ratio of the maximum and minimum singular values. 
This value varies with the method chosen for the computation of $\kappa$ as discussed in Ref.~\onlinecite{Scherer17}.
As seen from Fig.~\ref{fig:comp-CL-QC-case1}, if the QSVT parameter $\kappa_{\rm qsvt}$ is not sufficiently large, the calculation of the electric field using the QSVT does not converge.

The second case (Fig.~\ref{fig:comp-CL-QC-case2}) has a dielectric--dielectric interface at the center of the spatial domain, where the permittivity changes (as shown in Eq.~\eqref{eq:epsilon-rint}). 
When the left-propagating wave (incident wave) excited by the source $Q$ hits this interface, the wave is partially reflected and partially transmitted to the left half of the spatial domain ($r_x < r_{x, \rm int}$).
Hence, the right half of the spatial domain ($r_x > r_{x, \rm int}$) contains both incident and reflected waves.
Their interference changes the field amplitude and the wave-number spectrum, as seen in Fig.~\ref{fig:comp-CL-QC-case2}.
This will be analyzed in Secs.~\ref{sec:qft} and~\ref{sec:ae}.
The increase of the wave number in the right domain explains why one needs a larger number of spatial points, which results in a larger number of qubits, $n_x = 7$ instead of $n_x = 6$.
In turn, changing $n_x$ results in a larger condition number of $A$ (here, $\kappa \approx 400$), and that requires a higher value of $\kappa_{\rm qsvt}$ to compute $A^{-1}$ by the QSVT.

As seen from the green and blue lines in Fig.~\ref{fig:scan-signal-error}, if $\kappa_{\rm qsvt}$ is not large enough, then the decrease of $\epsilon_{\rm qsvt}$ does not improve the precision of the modeled signal (i.e., the electric field $E$). 
In other words, the parameter $\kappa_{\rm qsvt}$, which determines the number of the QSVT angles and their values, defines the maximum achievable precision of the simulated signals.
If $\kappa_{\rm qsvt}$ is sufficiently large (as mentioned in Sec.~\ref{sec:qsvt-scaling}, $\log_2\kappa_{\rm qsvt}$ scales as $\oO(n_x)$), then the resulting error changes linearly with $\epsilon_{\rm qsvt}$. 
% -----------------------------------------------------------------------
% --- Quantum Fourier Transform ---
% -----------------------------------------------------------------------
\subsection{Estimation of the wave numbers}\label{sec:qft}

% ******************************
% *** Circuit QFT-QSVT ***
\begin{figure}[t!]
\centering
\begin{quantikz}[row sep={0.6cm,between origins},column sep={0.2cm}]
\lstick{$\ket{0}_{m}$}                 &\qw        &\qw\yggr{4}{3}{$U_{\rm prep}$}{gray}{0.0}                  
                                                                         &\qw                  &\targ{}   &\gate[4]{\text{AAb}}&\meter{0/1} &\cwbend{3}      \\
\lstick{$\ket{0}_{a_{\rm qsvt}}$}      &\qwbundle{}&\qw                  &\gate[3]{\text{QSVT}}&\octrl{-1}&\qw         &\qw         &\qw              &\qw          &\qw \\
\lstick{$\ket{0}_{r_d}$}               &\qw        &\gate[2]{\text{INIT}}&\qw                  &\octrl{-1}&\qw         &\qw         &\qw              &\qw          &\qw \\
\lstick{$\ket{0}_{r_j}$}               &\qwbundle{}&\qw                  &\qw                  &\qw       &\qw         &\qw         &\gate{\text{QFT}}&\meter{0/1}  &\cw
\end{quantikz}
\caption{\label{circ:qft-qsvt} 
The circuit that measures the wave number of the stationary electric field $E$ in the whole spatial domain. 
The register $a_{\rm qsvt}$ includes all ancilla qubits necessary for the QSVT circuit (i.e., the registers $a_d$, $a_j$, $a_v$ and $q$).
The QSVT circuit is presented in Fig.~\ref{circ:qsvt-odd}. 
The QFT is a well-known circuit in QC and is described in detail in Ref.~\onlinecite{Nielsen10}.
The amplitude amplification procedure (denoted here as `AAb') for an unknown amplitude is described in Ref.~\onlinecite{Brassard02}.
% The QFT and AA circuits, which are common procedures in QC, can be found in Ref.~\onlinecite{Nielsen10} and in Ref.~\onlinecite{Brassard02}. 
% (As a reminder, in our work, the most bottom qubit is the least significant one.)
Here, the QFT is performed only if the qubit $m$ is measured in the unit state, otherwise the whole circuit should be restarted.
}
\end{figure}
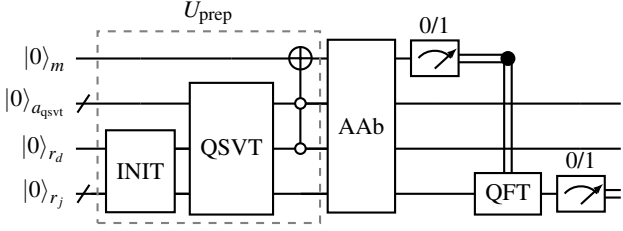
% ******************************

% ******************************
% *** qFT: results ***
\begin{figure}
\centering
\begin{tikzpicture}
\begin{axis}[
	ylabel={probability distribution},
	legend pos=north west,
	xmin = -3.0,
	xmax = 3.0,
 	ymax = 1.05,
	extra x ticks={},
	extra y ticks={},
	xmajorgrids=true, ymajorgrids=true, grid style=dashed,
	height = 0.3\textwidth, 
 	width  = 0.48\textwidth,
    title style={
        yshift=-3
    },
	ylabel style={yshift=-0.5ex},
    % yticklabel = {
    %         \pgfmathprintnumber[
    %             fixed,
    %             precision=0,
    %             zerofill,
    %         ]{\tick}
    % },
    xticklabel style={
        text width=2.5em,
        align=center
    },
    xticklabels=\empty,
	legend pos = north west,
	legend style={
	    fill=white, fill opacity=0.9, draw opacity=1,text opacity=1,
	    font=\small,  %\tiny, \small
	},
]

\draw [dotted, line width=2.0pt, black] (1.0,-1.0) -- (1.0,8.0);
\draw [dotted, line width=2.0pt, black] (-1.0,-1.0) -- (-1.0,8.0);

\addplot [blue,  mark=*, solid, mark options={solid}, line width=1.2pt] table [y=Y, x=X]{./data/fft_k500_n6.dat};
\addlegendentry{$$FFT$$};
\addplot [red,   mark=o, dashed, mark options={solid}, line width=1.2pt] table [y=Y, x=X]{./data/qft_k500_n6.dat};
\addlegendentry{$$QFT$$};

\node[black]      at (rel axis cs: 0.74, 0.90) {$k_{x,0}$};
% \node[black]      at (rel axis cs: 0.2, 0.75) {$\delta k_0 = 5.8\cdot10^{-2}$};

% \draw[draw=black, fill = white, opacity=0.9] (rel axis cs: 0.02,0.7) rectangle (rel axis cs: 0.60,0.42);
% \node[black]      at (rel axis cs: 0.3, 0.6) {$\epsilon_0 = \epsilon_1 = 1,\ \ n_x = 6$};
% \node[black]      at (rel axis cs: 0.3, 0.5) {$\kappa = 500,\ \epsilon_{\rm qsvt} = 10^{-5}$};

\draw[draw=black, fill = white, opacity=0.9] (rel axis cs: 0.02,0.7) rectangle (rel axis cs: 0.30,0.59);
\node[black]      at (rel axis cs: 0.16, 0.64) {$\epsilon_0 = \epsilon_1 = 1$};

\node[black] at (rel axis cs: 0.04, 0.20) {(a)};
\end{axis}
\end{tikzpicture}       
% ---------------------------------------------------------------------------------------
% ---------------------------------------------------------------------------------------
\begin{tikzpicture}
\begin{axis}[
	xlabel={$k_x$},
	ylabel={probability distribution},
	legend pos=north west,
 	xmin = -3.0,
	xmax = 3.0,
	ymax = 0.32,
	extra x ticks={},
	extra y ticks={},
	xmajorgrids=true, ymajorgrids=true, grid style=dashed,
	height = 0.3\textwidth, 
 	width  = 0.48\textwidth,
    title style={
        yshift=-3
    },
	ylabel style={yshift=-0.5ex},
    % yticklabel = {
    %         \pgfmathprintnumber[
    %             fixed,
    %             precision=0,
    %             zerofill,
    %         ]{\tick}
    % },
    xticklabel style={
        text width=2.5em,
        align=center
    },
	legend pos = north west,
	legend style={
	    fill=white, fill opacity=0.9, draw opacity=1,text opacity=1,
	    font=\small,  %\tiny, \small
	},
]

\draw [dotted, line width=2.0pt, black] (1.0,-1.0) -- (1.0,7.0);
\draw [dashed, line width=2.0pt, green] (2.0,-1.0) -- (2.0,7.0);
\draw [dotted, line width=2.0pt, black] (-1.0,-1.0) -- (-1.0,7.0);
\draw [dashed, line width=2.0pt, green] (-2.0,-1.0) -- (-2.0,7.0);

\addplot [blue,  mark=*, solid, mark options={solid}, line width=1.2pt] table [y=Y, x=X]{./data/fft_k600_n7.dat};
\addlegendentry{$$FFT$$};
\addplot [red,   mark=o, dashed, mark options={solid}, line width=1.2pt] table [y=Y, x=X]{./data/qft_k600_n7.dat};
\addlegendentry{$$QFT$$};

\node[black]      at (rel axis cs: 0.60, 0.90) {$k_{x,0}$};
\node[black]      at (rel axis cs: 0.92, 0.90) {$\sqrt{\epsilon_1}k_{x,0}$};

% \node[black]      at (rel axis cs: 0.2, 0.75) {$\delta k_0 = 8.2\cdot10^{-2}$};
% \node[black]      at (rel axis cs: 0.2, 0.65) {$\delta k_1 = 3.7\cdot10^{-2}$};

% \draw[draw=black, fill = white, opacity=0.9] (rel axis cs: 0.02,0.7) rectangle (rel axis cs: 0.60,0.42);
% \node[black]      at (rel axis cs: 0.3, 0.6) {$4 \epsilon_0 = \epsilon_1 = 4,\ \ n_x = 7$};
% \node[black]      at (rel axis cs: 0.3, 0.5) {$\kappa = 600,\ \epsilon_{\rm qsvt} = 10^{-7}$};

\draw[draw=black, fill = white, opacity=0.9] (rel axis cs: 0.02,0.7) rectangle (rel axis cs: 0.30,0.59);
\node[black]      at (rel axis cs: 0.16, 0.64) {$4 \epsilon_0 = \epsilon_1 = 4$};

\node[black] at (rel axis cs: 0.04, 0.20) {(b)};
\end{axis}
\end{tikzpicture}
% ---------------------------------------------------------------------------------------
% ---------------------------------------------------------------------------------------
\caption{
\label{fig:qft-fft} 
A comparison of the results obtained using classical FFT (solid blue lines/markers) and the QFT (dashed red lines/markers):
(a) $\epsilon_0 = \epsilon_1 = 1$, $n_x = 6$, $\kappa_{\rm qsvt} = 500$, $\epsilon_{\rm qsvt} = 10^{-5}$;
(b) $4\epsilon_0 = \epsilon_1 = 4$, $n_x = 7$, $\kappa_{\rm qsvt} = 600$, $\epsilon_{\rm qsvt} = 10^{-7}$.
The vertical black lines correspond to $k_x = \pm k_{x,0}$.
The vertical green lines correspond to $k_x = \pm \sqrt{\epsilon_1} k_{x,0}$.
Here, the wave-number grid is normalized to $k_{x,0} = \omega$.
}
\end{figure}
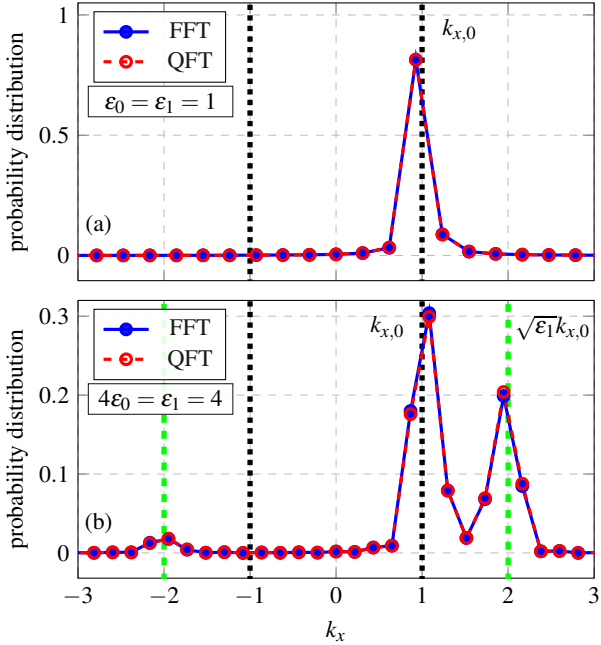
% ******************************

One can estimate the dominant wave number(s) of the electric field using the quantum Fourier transform (QFT).
A possible circuit is shown in Fig.~\ref{circ:qft-qsvt}. 
If one needs to measure the spectrum only in the left (right) half of the spatial domain, then the $X$ gate should be controlled also by the zero (unit) state of the uppermost qubit in the register $r_j$.
The zero-control node on the qubit $r_d$ entangles the unit state $\ket{1}_m$ in the qubit $m$ with the state returned by the QSVT subcircuit encoding the spatial distribution of the electric field.
% This gate converts the qubit $m$ into $\ket{1}$ only for the part of the output QSVT state encoding the spatial distribution of the electric field.
As seen from Eq.~\eqref{eq:qsvt-rescaling}, the amplitude of this state scales as $\oO(1/\kappa)$.
Thus, the state should be amplified by using the amplitude amplification (AA) procedure (`AAb')~\cite{Brassard02} for an unknown amplitude that requires $\oO(\kappa)$ repetitions of the circuit $U_{\rm prep}$ presented in Fig.~\ref{circ:qft-qsvt}.
After the `AAb', the probability to measure $\ket{1}_m$ becomes not less than $1/2$.

If the unit state $\ket{1}_m$ is measured, then the QFT, which requires $\oO(n_x^2)$ quantum gates, is performed in the register $r_j$.
Specifically, the spatial distribution of the electric field over the grid
\begin{equation}
    x = j\Delta x,\ j = [0, N_x),
\end{equation}
is transformed into its spectrum on the wave-number grid
\begin{equation}\label{sys:k-grid}
k_x = -k_{\max} + \Delta k j,\ j = [0, N_x),
\end{equation}
where $k_{\rm max} = \pi/\Delta x$, $\Delta x = 2h$, $\Delta k = 2k_{\rm max}/N_x$, and $N_x$ is an even integer (Sec.~\ref{sec:discr}). 
% If the unit state $\ket{1}_m$ is measured, then the QFT, which requires $\oO(n_x^2)$ quantum gates, is performed in the register $r_j$, thus, transforming the spatial distribution of the electric field over the grid
% \begin{equation}
%     x = j\Delta x,\ j = [0, N_x),
% \end{equation}
% into the wave spectrum over the wave number grid
% \begin{subequations}\label{sys:k-grid}
% \begin{eqnarray}
% &&k_x = -k_{x,\max} + \Delta k j,\ j = [0, N_x),\\
% &&k_{x,\max} = \frac{\pi}{\Delta x},\ \ \ \Delta k_x = \frac{2k_{x,\max}}{N_x},
% \end{eqnarray}
% \end{subequations}
% where $\Delta x = 2 h$ (Eq.~\eqref{eq:sigma-h}), and $N_x$ is an even integer as indicated in Sec.~\ref{sec:discr}.
More precisely, after the QFT, the bit strings in the register $r_j$ encode the wave number grid, and the state amplitudes encode the corresponding Fourier magnitudes.
The measurement of the qubits in the register $r_j$ outputs a bit string encoding a single wave number $k_x$ of a Fourier component with some amplitude $F_{k_x}$.
The probability to measure a certain $k_x$ equals $|F_{k_x}|^2$.
% When the state returned by the QFT is measured, it collapses to the bit string encoding the wave number of one of the strongest (the most probable) Fourier components of the electric field.

Since the circuit shown in Fig.~\ref{circ:qft-qsvt} is computationally expensive to emulate classically, we compute it in parts by modeling the INIT$+$QSVT circuit first and then transferring the computed QSVT state directly to the QFT as an input.
The comparison between the classical FFT and QFT is shown in Fig.~\ref{fig:qft-fft}, where both FFT and QFT are defined over the same wave-number grid~\eqref{sys:k-grid}.
Specifically, shown in Fig.~\ref{fig:qft-fft} is the distribution of the probability with which a measurement returns a given $k_x$.
According to the figure, each measurement most likely returns the wave numbers $k_{x,0}$ or $\sqrt{\epsilon_1}k_{x,0}$, which is in agreement with Eq.~\eqref{eq:kx-epsilon}.

This QFT-based procedure is particularly useful when the wave spectrum consists mainly of just a few dominant modes, which can be identified with $\oO(1)$ measurements. 
A relevant application could be to the modeling of radiofrequency modes in fusion plasmas\cite{Stix92}, where the local spectrum often consists of a mode launched by an antenna and, possibly, few other modes generated by reflection and (or) mode conversion.~\cite{Tracy14}
The corresponding wave vectors vary in space, and the above procedure can be used to identify these wave vectors at a given location of interest.

% % During a single measurement, the circuit~\ref{circ:qft-qsvt} returns a bitstring in the register $r_j$. 
% % This bitstring encodes a wave number $k_{x, \rm meas}$ from the $k$-grid, and the probability to measure one or another $k_{x, \rm meas}$ is indicated in Fig.~\ref{fig:qft-fft}.
% % By using the QFT, which requires $\oO(n_x^2)$ operations, and performing $N_{\rm meas}$ measurements, one can estimate $k_x$ related to the most pronounced Fourier components in the field spectrum.

% It makes sense to apply the QFT only if there are a few dominant modes in the spectrum.
% In this case, the number of measurements (and as a result, the number of repetitions of the whole quantum circuit) is limited.
% % This takes place, for instance, in plasma systems with the mode conversion, where in two different spatially separated domains there is only a single dominant mode and the wave numbers of these modes are different in different spatial domains.
% For instance, this takes place in plasma systems with the mode conversion, where in each of several spatially separated domains there is only a single dominant mode, and the wave numbers of these modes are different in different spatial domains.
% -----------------------------------------------------------------------
% --- Amplitude Estimation ---
% -----------------------------------------------------------------------
\subsection{Amplitude estimation of the field energy}\label{sec:ae}
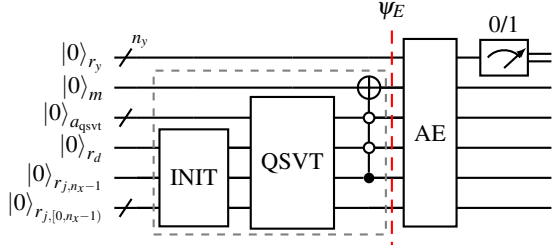
\begin{figure}[t!]
\centering
\begin{quantikz}[row sep={0.4cm,between origins},column sep={0.3cm}]
\lstick{$\ket{0}_{r_y}$}                &\qwbundle{n_y}&\qw                              &\qw                  &\qw\slice{$\psi_E$}&\gate[6]{\text{AE}}&\meter{0/1}&\cw\\
\lstick{$\ket{0}_{m}$}                  &\qw           &\qw\yggr{5}{3}{}{gray}{-1.0}     &\qw                  &\targ{}            &\qw                &\qw        &\qw\\
\lstick{$\ket{0}_{a_{\rm qsvt}}$}       &\qwbundle{}   &\qw                              &\gate[4]{\text{QSVT}}&\octrl{-1}         &\qw                &\qw        &\qw\\
\lstick{$\ket{0}_{r_d}$}                &\qw           &\gate[3]{\text{INIT}}            &\qw                  &\octrl{-1}         &\qw                &\qw        &\qw\\
\lstick{$\ket{0}_{r_{j, n_x-1}}$}       &\qw           &\qw                              &\qw                  &\ctrl{-1}          &\qw                &\qw        &\qw\\
\lstick{$\ket{0}_{r_{j, [0,n_x-1)}}$}   &\qwbundle{}   &\qw                              &\qw                  &\qw                &\qw                &\qw        &\qw
\end{quantikz}\caption{\label{circ:AE-E-general} 
The circuit for measuring the electric-field energy in the right spatial domain by using the AE.
The state $\psi_E$ is described in Eq.~\eqref{eq:psi-E} and prepared by the subcircuit $U_{\rm prep}$ indicated here by the dashed box.
}
\end{figure}

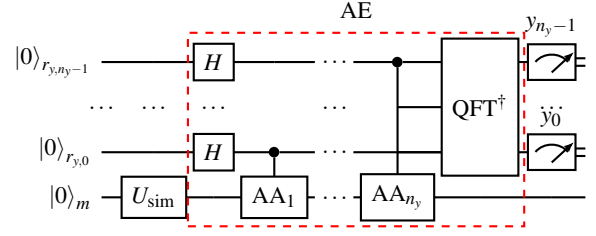
\begin{figure}[t!]
\centering
\begin{quantikz}[row sep={0.6cm,between origins}, column sep={0.1cm}]
\lstick{$\ket{0}_{r_{y,n_y-1}}$}&\qw               &\gate{H}\yggr{4}{6}{AE}{red}{-1.0}
                                                            &\qw                   &\qw &\dots\ &\ctrl{3}                    &\gate[3]{\text{QFT}^\dagger}&\meter{$y_{n_y-1}$}&\cw\\
\dots                           &\dots             &\dots   &                      &    &\dots\ &                            &                            &\dots                  \\
\lstick{$\ket{0}_{r_{y,0}}$}    &\qw               &\gate{H}&\ctrl{1}              &\qw &\dots\ &\qw                         &\qw                         &\meter{$y_0$}      &\cw\\
\lstick{$\ket{0}_m$}            &\gate{U_{\rm sim}}&\qw     &\gate{\text{AA}_1}    &\qw &\dots\ &\gate{\text{AA}_{n_y}}&\qw    &\qw                         &\qw   
\end{quantikz}
\caption{\label{circ:AE-simplified} 
The circuit used for estimating $\tilde p_{m,1}$.
The subcircuit $U_{\rm sim}$ is a reduced version of the circuit $U_{\rm prep}$ shown in the dashed box in Fig.~\ref{circ:AE-E-general}.
The circuit within the red dashed box corresponds to the AE block in Fig.~\ref{circ:AE-E-general}. 
The notation AA$_m$ denotes $2^{m-1}$ applications of the AA operator (Appendix~\ref{app:two-gaussians}).
}
\end{figure}

To measure the field energy summed over a chosen spatial domain, we can use the so-called amplitude estimation (AE) procedure.
For example, it can be implemented using the standard algorithm described in Ref.~\onlinecite{Brassard02}.
% There is a variety of AE techniques, but for the matter of demonstration, we use the standard algorithm described in Ref.~\onlinecite{Brassard02}.
% By using the same as in Ref.~\onlinecite{Novikau22}, we can compute e.g. the electric energy integrated in $x>0$ domain.
The corresponding circuit is shown in Fig.~\ref{circ:AE-E-general}, where the register $r_y$ has $n_y$ qubits.
There, the subcircuit $U_{\rm prep}$ entangles the unit state $\ket{1}_m$ with the superposition of states encoding the spatial distribution of the electric field in the right domain ($r_x > r_{x, \rm int}$):
\begin{equation}\label{eq:psi-E}
    \psi_E = \ket{1}_m\ylb\ket{0}_{r_d}\sum_{j=N_x/2}^{N_x-1} E_j\ket{j}_{r_j}\yrb + \ket{0}_m (\dots).
\end{equation}
(Here, we omit the register $r_y$, which remains in the zero state. 
We also omit the multiplication factor from Eq.~\eqref{eq:qsvt-rescaling} and the QSVT ancillae.)
The electric-field energy can be measured as the probability amplitude of the state $\ket{1}_m$:
\begin{equation}
    p_{m,1} = \sum_{j=N_x/2}^{N_x-1} |E_j|^2.
\end{equation}
Instead of emulating the circuit from Fig.~\ref{circ:AE-E-general} directly, which is computationally expensive, we emulate a simplified circuit shown in Fig.~\ref{circ:AE-simplified}. 
There, the operator $U_{\rm prep}$ is reduced to the one-qubit rotation $R_y(\theta_E)$, so
\begin{equation}
    U_{\rm sim}\ket{0}_m = \cos(\theta_E)\ket{0}_m + \sin(\theta_E)\ket{1}_m,
\end{equation}
where $\sin^2(\theta_E) = p_{m,1}$.
(This defines $\theta_E$ up to a modulo $\pi$, which is not important since we are only interested in $\sin^2(\theta_E)$.) 
More precisely, we emulate the actual circuit $U_{\rm prep}$ to compute the state $\psi_E$, from which we calculate the probability $p_{m,1}$ numerically.
As a result, we can find the angle $\theta_E$. 
% (the ambiguity in the value of the angle is not important since we need only to guarantee that $\sin^2(\theta_E) = p_{m,1}$).
After that, we model the circuit shown in Fig.~\ref{circ:AE-simplified} using $U_{\rm sim}$ with the pre-computed $\theta_E$.
To demonstrate the operation of the AE, we also report emulation of the non-simplified AE circuit (as in Fig.~\ref{circ:AE-E-general}) for a Gaussian field profile in Appendix~\ref{app:two-gaussians}.
% In Appendix~\ref{app:two-gaussians}, for the matter of demonstration, an AE circuit for a Gaussian function is simulated without the described simplification.

% **************************************************
% *** AE circuit of electric field in the right boundary ***

The bit string measured after the AE encodes an integer $i_y$ from which one can estimate the desired probability amplitude:\cite{Brassard02}
\begin{equation}
    \tilde p_{m,1} = 1 - \sin^2\ylb\frac{\pi i_y}{N_{\rm AA}}\yrb,
\end{equation}
where $N_{\rm AA} = 2^{n_y}$, and $\tilde p_{m,1}$ is the estimated probability of the state $\ket{1}_m$.
With a probability more than $0.81$, the measurement outcomes $\tilde{p}_{m,1}$ with the following absolute error:\cite{Brassard02}
\begin{equation}\label{eq:ae-delta}
    \delta \equiv |p_{m,1} - \tilde p_{m,1}| \leq 2\pi\frac{\sqrt{p_{m,1}(1-p_{m,1})}}{N_{\rm AA}} + \frac{\pi^2}{N_{\rm AA}^2}.
\end{equation}
Since $p_{m,1}$ is unknown in advance, in practice one can estimate $\delta$ by replacing $p_{m,1}$ with $\tilde{p}_{m,1}$ (obtained from measurements) in the right-hand side of Eq.~\eqref{eq:ae-delta}.
Then, the field energy stored in a given domain spanning $N_{x,\rm area}$ points can be estimated as  
\begin{equation}
    \tilde{\mathcal{E}} = \frac{(\beta_{\rm sc}\kappa)^2}{N_{x,\rm area}} \ylb\tilde{p}_{m,1} \pm \delta\yrb.
\end{equation}

The AE circuit requires $\oO(N_{\rm AA})$ repetitions of the AA operator, which is described in Appendix~\ref{app:two-gaussians} and includes two calls to the QSVT.
Because $\delta$ scales as $\oO(1/N_{\rm AA})$, to measure the energy with an absolute error not larger than $\delta$, one needs $\oO(1/\delta)$ queries to the QSVT circuit.
Since the amplitude of the QSVT resulting state, where the system fields are encoded, is $\oO(1/\kappa)$,  $\delta$ should also be at least as small as $\oO(1/\kappa)$.
Thus, the AE requires $\oO(\kappa)$ queries to the QSVT circuit.

% Finally, the field energy can be estimated in the following way:  
% \begin{equation}
%     \tilde{\mathcal{E}} = \frac{(\beta_{\rm sc}\kappa)^2}{N_{x,\rm area}} \ylb\tilde{p}_{m,1} \pm \delta\yrb,
% \end{equation}
% where $N_{x,\rm area}$ is the number of points in the chosen spatial domain, where the energy is summed.

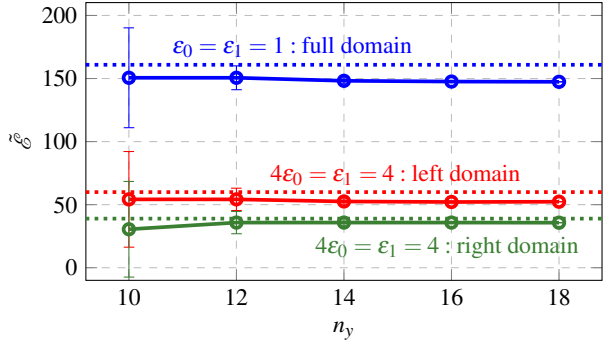
\begin{figure}
\centering
% ---------------------------------------------------------------------------------------
% -------------------- ------------------------------------------------------------------
\begin{tikzpicture}
\begin{axis}[
 	xlabel={$n_y$},
 	ylabel={$\tilde{\mathcal{E}}$},
	legend pos=north west,
  	ymin = -10,
	extra x ticks={},
	extra y ticks={},
	xmajorgrids=true, ymajorgrids=true, grid style=dashed,
	height=0.30\textwidth, 
 	width=0.48\textwidth,
    title style={
        yshift=-3
    },
	ylabel style={yshift=-0.5ex},
    xticklabel style={
        text width=2.5em,
        align=center
    },
	legend pos = south west,
	legend style={
	    fill=white, fill opacity=0.9, draw opacity=1,text opacity=1,
	    font=\small,  %\tiny, \small
	},
]

\addplot [blue, mark=o, solid, mark options={solid}, line width=1.4pt, error bars/.cd, y fixed, y dir=both, y explicit] table [y=Y, x=X, y error=Y_ERROR]{./data/AE_scan_k500_energy.dat};
% \addlegendentry{$\epsilon_0 = \epsilon_1 = 1: \text{full domain}$};
\addplot [red, mark=o, solid, mark options={solid}, line width=1.4pt, error bars/.cd, y fixed, y dir=both, y explicit] table [y=Y, x=X, y error=Y_ERROR]{./data/AE_scan_k600_left_energy.dat};
% \addlegendentry{$4\epsilon_0 = \epsilon_1 = 4: \text{left domain}$};
\addplot [OliveGreen, mark=o, solid, mark options={solid}, line width=1.4pt, error bars/.cd, y fixed, y dir=both, y explicit] table [y=Y, x=X, y error=Y_ERROR]{./data/AE_scan_k600_right_energy.dat};
% \addlegendentry{$4\epsilon_0 = \epsilon_1 = 4: \text{right domain}$};

\draw [dotted, line width=1.4pt, blue]       (0.0, 1.609e+02) -- (30, 1.609e+02);
\draw [dotted, line width=1.4pt, red]        (0.0, 5.996e+01) -- (30, 5.996e+01);
\draw [dotted, line width=1.4pt, OliveGreen] (0.0, 3.893e+01) -- (30, 3.893e+01);

\node[blue]         at (rel axis cs: 0.4, 0.84) {$\epsilon_0 = \epsilon_1 = 1: \text{full domain}$};
\node[red]          at (rel axis cs: 0.6, 0.38) {$4\epsilon_0 = \epsilon_1 = 4: \text{left domain}$};
\node[OliveGreen]   at (rel axis cs: 0.7, 0.12) {$4\epsilon_0 = \epsilon_1 = 4: \text{right domain}$};

\end{axis}
\end{tikzpicture}
% ---------------------------------------------------------------------------------------
% -------------------- ------------------------------------------------------------------
\caption{
\label{fig:AE-scans} 
Change of the estimated energy $\tilde{\mathcal{E}}$ with the number of qubits in the register $r_y$.
The AE of the energy summed over the full spatial domain in the case with $\epsilon_0 = \epsilon_1 = 1$ is indicated by the blue markers.
The AE of the energy summed over the left domain $r_x<r_{x,\rm int}$ or the right domain $r_x>r_{x,\rm int}$ in the case with $4 \epsilon_0 = \epsilon_1 = 4$ is indicated by the red and green markers, respectively.
The horizontal dotted lines indicate the corresponding energies computed in classical simulations.
}
\end{figure}

The results of our emulations of the above AE procedure are summarized in Fig.~\ref{fig:AE-scans}.
By increasing the number of qubits in the register $r_y$, one can reduce the measurement error $\delta$ so that the QSVT approximation error dominates over $\delta$.
In turn, the QSVT error itself can be reduced by increasing $\kappa_{\rm qsvt}$ and (or) decreasing $\epsilon_{\rm qsvt}$ as described in Sec.~\ref{sec:comparison-common}.

% **************************************************
% *** AE: scans on ny ***
% **************************************************
% -----------------------------------------------------------------------------------
% --- Measurement of the Wigner function ---
% -----------------------------------------------------------------------------------
\subsection{Measurement of the absorbed wave power}\label{sec:JE}

% ******************************************
% *** Circuit to compute the product jE ***
\newcommand{\yINIT}{\gate[2]{\rm INIT}}
\newcommand{\yneb}{\qwbundle{n_{\rm EB}}}
\newcommand{\ynwb}{\qwbundle{n_{\rm w}}}
\newcommand{\yne}{n_{\rm EB}}
\newcommand{\ynw}{n_{\rm w}}
\newcommand{\yE}{{\rm E}}
\newcommand{\yB}{{\rm B}}
\newcommand{\yqwb}{\qwbundle{}}
\newcommand{\yMI}{\gate[2]{\rm MI}}
\newcommand{\ySKB}{\gate[3]{S_{k_{\rm B}}}}
\newcommand{\ySIL}{\gate[5,label style={yshift=-0.0cm}]{S_{I-r_{\rm EB}}}}
\newcommand{\ylth}{\qw}
\newcommand{\yCNEB}{\gate[4]{C_{N_{\rm EB}}}}
\newcommand{\yANHW}{\gate[3]{A_{N_{\rm h,w}}}}
\newcommand{\yCNW}{\gate[4]{C_{N_{\rm w}}}}
\newcommand{\yGA}{\gate{\rm GA}}
\begin{figure*}[t!]
\centering
% \begin{quantikz}[transparent, row sep={0.6cm,between origins},column sep={0.15cm}]
\begin{quantikz}[transparent, row sep=0.01cm,column sep={0.15cm}]
\lstick{$m$}                &\qw  &\qw   &\qw\slice{$\ket{\xi_1}$}     
                                                  &\qw\slice{$\ket{\xi_2}$}  
                                                        &\qw  &\qw   &\qw   &\qw      &\qw\slice{$\ket{\xi_3}$}       
                                                                                                 &\qw     &\qw     &\qw      &\qw\slice{$\ket{\xi_4}$}     
                                                                                                                                      &\targ{}  &\meter{}&\cw\\
\lstick{$r_{\rm sel}$}      &\qw  &\qw   &\qw     &\qw  &\qw  &\qw   &\qw   &\qw      &\targ{}   &\qw     &\ctrl{1}&\ctrl{1} &\qw     &\ctrl{-1}&\qw     &\qw\\
\lstick{$r_{\rm swap}$}     &\qw  &\qw   &\qw     &\qw  &\qw  &\qw   &\qw   &\qw      &\qw       &\gate{H}&\ctrl{5}&\ctrl{10}&\gate{H}&\ctrl{-1}&\qw     &\qw\\
\lstick{${\rm com}_{II}$}   &\qw  &\qw   &\qw     &\qw  &\qw  &\qw   &\yCNEB&\qw      &\ctrl{-2} &\qw     &\qw     &\qw      &\qw     &\qw      &\qw     &\qw\\
\lstick{$a_{II,{\rm sign}}$}&\qw  &\qw   &\qw     &\qw  &\ySKB&\qw   &\qw   &\qw      &\octrl{-1}&\qw     &\qw     &\qw      &\qw     &\qw      &\qw     &\qw\\
\lstick{$II_{[\yne, n_x)}$} &\yqwb&\yINIT&\yMI    &\qw  &\qw  &\qw   &\qw   &\qw      &\qw       &\qw     &\qw     &\qw      &\qw     &\qw      &\qw     &\qw\\
\lstick{$II_{[0,\yne)}$}    &\yqwb&\qw   &\qw     &\qw  &\qw  &\qw   &\qw   &\qw      &\qw       &\qw     &\swap{1}&\qw      &\qw     &\qw      &\qw     &\qw\\
\lstick{$r_{\rm EB}$}       &\yneb&\qw   &\gate{H}&\ySIL&\qw  &\qw   &\qw   &\qw      &\qw       &\qw     &\targX{}&\qw      &\qw     &\qw      &\qw     &\qw\\
\lstick{${\rm com}_{I}$}    &\qw  &\qw   &\qw     &\ylth&\qw  &\qw   &\yCNW &\ctrl{1} &\ctrl{-4} &\qw     &\qw     &\qw      &\qw     &\qw      &\qw     &\qw\\
\lstick{$a_{I,{\rm sign}}$} &\qw  &\qw   &\qw     &\qw  &\ySKB&\yANHW&\qw   &\octrl{2}&\octrl{-1}&\qw     &\qw     &\qw      &\qw     &\qw      &\qw     &\qw\\
\lstick{$I_{[\ynw, n_x)}$}  &\yqwb&\yINIT&\yMI    &\qw  &\qw  &\qw   &\qw   &\qw      &\qw       &\qw     &\qw     &\qw      &\qw     &\qw      &\qw     &\qw\\
\lstick{$I_{[0,\ynw)}$}     &\yqwb&\qw   &\qw     &\qw  &\qw  &\qw   &\qw   &\yGA     &\qw       &\qw     &\qw     &\swap{1} &\qw     &\qw      &\qw     &\qw\\
\lstick{$r_{\rm w}$}        &\ynwb&\qw   &\gate{H}&\qw  &\qw  &\qw   &\qw   &\qw      &\qw       &\qw     &\qw     &\targX{} &\qw     &\qw      &\qw     &\qw
\end{quantikz} 
\caption{\label{circ:jE} 
A possible circuit that can be used to compute $p_{1,r_{\rm swap}}$ (Eq.~\eqref{eq:p1swap}), which is the probability to have both qubits $r_{\rm swap}$ and $r_{\rm sel}$ in the unit state.
All qubits are initialized in the zero state.
% To indicate explicitly the target qubits of the `GA' and the swapping operators, we separate the $I$ and $II$ registers into two parts, where for each register, the subindices $[0,n_1)$ and $[n_1, n_2)$ indicate the first $n_1$ and the last $n_2 - n_1$ qubits, correspondingly.
The subcircuits `MI' and `GA' are the QSVTs to compute the electromagnetic field $E$ and the Gaussian, respectively.
For clarity, we do not show here all the ancillae qubits necessary for the QSVT procedures and the register $d$ assuming that the `MI' outputs  a state encoding only the electric field.
% The register $r_{\rm EB}$ has $n_{\rm EB}$ qubits and is used to couple the electric field encoded in the register $II$ with those encoded in $I$.
% The register $r_{\rm w}$ has $n_{\rm w}$ qubits.
The operators $S_{I-r_l}$ and $S_{k_{B}}$ are two types of subtractors described in Appendix~\ref{app:arithmetics}.
The adder $A_i$ and comparator $C_i$ (for some integer $i$) are also presented in Appendix~\ref{app:arithmetics}.
The qubit ${\rm com}_{I}$ is not used by the subtractor $S_{I-r_l}$.
To compute $p_{0,r_{\rm swap}}$, the rightmost Toffoli gate must be controlled by the zero state of the qubit $r_{\rm swap}$.
}
\end{figure*}
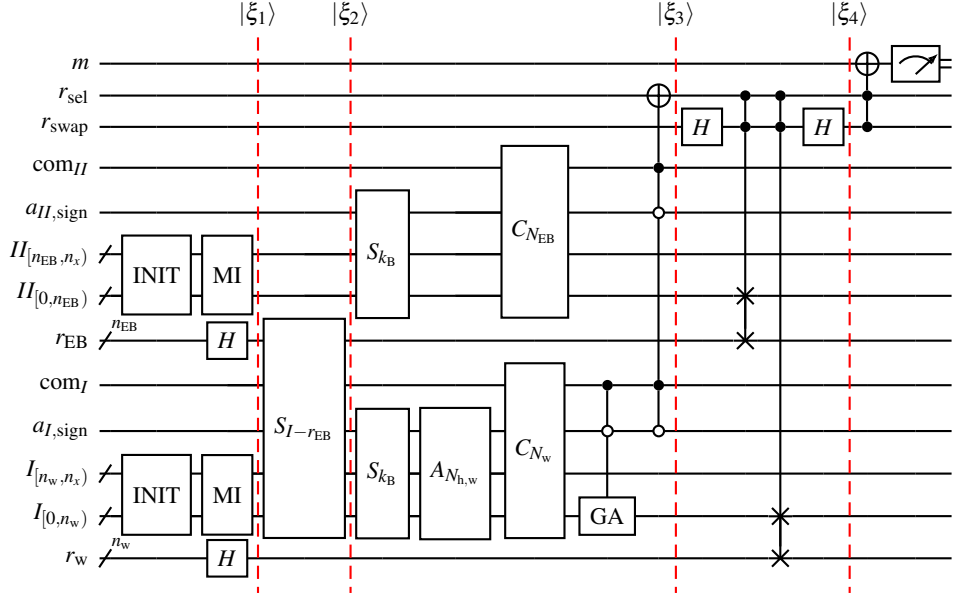
\let\yINIT\undefined
\let\yneb\undefined
\let\ynwb\undefined
\let\ynw\undefined
\let\yE\undefined
\let\yB\undefined
\let\yqwb\undefined
\let\yMI\undefined
\let\ySKB\undefined
\let\ySIL\undefined
\let\ylth\undefined
\let\yCNEB\undefined
\let\yANHW\undefined
\let\yCNW\undefined
\let\yGA\undefined

Another important parameter to measure in simulations of stationary waves is the absorbed power.
In a given region $(x_{\rm B}, x_{\rm E})$, this power can be calculated as\cite{Stix92}
% \begin{equation}\label{eq:jE-init-continuous}
% \begin{split}
%     \mathcal{P} &= \int_{x_{\rm B}}^{x_{\rm E}}\diff x\ \text{Re}[J(x)] \text{Re}[E(x)] \\
%     &= \frac{1}{2}\int_{x_{\rm B}}^{x_{\rm E}} \diff x\  E^*(x)
%         \int_{-\infty}^{+\infty}\diff x^\prime \sigma(x, x^\prime) E(x^\prime),
% \end{split}
% \end{equation}
\begin{equation}\label{eq:jE-init-continuous}
    \mathcal{P} = \frac{1}{2}\int_{x_{\rm B}}^{x_{\rm E}} \diff x\  E^*(x)
        \int_{-\infty}^{+\infty}\diff x^\prime \sigma(x, x^\prime) E(x^\prime),
\end{equation}
% where $J(x)$ is the electric current induced by the wave electric field, 
where $\sigma$ is the conductivity at a given frequency, and the symbol $^*$ denotes the complex conjugation.
The integration over $x^\prime$ can be replaced by integration over $\zeta = x^\prime - x$:
\begin{equation}\label{eq:jE-init-continuous-2}
\mathcal{P} = \frac{1}{2}\int_{x_{\rm B}}^{x_{\rm E}} \diff x\  E^*(x)
        \int_{-\zeta_{\rm w}}^{+\zeta_{\rm w}}\diff\zeta\ \sigma(x, x+\zeta) E(x + \zeta),
\end{equation}
% (For example, in the absence of spatial dispersion, one has $\bar{\sigma}(\bar{x}, \zeta) \propto \delta(\zeta)$.) 
% Then, one can rewrite Eq.~\eqref{eq:jE-init-continuous} as
% \begin{equation}
% \mathcal{P} = \frac{1}{2}\int_{x_{\rm B}}^{x_{\rm E}} \diff x\  E^*(x)
%         \int_{-\infty}^{+\infty}\diff x^\prime \bar{\sigma}(\bar{x}, \zeta) E(x^\prime),
% \end{equation}
where the domain in the second integral can be finite and arbitrary as long as it is much larger than the localization scale of $\sigma$ with respect to $\zeta$ at a given $x$. 
Let us also adopt, for simplicity, that the interval $(x_{\rm B}, x_{\rm E})$ is small enough so the dependence of $\sigma$ on $x$ on this interval can be neglected.
In other words, let us assume, say, $\sigma(x, x+\zeta) \approx \overline{\sigma}(x_{\rm C}, \zeta)$, where $x_{\rm C} = (x_{\rm B} + x_{\rm E})/2$ is the center of the interval. 
(Further, for simplicity, we omit $x_{\rm C}$ and the bar in $\overline{\sigma}(x_{\rm C}, \zeta)$.)
Then, after discretization, Eq.~\eqref{eq:jE-init-continuous-2} becomes
\begin{subequations}\label{sys:jE}
\begin{eqnarray}
&&\mathcal{P} = \frac{D}{2N_{\rm w}N_{\rm EB}},\\
&&D = \sum_{k = k_{\rm B}}^{k_{\rm E}} E_k^* \sum_{j = -N_{\rm h,w}}^{N_{\rm h,w}} 
    \sigma_j E_{k+j},\label{sys:jE-b}
\end{eqnarray}
\end{subequations}
where the indices $k_{\rm B}$ and $k_{\rm E}$ correspond to the spatial points $x_{\rm B}$, $x_{\rm E}$, respectively.
The spatial interval has $N_{\rm EB} = k_{\rm E} - k_{\rm B} + 1$ points, and the electric current $\sigma_j E_{k+j}$ at the point $x_k$ is induced by the electric field within the window of  
$N_{\rm w} = 2N_{\rm h,w}+1$ points centered at $x_k$.
To encode $N_{\rm EB}$ and $N_{\rm w}$ spatial points, one needs $n_{\rm EB} = \lceil\log_2N_{\rm EB}\rceil$ and $n_{\rm w}= \lceil\log_2N_{\rm w}\rceil$ qubits, respectively.
Positive $\mathcal{P}$ corresponds to the wave power absorbed by the environment e.g., plasma.

In practice, $\sigma_j$ depends on the given medium and is supposed to be calculated analytically or numerically in kinetic simulations.\cite{Svidzinski16}
Here, we assume that the conductivity is a Gaussian function, $\sigma_j = G_j$, and rewrite Eq.~\eqref{sys:jE-b} as follows:
\begin{equation}\label{eq:jE-disc-theory-alt}
D = \sum_{k = 0}^{N_x-1}\sum_{p = 0}^{N_x-1} M_{k,p} E_k^* E_p,
\end{equation}
with a matrix
\begin{equation}
M_{k,p} = \Theta_{k-k_{\rm B}}\Theta_{k_{\rm E} - k}G_{p-k}\Theta_{p - (k-N_{\rm h,w})}
    \Theta_{(k+N_{\rm h,w})-p},
\end{equation}
and Heaviside functions $\Theta$.
The sum~\eqref{eq:jE-disc-theory-alt} has the form of the inner product $\bra{E}M\ket{E}$ and, therefore, its circuit representation can be computed by applying the Hadamard test,\cite{Ortiz01, Mitarai19}
provided that the matrix $M$ is encoded into a unitary one.
% where the electric field is calculated using the QSVT and the matrix $M$ should be encoded into a unitary one.
Yet, the block-encoding of $M$ into a unitary can be difficult.
Therefore, we propose an alternative way based on the SWAP test\cite{Buhrman01} and quantum arithmetic operators.

The idea is to keep the simple vector form $G_j$ of the conductivity (instead of its matrix form $G_{p-k}$ as in Eq.~\eqref{eq:jE-disc-theory-alt}).
% This can be done by letting the electric fields have coupled spatial indices $k$ and $k+j$. 
Hence, we rewrite $D$ in Eq.~\eqref{sys:jE-b} as 
\begin{equation}\label{eq:sum-jE}
D = \sum_{k = 0}^{N_{\rm EB}-1} E_{k_{\rm B}+k}^* 
    \sum_{j = 0}^{N_{\rm w}-1} G_j E_{k_{\rm B} + k + j - N_{\rm h,w}}.
\end{equation}
Here, the Gaussian $G_j$ depends on a single index and can be calculated by the QSVT as explained in Appendix~\ref{app:qsvt-gaussian}.
However, now the electric fields are summed over different spatial intervals and should be encoded into states on two different registers, $I$ and $II$.
These intervals can be set by applying various arithmetic operators described in Appendix~\ref{app:arithmetics}.
% as will be explained later.
The coupling of the electric fields in Eq.~\eqref{eq:sum-jE} through the index $k$ can be implemented by using a quantum subtractor that performs the transformation $\ket{k}\ket{p} \to \ket{k}\ket{p-k}$.

The SWAP test computes the product of the form $|\itimes{\psi}{\lambda}|^2$ of two quantum states $\ket\psi$ and $\ket\lambda$ calculated in two separate qubit registers.
In other words, one finds the square of the sum $\sum_j \psi_j^*\lambda_j$. 
% In our case, one can take $\ket\psi \propto \ket{E_{k_{\rm B}+k}}$ and $\ket\lambda \propto \ket{G_j E_{k_{\rm B} + k + j - N_{\rm h,w}}}$.
To find the double sum as~\eqref{eq:sum-jE}, we introduce supplemental registers, $r_{\rm EB}$ and $r_{\rm w}$, and perform one SWAP test between the registers $II$ and $r_{\rm EB}$, and another SWAP test between $I$ and $r_{\rm w}$.
The circuit for the computation of the double sum~\eqref{eq:sum-jE} is shown in Fig.~\ref{circ:jE}, where the electric fields are computed using the QSVT subcircuits denoted `MI'.
(`MI' is the same QSVT circuit as the one used in Figs.~\ref{circ:qft-qsvt} and~\ref{circ:AE-E-general}.)
Further, we consider the circuit shown in Fig.~\ref{circ:jE} step by step.

First of all, we compute the spatial distribution of the electric fields and initialize the registers $r_{\rm w}$ and $r_{\rm EB}$:
\begin{equation}
\ket{\xi_1}  = \sum_{k = 0}^{N_x - 1} E_k \ket{k}_{II} \ket{\phi^{\rm EB}_{\{l\}}}_{r_{\rm EB}} 
    \sum_{p = 0}^{N_x-1} E_p\ket{p}_{I} \ket{\phi^{\rm w}_{\{i\}}}_{r_{\rm w}},
\end{equation}
where all other qubits are in the zero state,
%$N_x$ is the spatial size of the whole system (Sec.~\ref{sec:discr}), 
and the state $\ket{\xi_1}$ is indicated in Fig.~\ref{circ:jE}.
The registers $r_{\rm EB}$ and $r_{\rm w}$ encode the following superpositions:
\begin{equation}\label{eq:phi:EB:w}
    \ket{\phi^{\rm EB}_{\{l\}}} = \eta_{\rm EB}\sum_{l = 0}^{N_{\rm EB}-1}\ket{l},\quad
    \ket{\phi^{\rm w}_{\{i\}}} = \eta_{\rm w}\sum_{i = 0}^{N_{\rm w}-1}\ket{i},
\end{equation}
with the constants
\begin{equation}
    % \eta_{\rm EB} = \frac{1}{2^{n_{\rm EB}/2}},\quad \eta_{\rm w} = \frac{1}{2^{n_{\rm w}/2}}.
    \eta_{\rm EB} = 2^{-n_{\rm EB}/2},\quad \eta_{\rm w} = 2^{-n_{\rm w}/2}.
\end{equation}
In Eqs.~\eqref{eq:phi:EB:w}, the subindices indicating the qubit registers are skipped, since the states $\ket{\phi^{\rm EB}_{\{l\}}}$ and $\ket{\phi^{\rm w}_{\{i\}}}$ will also appear in other registers.

After that, the subtractor $S_{I-r_{\rm EB}}$ couples the indices in the registers $I$ and $r_{\rm EB}$:
% \begin{equation}\label{eq:xi-2}
% \ket{\xi_2} = \sum_{k = 0}^{N_x - 1} E_k \ket{k}_{II} \ket{\phi^{\rm EB}_{\{l\}}}_{r_{\rm EB}} 
%     \sum_{p = 0}^{N_x-1} E_p\ket{p-l}_{I}\ket{\phi^{\rm w}_{\{i\}}}_{r_{\rm w}},
% \end{equation}
\begin{equation}\label{eq:xi-2}
\ket{\xi_2} = \eta_{\rm EB}\sum_{k = 0}^{N_x - 1} E_k \ket{k}_{II} 
    \sum_{l = 0}^{N_{\rm EB}-1}\ket{l}_{r_{\rm EB}} 
    \sum_{p = 0}^{N_x-1} E_p\ket{p-l}_{I}\ket{\phi^{\rm w}_{\{i\}}}_{r_{\rm w}}.
\end{equation}
% where $\ket{\phi^{\rm EB}_{\{l\}}}$ is now entangled with the state $\ket{p-l}$.
The expression in the rightmost sum can be recasted as $E_{l+j}\ket{j}_{I}$ by setting $j = p-l$, where $j = [-l, N_x - l)$.
The subtractor $S_{I-r_{\rm EB}}$ uses the ancilla $a_{I,\rm sign}$ as a flag (or as a sign bit) which is set into the unit state for negative values of the index $j$.
For simplicity, this ancilla is skipped in Eq.~\eqref{eq:xi-2}.
% , and the sum $\sum_j$ covers both positive and negative values of $j$.
Later, the state of this ancilla will be taken into account to deal only with positive values of $j$.

The next step is to encode the electric field amplitudes at the spatial points with indices $[k_{\rm B}, k_{\rm E}]$ into the first $n_{\rm EB}$ least-significant qubits in the register $II$.
To do that, we apply the subtractor $S_{k_{\rm B}}$ to the register $II$.
This operator subtracts the (unsigned) integer $k_{\rm B}$ from the integers encoded within the register $II$ and entangles all indices not less than $k_{\rm B}$ with the zero state of the qubit $a_{II,{\rm sign}}$.
After that, the comparator $C_{N_{\rm EB}}$ is applied to set the correct upper limit in the sum $\sum_k$ by entangling all indices $k$ less than $N_{\rm EB}$ with the unit state $\ket{1}_{{\rm com}_{II}}$.

Similar arithmetic operators are applied to the register $I$ to entangle the intervals $[x_{k_{\rm B} + l - N_{\rm h,w}}, x_{k_{\rm B} + l  + N_{\rm h,w}}]$ for $l = [0, N_{\rm EB})$ with the state $\ket{1}_{{\rm com}_I}\ket{0}_{a_{I,{\rm sign}}}$.

After that, the QSVT circuit is applied to the first $n_{\rm w}$ least-significant qubits of the register $I$ to multiply the electric field by the Gaussian as described in Appendix~\ref{app:filter-gaussian}. 
To center correctly the Gaussian within the window $[x_{k_{\rm B} + l - N_{\rm h,w}}, x_{k_{\rm B} + l  + N_{\rm h,w}}]$, one can use Eqs.~\eqref{sys:alphas-gauss}.
The resulting data are entangled with the unit state $\ket{1}_{r_{\rm sel}}$:
\begin{equation}\label{eq:xi3}
\bra{1}_{r_{\rm sel}}\ket{\xi_3} = 
    \ket{\phi^E}_{II}\ket{\phi^{\rm EB}_{\{l\}}}_{r_{\rm EB}}\ket{\phi^G_l}_I\ket{\phi^{\rm w}_{\{i\}}}_{r_{\rm w}}
\end{equation}
where the following states are used:
\begin{subequations}
\begin{eqnarray}
    &&\ket{\phi^E} = \sum_{k=0}^{N_{\rm EB}-1}E_{k_{\rm B} + k}\ket{k},\\
    &&\ket{\phi^G_l} = \sum_{j = 0}^{N_{\rm w}-1} G_j E_{k_{\rm B} + l + j - N_{\rm h,w}}\ket{j},
\end{eqnarray}
\end{subequations}
and we are not interested in the state $\ket{0}_{r_{\rm sel}}$. 
Note that in Eq.~\eqref{eq:xi3}, the double sum $\sum_l\ket{l}_{r_{\rm EB}}\sum_j G_j E_{k_{\rm B} + l + j - N_{\rm h,w}}\ket{j}_I$ is written as $\ket{\phi^{\rm EB}_{\{l\}}}_{r_{\rm EB}}\ket{\phi^G_l}_I$, where the subindex $l$ in $\ket{\phi^G_l}$ indicates the coupling to the state $\ket{\phi^{\rm EB}_{\{l\}}}$.

To compute the absolute value of $D$ from Eq.~\eqref{eq:sum-jE}, we perform the SWAP test:
\begin{equation}\label{eq:xi-4}
\begin{split}
\bra{1}_{r_{\rm swap}}\bra{1}_{r_{\rm sel}}&\ket{\xi_4} =
    \frac{1}{2}\bigg(\ket{\phi^E}_{II}\ket{\phi^{\rm EB}_{\{l\}}}_{r_{\rm EB}}\ket{\phi^G_l}_I\ket{\phi^{\rm w}_{\{i_1\}}}_{r_{\rm w}}\\
    - &\ket{\phi^{\rm EB}_{\{p\}}}_{II}\ket{\phi^E}_{r_{\rm EB}}\ket{\phi^{\rm w}_{\{i_2\}}}_I\ket{\phi^G_p}_{r_{\rm w}}
    \bigg).    
\end{split}
\end{equation}
By applying the corresponding Toffoli gate, one can entangle $\bra{1}_{r_{\rm swap}}\bra{1}_{r_{\rm sel}}\ket{\xi_4}$ with the unit state $\ket{1}_m$ and measure its probability $p_{1,r_{\rm swap}}$ by using the AE.
The probability of the state~\eqref{eq:xi-4} is 
\begin{equation}\label{eq:p1swap}
    p_{1,r_{\rm swap}} = \frac{\eta_{\rm EB}^2 \eta_{\rm w}^2}{2}(|c_0|^2 - S),\\
\end{equation}
with an unknown $|c_0|^2$.
The positive real value $S$ is computed as a product of the sum:
\begin{equation}\label{eq:sum-1-S}
\begin{split}
\itimes{\phi_{\{p\}}^{\rm EB}}{\phi^E}_{II}&\itimes{\phi^G_p}{\phi^{\rm w}_{\{i_1\}}}_{r_{\rm w}}\\
    &= \sum_{p = 0}^{N_{\rm EB}-1}E_{k_{\rm B}+p}
       \sum_{i_1=0}^{N_{\rm w}-1}G^*_{i_1}E^*_{k_{\rm B} + p + i_1 - N_{\rm h,w}},
\end{split}
\end{equation}
with the sum
\begin{equation}\label{eq:sum-2-S}
\begin{split}
\itimes{\phi^E}{\phi_{\{l\}}^{\rm EB}}_{r_{\rm EB}}&\itimes{\phi^{\rm w}_{\{i_2\}}}{\phi^G_l}_I\\
    &= \sum_{l = 0}^{N_{\rm EB}-1}E^*_{k_{\rm B}+l}
       \sum_{i_2=0}^{N_{\rm w}-1}G_{i_2}E_{k_{\rm B} + l + i_2 - N_{\rm h,w}}.
\end{split}
\end{equation}
By comparing Eq.~\eqref{eq:sum-jE} with the Eqs.~\eqref{eq:sum-1-S} and~\eqref{eq:sum-2-S}, one can see that $S\equiv |D|^2$.

The constant $|c_0|^2$ can be eliminated by measuring also $p_{0,r_{\rm swap}}$, which is the probability of the state $\bra{0}_{r_{\rm swap}}\bra{1}_{r_{\rm sel}}\ket{\xi_4}$.
Because this state is the same as~\eqref{eq:xi-4} but with the `+' sign in the brackets, its probability is
\begin{equation}
    p_{0,r_{\rm swap}} = \frac{\eta_{\rm EB}^2\eta_{\rm w}^2}{2}(|c_0|^2 + |D|^2).
\end{equation}
Thus, the absolute value of the double sum~\eqref{eq:sum-jE} can be calculated as 
\begin{equation}
    |D| = \frac{1}{\eta_{\rm EB}\eta_{\rm w}}\sqrt{p_{0,r_{\rm swap}} - p_{1,r_{\rm swap}}}.
\end{equation}
As mentioned before, the amplitude of the QSVT output state (returned by `MI') is $\oO(1/\kappa)$.
To compute the sum~\eqref{eq:sum-jE}, one launches the `MI' twice in parallel to compute the product $E^* E$, which is $\oO(1/\kappa^2)$.
Thus, to estimate $p_{0,r_{\rm swap}}$ or $p_{1,r_{\rm swap}}$, one will need $\oO(\kappa^2)$ queries to the QSVT circuit.
However, if $N_{\rm EB}$ is comparable with the system size, then the number of the queries can be reduced at least to $\oO(\kappa)$ for two- and three-dimensional spatial systems.

\section{Conclusions}\label{sec:conclusions}
In this paper, we propose a quantum algorithm for simulating dissipative waves in inhomogeneous linear media as a boundary-value problem. 
Our algorithm is based on the QSVT, which is a state-of-the-art technique that was previously proposed for other problems but is applied here to a boundary-value wave problem for the first time. 
Specifically, we model an EM wave that is excited by a prescribed source and propagates in a dielectric medium with a piecewise-continuous dielectric function. 
We show how to encode the corresponding non-Hermitian Hamiltonian in a quantum circuit, calculate the field distribution using this circuit, and perform measurements of the spatial spectrum, wave energy, and power dissipation for this and similar wave problems. 
We emulate quantum simulations of this circuit on a classical computer and show that the numerical results are in agreement with theory and usual classical simulations. 
Most of the quantum circuits presented in this paper are constructed and computed using our computational framework.~\cite{QSP-code}

We also show that the overall quantum simulations of dissipative waves based on the QSVT scale favorably compared to classical simulations in multi-dimensional systems. 
We expect the gain to be particularly efficient in kinetic plasma problems, where the wave modeling is done in phase space with six or even more dimensions. 
Still, there are several potential problems for practical applications of the QSVT, which should be pointed out.
% Still, the following potential problems for practical applications of the QSVT should be pointed out.
First of all, the encoding of the considered wave classical system requires at least $4+n_x$ qubits and the resulting QSVT circuit involves hundreds of calls to the block-encoding oracle, even without taken into account the measurements.
Quantum hardware with such specifications is unlikely to appear in the foreseeable future.
Another issue is that typical classical wave systems are characterized by matrices with large condition numbers.
Since the QSVT used for the inversion of such matrices returns a state whose amplitude scales as $\oO(1/\kappa)$, the measurement of the state requires $\oO(\kappa)$ queries to the QSVT circuit.
This can significantly reduce the quantum speedup.
Apart from that, to invert a matrix with a large condition number, one needs to precompute classically a significant number of QSVT angles.
This is difficult to do using the codes that are currently available for calculating such angles. 
In our case, the GPU parallelization of the code responsible for the polynomial approximation of the inverse function allowed us to compute the QSVT angles only for $\kappa_{\rm qsvt} \lesssim 1000$ (Sec.~\ref{sec:qsvt-angles}). 
This means that quantum modeling of classical wave systems is limited and, in practice, will likely require efficient preconditioning of the dispersion matrices. (A similar problem is known for classical modeling of large wave systems.\cite{Green20})
For instance, due to the sparsity of the matrices associated with the computation of EM waves (such as those in cold plasmas\cite{Novikau22}), a quantum version of the sparse approximate inverse preconditioner proposed in Ref.~\onlinecite{Clader13} may be useful.

\begin{acknowledgments}
The research described in this paper was conducted under the Laboratory Directed Research and Development (LDRD) Program at Princeton Plasma Physics Laboratory, a national laboratory operated by Princeton University for the U.S. Department of Energy under Prime Contract No. DE-AC02-09CH11466.
The author(s) are pleased to acknowledge that the work reported on in this paper was substantially performed using the Princeton Research Computing resources at Princeton University which is consortium of groups led by the Princeton Institute for Computational Science and Engineering (PICSciE) and Office of Information Technology's Research Computing.
The authors also thank Yulong Dong for valuable discussions.
% The authors also thank John M. Martyn and Yulong Dong for valuable discussions.
\end{acknowledgments}

\appendix
% --------------------------------------------------------------------------
% --- Rotation gates ---
% --------------------------------------------------------------------------
\section{Basic and supplemental gates}\label{app:basics}

% ******************************************
% *** Block-encoding of the x-coordinate ***
\begin{figure*}[t!]
\centering
\begin{quantikz}[row sep={0.8cm,between origins},column sep={0.3cm}]
\lstick{$\ket{0}_{\rm sign}$}          &\gate{X} &\gate[4]{\rm QFT}\yggr{4}{6}{}{gray}{0.0}
                                                       &\gate{b^{\rm sub}_0 P_1}&\qw                     &\qw                     &\qw                     &\gate[4]{\rm QFT^\dagger}&\gate{X} &\qw\\
\lstick{$\ket{b^{\rm targ}_2}_{t,2}$}  &\gate{X} &\qw  &\gate{b^{\rm sub}_0 P_2}&\gate{b^{\rm sub}_1 P_1}&\qw                     &\qw                     &\qw                      &\gate{X} &\qw\\
\lstick{$\ket{b^{\rm targ}_1}_{t,1}$}  &\gate{X} &\qw  &\gate{b^{\rm sub}_0 P_3}&\gate{b^{\rm sub}_1 P_2}&\gate{b^{\rm sub}_2 P_1}&\qw                     &\qw                      &\gate{X} &\qw\\
\lstick{$\ket{b^{\rm targ}_0}_{t,0}$}  &\gate{X} &\qw  &\gate{b^{\rm sub}_0 P_4}&\gate{b^{\rm sub}_1 P_3}&\gate{b^{\rm sub}_2 P_2}&\gate{b^{\rm sub}_3 P_1}&\qw                      &\gate{X} &\qw
\end{quantikz}
\caption{\label{circ:subtractor} 
The circuit for the subtractor $S_{k_{\rm sub}}$ for the case with $n_t = 3$ target qubits, where the unsigned integer $k_{\rm sub}$ is represented as the bit string $\left[b^{\rm sub}_3b^{\rm sub}_2b^{\rm sub}_1b^{\rm sub}_0\right]$ ($b^{\rm sub}_0$ is the least significant bit). 
The operator subtracts $k_{\rm sub}$ from the unsigned integer, $k_{\rm targ}$, encoded as the bit string $\ket{b^{\rm targ}_2 b^{\rm targ}_1 b^{\rm targ}_0}_t$ into the target register $t$. 
The difference $\Delta = k_{\rm targ} - k_{\rm sub}$ is written back to the register $t$, and the `sign' qubit is inverted if the result is negative.
Here, $b^{\rm sub}_j P_k$ applies the phase gate $P(2\pi/2^k)$ (Eq.~\eqref{eq:phase-gate}) if $b^{\rm sub}_j = 1$. 
The inner dashed box is the adder, $k_{\rm targ} + k_{\rm sub}$, where the `sign' qubit stores the carry bit of the sum.
}
\end{figure*}
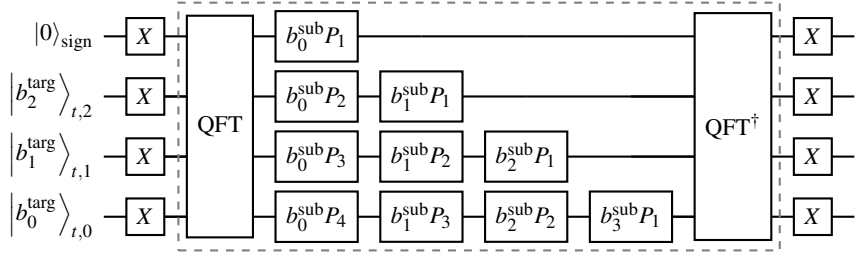

\subsection{Rotation gates}\label{sec:rotations}
The following rotation gates are used throughout the paper:
\begin{subequations}\label{eq:rotations}
\begin{eqnarray}
&&R_x(\theta) = 
    \begin{pmatrix}
        \cos\frac{\theta}{2} & -\yi\sin\frac{\theta}{2} \\
        -\yi\sin\frac{\theta}{2} & \cos\frac{\theta}{2} 
    \end{pmatrix},\label{eq:rx}\\
&&R_y(\theta) = 
    \begin{pmatrix}
        \cos\frac{\theta}{2} & -\sin\frac{\theta}{2} \\
        \sin\frac{\theta}{2} &  \cos\frac{\theta}{2} 
    \end{pmatrix},\label{eq:ry}\\
&&R_z(\theta) = 
    \begin{pmatrix}
        e^{-\yi\theta/2} & 0 \\
        0                & e^{\yi\theta/2}
    \end{pmatrix}.\label{eq:rz}
\end{eqnarray}
\end{subequations}
Each of these operators satisfies
\begin{equation}\label{eq:app-two-angles}
    R_a(\beta)R_a(\alpha) = R_a(\alpha + \beta),\quad a=x,y,z,
\end{equation}
and the same applies to
\begin{equation}\label{eq:phase-gate}  
P(\theta) = 
    \begin{pmatrix}
        1 & 0 \\
        0 & e^{\yi\theta}
    \end{pmatrix}.
\end{equation}
% where each operator satisfies the expression
% \begin{equation}\label{eq:app-two-angles}
%     R_a(\beta)R_a(\alpha) = R_a(\alpha + \beta)
% \end{equation}
% A complex value, $v = v_0 \exp(\yi\arg(v))$, can be computed by using the rotation
The rotation
\begin{equation}\label{eq:Rc}
    R_c(v) = R_y(\theta_{v,2}) R_z(\theta_{v,1}),
\end{equation}
can be used to compute a complex value, $v = |v| \exp(\yi\arg(v))$, by acting on the zero state:
\begin{equation}
    R_c(v)\ket{0} = \cos(\theta_{v,2}/2)e^{-\yi\theta_{v,1}/2}\ket{0} + \sin(\theta_{v,2}/2)e^{-\yi\theta_{v,1}/2}\ket{1},
\end{equation}
where $v$ can be encoded as a complex amplitude of either the zero state, $v = \cos(\theta_{v,2}/2)e^{-\yi\theta_{v,1}/2}$, or the unit state, $v=\sin(\theta_{v,2}/2)e^{-\yi\theta_{v,1}/2}$.
% The angles $\theta_{v,1}$ and $\theta_{v,2}$ are calculated according to
% \begin{subequations}\label{sys:Rc-angles}
% \begin{eqnarray}
% &&\theta_{v,1} = -2\arg(v),\\
% &&\theta_{v,2} = 2\arccos(v_0).
% \end{eqnarray}
% \end{subequations}

% ******************************************
% *** Block-encoding of the x-coordinate ***
\begin{figure}[b]
\centering
\begin{quantikz}[row sep={0.5cm,between origins},column sep={0.2cm}]
\lstick{$\ket{0}_a$}   &\gate{R_y(2\alpha_0)} &\gate{R_{n_x-1}} &\gate{R_{n_x-2}}&\qw    &\ldots\ \  &\gate{R_0}&\gate{X} &\qw       \\
\lstick{$r_{x,n_x-1}$} &\qw                   &\qw              &\qw             &\qw    &\ldots\ \  &\ctrl{-1} &\qw      &\qw       \\
\dots                  &\dots                 &                 &                &       &\ldots\ \  &          &\ldots   &          \\
\lstick{$r_{x,1}$}     &\qw                   &\qw              &\ctrl{-3}       &\qw    &\ldots\ \  &\qw       &\qw      &\qw       \\
\lstick{$r_{x,0}$}     &\qw                   &\ctrl{-4}        &\qw             &\qw    &\ldots\ \  &\qw       &\qw      &\qw          
\end{quantikz}
\caption{\label{circ:x-block-encoding} 
The circuit implementing the matrix $U_A$ for the encoding of the coordinate $\psi=\sin(x)$ according to Eqs.~\eqref{sys:x-phi-encoding}. 
Here, $R_k = R_y(2\alpha/2^k)$.
}
\end{figure}
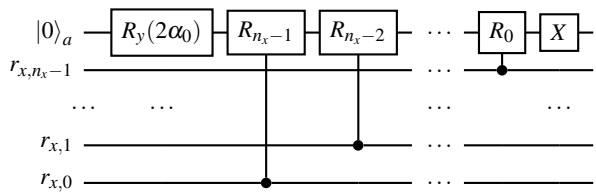
% ******************************************

\subsection{Parameters for the oracle $O_H$}
The rotations from Sec.~\ref{sec:rotations} are used in the oracle $O_H$ (Figs.~\ref{circ:OH} and~\ref{circ:OH-parts}) with the following angles:
\begin{subequations}\label{sys:angles-OH}
\begin{eqnarray}
\theta_{\omega,\epsilon_L}       &=& 2 \arcsin(- \omega\epsilon_L d_H), \quad L \in [0,1],\label{eq:theta-omega}\\
\theta_{\omega}                  &=& 2 \arcsin(- \omega d_H),\\
\theta_{\omega,\epsilon_0,\rm e} &=& 2 \arcsin(- \omega\epsilon_0 d_H^{3/2}),\\
\theta_{\omega,\rm e}            &=& 2 \arcsin(- \omega d_H^{3/2}),\\
\theta_{\eta_\pm, 1}             &=& -2 \arg(\eta_\pm),\\
\theta_{\eta_+, 2}               &=& 2 \arcsin(|\eta_+| d_H^{3/2}),\\
\theta_{\eta_-, 2}               &=& 2 \arcsin(|\eta_-| d_H^{2}),\\
\theta_{\pm\sigma}               &=& 2 \arcsin(\pm\sigma d_H^{2}),\\
\theta_{\pm\sigma,\rm e}         &=& 2 \arcsin(\pm\sigma d_H^{3/2}) - \theta_{\pm\sigma},\\
\theta_{\pm\pi}                  &=& - \theta_{\pm\sigma}.
\end{eqnarray}
\end{subequations}
For instance, the angles $\theta_{\omega,\epsilon_0}$ are used in the oracle $O_{H,\omega}$ shown in Fig.~\ref{circ:OH-parts} to compute the elements $\yi\omega\epsilon_0$ of the matrix \eqref{eq:A-bulk} at the rows with $k \in [2, 2N_x-3]$.
According to Eq.~\eqref{eq:dh-matrix}, for these elements, the oracle $O_S$ returns the multiplication factor $d_H^{-1}$, which is taken into account in Eq.~\eqref{eq:theta-omega}.
The elements $\yi\omega\epsilon_0$ also appear at $k = 1$ and $k = N_x-2$ in the matrix \eqref{eq:A-bulk}.
However, for these indices, the oracle $O_S$ returns $d_H^{-3/2}$ that is taken into account in the computation of the angle $\theta_{\omega,\epsilon_0,\rm e}$, where the subindex `e' stands for `edge' (boundary).
The same principle applies to the calculation of the angles $\theta_{\omega}$ and $\theta_{\omega,\rm e}$, also $\theta_{\pm\sigma}$ and $\theta_{\pm\sigma,\rm e}$.

% ******************************************
% *** Gaussians: scan ***
\begin{figure*}
\centering
% ---------------------------------------------------------------------------------------
% ---------------------------------------------------------------------------------------
\begin{tikzpicture}
    \begin{axis}[
      	title={Gaussian$(x)$},
    	xlabel={$x$},
    	ylabel={},
    	legend pos=north west,
     	xmin = -1.0,
     	xmax =  1.0,
    	extra x ticks={},
    	extra y ticks={},
    	xmajorgrids=true, ymajorgrids=true, grid style=dashed,
    	height=0.3\textwidth, 
     	width=0.33\textwidth,
        title style={yshift=-4},
    	ylabel style={yshift=-0.5ex},
    	yticklabel = {
            \pgfmathprintnumber[
                fixed,
                precision=2,
                zerofill,
            ]{\tick}
        },
        xticklabel style={text width=2.5em,  align=center},
    	legend pos = north east,
    	legend style={
    	    fill=white, fill opacity=0.9, draw opacity=1,text opacity=1,
    	    font=\small,  %\tiny, \small
    	},
    ]
    
    \addplot [blue,  solid, line width=1.2pt] table [y=Y, x=X]{./data/gauss_centered.dat};
    \addplot [red,   dashed, line width=1.2pt] table [y=Y, x=X]{./data/gauss_left.dat};
    
    \node[black] at (rel axis cs: 0.08, 0.90) {(a)};
    \end{axis}
\end{tikzpicture}
% ---------------------------------------------------------------------------------------
% ---------------------------------------------------------------------------------------
\begin{tikzpicture}
\begin{axis}[
 	title={$N_{\rm angles}$},
	xlabel={$\log_{10}(1/\epsilon_{\rm})$},
	ylabel={},
	legend pos=north west,
	extra x ticks={},
	extra y ticks={},
	xmin = 4,
	ymin = -10,
	ymax = 1900,
	xmajorgrids=true, ymajorgrids=true, grid style=dashed,
	height=0.3\textwidth, 
	width=0.33\textwidth,
    title style={yshift=-4},
	ylabel style={yshift=-1.5ex},
	xlabel style={yshift=1ex},
    xticklabel style={
        text width=2.5em,
        align=center
    },
	legend pos = north east,
	legend style={
	    fill=white, fill opacity=0.9, draw opacity=1,text opacity=1,
	    font=\small,  %\tiny, \small
	},
]

\addplot [blue, solid, mark=o, line width=1.0pt] table [y=Y, x=X]{./data/scan_gauss_mu001_eps.dat};
\addplot [red, solid, mark=o, line width=1.0pt] table [y=Y, x=X]{./data/scan_gauss_mu01_eps.dat};
\addplot [OliveGreen, solid, mark=o, line width=1.0pt] table [y=Y, x=X]{./data/scan_gauss_mu025_eps.dat};
\addplot [gray, solid, mark=o, line width=1.0pt] table [y=Y, x=X]{./data/scan_gauss_mu04_eps.dat};
\addplot [orange, solid, mark=o, line width=1.0pt] table [y=Y, x=X]{./data/scan_gauss_mu05_eps.dat};

\node[blue]        at (rel axis cs: 0.60, 0.40) {$\mu = 0.01$};
\node[red]         at (rel axis cs: 0.50, 0.10) {$\mu = 0.10$};
\node[OliveGreen]  at (rel axis cs: 0.70, 0.22) {$\mu = 0.25$};
\node[gray]        at (rel axis cs: 0.62, 0.60) {$\mu = 0.40$};
\node[orange]      at (rel axis cs: 0.20, 0.75) {$\mu = 0.50$};

\node[black] at (rel axis cs: 0.08, 0.90) {(b)};
\end{axis}
\end{tikzpicture}
% ---------------------------------------------------------------------------------------
% ---------------------------------------------------------------------------------------
\begin{tikzpicture}
\begin{axis}[
 	title={$N_{\rm angles}$},
	xlabel={$\mu$},
	ylabel={},
	legend pos=north west,
	extra x ticks={},
	extra y ticks={},
	xmajorgrids=true, ymajorgrids=true, grid style=dashed,
	height=0.3\textwidth, 
	width=0.33\textwidth,
    title style={yshift=-4},
	ylabel style={yshift=-0.2ex},
 	xlabel style={yshift=0.5ex},
    xticklabel style={
        text width=2.5em,
        align=center
    },
	legend pos = north east,
	legend style={
	    fill=white, fill opacity=0.9, draw opacity=1,text opacity=1,
	    font=\small,  %\tiny, \small
	},
]

\addplot [blue, solid, mark=o, line width=1.0pt] table [y=Y, x=X]{./data/scan_gauss_eps7_mu.dat};
\addplot [red, dashed, mark=o, mark options={solid}, line width=1.0pt] table [y=Y, x=X]{./data/scan_gauss_eps10_mu.dat};

\node[blue] at (rel axis cs: 0.68, 0.86) {$\epsilon_{\rm qsvt} \approx 10^{-7}$};
\node[red]  at (rel axis cs: 0.34, 0.50) {$\epsilon_{\rm qsvt} \approx 10^{-10}$};

\node[black] at (rel axis cs: 0.08, 0.90) {(c)};
\end{axis}
\end{tikzpicture}
% ---------------------------------------------------------------------------------------
% ---------------------------------------------------------------------------------------
\caption{\label{fig:gauss} 
(a) The Gaussian functions computed by the QSVT and centered around $x_c = 0.0$ (solid blue line) and $x_c = -0.33$ (dashed red line). 
Here, $\beta_{\rm sc} = 0.98$, $\mu = 0.25$, and $n_x = 6$. 
The peak amplitude is downscaled by the factor $2^{n_x/2}$ due to the initialization, which is the same as in the circuit shown in Fig.~\ref{circ:two-gaussians}.  
(b) The dependence of the number of the QSVT angles, $N_{\rm angles}$, on the QSVT approximation error $\epsilon_{\rm qsvt}$ for various widths $\mu$.
(c) The dependence of $N_{\rm angles}$ on $\mu$ for $\epsilon_{\rm qsvt} = 10^{-7}$ and $\epsilon_{\rm qsvt} = 10^{-10}$.
}
\end{figure*}
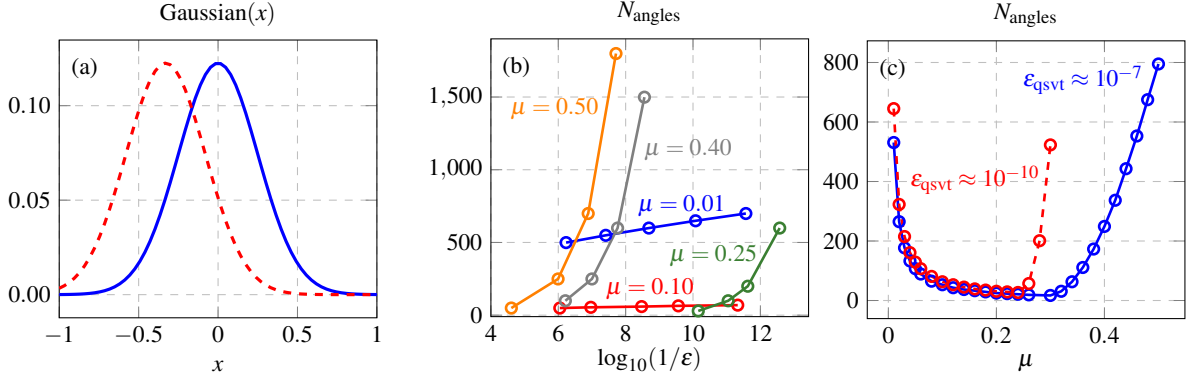
% ******************************************

\subsection{Arithmetic operators}\label{app:arithmetics}
Apart from the incrementor and decrementor used in Fig.~\ref{circ:UA} and described in Ref.~\onlinecite{Novikau22}, we use also a subtractor $S_{k_{\rm sub}}$ and a comparator $C_{k_{\rm com}}$.
The former operator (Fig.~\ref{circ:subtractor}) subtracts the predefined unsigned integer $k_{\rm sub}$ from the integer(s) encoded in the $n_t$ target qubits $t$. 
Its implementation is based on the circuit described in Ref.~\onlinecite{Suau21}.
In our realization, the most significant (the uppermost) qubit is used to store the sign of the resulting integer, and the absolute value of the output integer is written back to the target qubits.

By removing the $X$ gates from Fig.~\ref{circ:subtractor}, one obtains the adder $A_{k_{\rm sub}}$ that adds the integer $k_{\rm sub}$ to the integer(s) encoded in the register $t$.

The subtractor is used to construct the comparator~\cite{Suau21} $C_{k_{\rm com}}$ that inverts the `com' ancilla qubit (Fig.~\ref{circ:jE}) if the predefined unsigned integer $k_{\rm com}$ is strictly larger than the integer encoded in the target qubits.
The comparator leaves the target qubits and the `sign' qubit (used for the intermediate computations within the comparator) untouched. 
More details can be found in Ref.~\onlinecite{Suau21}. 
(Keep in mind that the least significant qubit in all our circuits is the lowermost one.)

As shown in Ref.~\onlinecite{Draper00}, the circuit presented in Fig.~\ref{circ:subtractor} can be modified to subtract the integer $k_{\rm sub}$ which is not predefined but encoded as a bit string in another register $r_{\rm sub}$.
For that, the phase gates $P_k$ in the subtractor should be controlled by the corresponding qubits of the register $r_{\rm sub}$. 
This operator acts then on two registers, $t$ and $r_{\rm sub}$, and is denoted here as $S_{t - r_{\rm sub}}$. The resulting difference is written back to the target register $t$.

The circuits for the subtractors and the comparator require $\oO(n^2)$ quantum operations.

\section{QSVT computation of a Gaussian function}\label{app:ae-gaussian}

\subsection{General algorithm}\label{app:qsvt-gaussian}
To calculate the absolute value of the absorption power given by \eqref{eq:sum-jE}, one needs to compute the Gaussian function $G$ in the quantum circuit.
This function can be approximated by a polynomial by using the Fourier approach, Eq.~\eqref{eq:fourier-ck}. 
The circuit for the polynomial is constructed by using the QSVT (Fig.~\ref{circ:qsvt-odd}).
% An arbitrary polynomial can be computed as a sum of two polynomials of opposite parity by using the Linear Combination of Unitaries (LCU) as shown in Ref.~\onlinecite{Dong21}.
The QSVT finds the polynomial as a function of singular values of some given matrix $M$.
If the matrix is diagonal, and the spatial coordinate grid is placed at the diagonal, $M = \text{diag}(x)$, then the QSVT computes the polynomial of the coordinate $x$.
% In such a way, we can construct a function e.g. of the spatial coordinate $x$.
% In this case, the matrix $A$ for the QSVT algorithm might be formed by placing the coordinates $x_j$ at the matrix diagonal: $A = \text{diag}(x_j)$. 
After that, $M$ is block-encoded into the unitary matrix $U_A$, which is used within the QSVT circuit as shown in Fig.~\ref{circ:qsvt-odd}.
However, since the $x$ grid, $x_j$, is a linear function of the index $j$, it is problematic to encode the $x$ points directly into $U_A$ by using quantum gates, which are more naturally suited to represent rotations.
Instead, one can encode $\psi = \sin(x)$ and consider the Gaussian as a function of $\arcsin(\psi)$:
\begin{equation}\label{eq:gaussian}
    G(\psi) = \beta_{\rm sc} \exp\ylb -\frac{\arcsin^2(\psi)}{2\mu^2} \yrb,
\end{equation}
where $\beta_{\rm sc}$ is a rescaling factor, $\mu$ is the Gaussian width.
To block-encode $\psi_j=\sin(x_j)$, the circuit shown in Fig.~\ref{circ:x-block-encoding} is used.
This circuit generates the sine of $x_j$:
\begin{equation}\label{sys:x-phi-encoding}
    x_j = \alpha_0 + j\Delta x,\quad j=[0,N_x),\quad \Delta x = 2\alpha/N_x.
\end{equation}
Similar to Sec.~\ref{sec:discr}, we have $N_x = 2^{n_x}$, and the register $r_x$ in Fig.~\ref{circ:x-block-encoding} has $n_x$ qubits.
For the given integer $j$ encoded as a bit string in the register $r_x$, the circuit returns $\sin(x_j)$ as the amplitude of the zero state of the ancilla $a$. 
It can be shown using Eq.~\eqref{eq:ry} that the $x$-grid from $-1.0 - x_c$ to $1.0 - x_c$ for some real $x_c$ is generated by the following parameters:
\begin{equation}\label{sys:alphas-gauss}
    \alpha_0 = -1.0 - x_c,\quad \alpha = N_x/(N_x-1).
\end{equation}
The Gaussian generated on this grid is centered at $x_c$ (Fig.~\ref{fig:gauss}).
% The Gaussian generated on this grid can be interpreted as the profile with the width $\mu$ on $x\in [-1.0, 1.0]$ with the peak at $x_c$ (Fig.~\ref{fig:gauss}).

The Gaussian \eqref{eq:gaussian} is approximated by a polynomial of definite parity by using the Fourier approach~\eqref{eq:fourier-ck}, and the QSVT angles are computed by using the minimization procedure.\cite{Dong21}
(By setting $\beta_{\rm sc} < 1$ in Eq.~\eqref{eq:gaussian}, one can speed up the calculation of the QSVT angles.)
Since the Gaussian is an even function, the resulting QSVT circuit does not include the last three operators indicated by the dashed box in Fig.~\ref{circ:qsvt-odd}.
The QSVT circuit uses the oracle $U_A$ shown in Fig.~\ref{circ:x-block-encoding} to encode $\psi(x)$.
As shown in Fig.~\ref{fig:gauss}, the number of the QSVT angles depends logarithmically on $\epsilon_{\rm qsvt}$ for small $\mu$, and the dependence becomes polynomial for $\mu \gtrapprox 0.25$ (Fig.~\ref{fig:gauss}). 
If the Gaussian peak is narrow ($\mu < 0.05$) or becomes comparable with the length of the spatial domain $x = [-1.0, 1.0]$ ($\mu > 0.25$), the number of angles grows exponentially with $\mu$ for a fixed $\epsilon_{\rm qsvt}$.

% As shown in Fig.~\ref{fig:gauss}, the length of the QSVT (number of the QSVT angles or, in other words, number of queries to the block-encoding oracle) depends logarithmically on $\epsilon_{\rm qsvt}$ only in a certain interval of the Gaussian width, $\mu \approx [0.01, 0.25)$.
% Beyond these limits, where the Gaussian peak is narrow or becomes comparable with the length of the spatial domain $x = [-1.0, 1.0]$, the number of queries in the QSVT circuit grows exponentially with $\mu$ for a fixed $\epsilon_{\rm qsvt}$.
% The number of the QSVT angles depends logarithmically on $\epsilon_{\rm qsvt}$ for small $\mu$, and the dependence becomes polynomial for wide Gaussians (Fig.~\ref{fig:gauss}). 

% ******************************************
% *** Gaussians: circuit ***

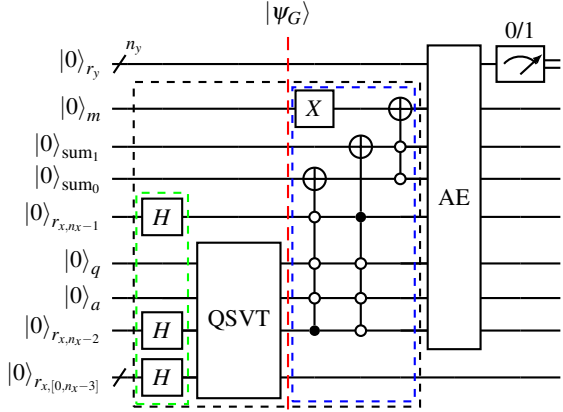
\begin{figure}[t!]
\centering
\begin{quantikz}[row sep={0.1cm},column sep={0.2cm}]
\lstick{$\ket{0}_{r_y}$}                  &\qwbundle{n_y}&\qw     &\qw\slice{$\ket{\psi_G}$}                  
                                                                                           &\qw        &\qw             &\qw          &\gate[8]{\text{AE}} &\meter{0/1} &\cw\\      
\lstick{$\ket{0}_{m}$}                    &\qw           &\qw\yggr{8}{5}{}{black}{0}     
                                                                  &\qw                     &\gate{X}\yggr{8}{3}{}{blue}{-2}        
                                                                                                       &\qw             &\targ{}      &\qw                 &\qw         &\qw\\ 
\lstick{$\ket{0}_{\rm sum_1}$}            &\qw           &\qw     &\qw                     &\qw        &\targ{}         &\octrl{-1}   &\qw                 &\qw         &\qw\\    
\lstick{$\ket{0}_{\rm sum_0}$}            &\qw           &\qw     &\qw                     &\targ{}    &\qw             &\octrl{-1}   &\qw                 &\qw         &\qw\\
\lstick{$\ket{0}_{r_{x,n_x-1}}$}          &\qw           &\gate{H}\yggr{5}{1}{}{green}{-1}            
                                                                  &\qw                     &\octrl{-1} &\ctrl{-2}       &\qw          &\qw                 &\qw         &\qw\\
\lstick{$\ket{0}_q$}                      &\qw           &\qw     &\gate[4]{\text{QSVT}}   &\octrl{-1} &\octrl{-1}      &\qw          &\qw                 &\qw         &\qw\\       
\lstick{$\ket{0}_a$}                      &\qw           &\qw     &\qw                     &\octrl{-1} &\octrl{-1}      &\qw          &\qw                 &\qw         &\qw\\
\lstick{$\ket{0}_{r_{x,n_x-2}}$}          &\qw           &\gate{H}&\qw                     &\ctrl{-1}  &\octrl{-1}      &\qw          &\qw                 &\qw         &\qw\\
\lstick{$\ket{0}_{r_{x,[0, n_x-3]}}$}     &\qwbundle{}   &\gate{H}&\qw                     &\qw        &\qw             &\qw          &\qw                 &\qw         &\qw  
\end{quantikz}
\caption{\label{circ:two-gaussians}
The circuit for the construction of one Gaussian in the left spatial domain ($x < 0$) and another Gaussian in the right domain ($x > 0$) as shown in Fig.~\ref{fig:two-gaussians}.
The circuit also integrates the function in the second and third quarters of the spatial area, $x = (-0.5, 0.5)$.
The QSVT circuit uses the subcircuit presented in Fig.~\ref{circ:x-block-encoding} as the block-encoding oracle.
The green box indicates the initialization subcircuit.
The blue box entangles the unit state $\ket{1}_m$ with the spatial distribution of the Gaussians in $x=(-0.5, 0.5)$ by using the ancilla register `sum'.
The most significant qubit of the register $r_x$ is separated from the rest of the register qubits to show correctly the location of the QSVT subcircuit.
The black dashed box indicates the circuit $U_{\rm prep}$.
}
\end{figure}

% ******************************************

\subsection{Two Gaussians}\label{app:two-gaussians}
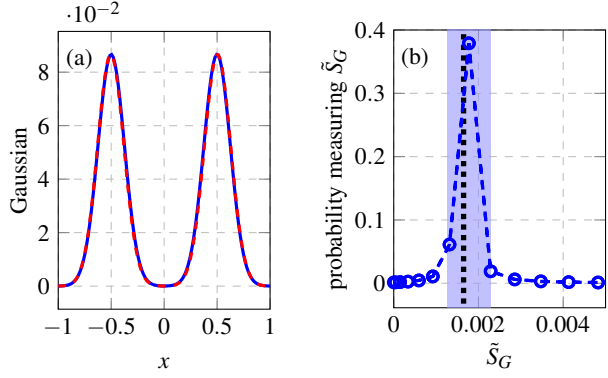
\begin{figure}
\centering
% ---------------------------------------------------------------------------------------
% ---------------------------------------------------------------------------------------
\begin{tikzpicture}
    \begin{axis}[
    	xlabel={$x$},
    	ylabel={Gaussian},
    	legend pos=north west,
     	xmin = -1.0,
     	xmax =  1.0,
    	extra x ticks={},
    	extra y ticks={},
    	xmajorgrids=true, ymajorgrids=true, grid style=dashed,
    	height=0.30\textwidth, 
     	width=0.25\textwidth,
        title style={yshift=-3},
    	ylabel style={yshift=-0.5ex},
    	xlabel style={yshift=-0.5ex},
        xticklabel style={
            text width=2.5em,
            align=center
        },
    	legend pos = north east,
    	legend style={
    	    fill=white, fill opacity=0.9, draw opacity=1,text opacity=1,
    	    font=\small,  %\tiny, \small
    	},
    ]
    
    \addplot [blue,  solid, line width=1.2pt] table [y=Y, x=X]{./data/two_gauss_cl.dat};
    \addplot [red,   dashed, line width=1.2pt] table [y=Y, x=X]{./data/two_gauss_qc.dat};
    
    % \node[blue]      at (rel axis cs: 0.9, 0.90) {CL};
    % \node[red]       at (rel axis cs: 0.9, 0.60) {QC};
    
    \node[black] at (rel axis cs: 0.1, 0.90) {(a)};
    \end{axis}
\end{tikzpicture}
% ---------------------------------------------------------------------------------------
% ---------------------------------------------------------------------------------------
\begin{tikzpicture}
    \begin{axis}[
    	xlabel={$\tilde{S}_G$},
    	ylabel={probability measuring $\tilde{S}_G$},
    	legend pos=north west,
     	ymax =  0.4,
    	xmin = 0,
     	xmax = 5e-3,
    	extra x ticks={},
    	extra y ticks={},
    	xmajorgrids=true, ymajorgrids=true, grid style=dashed,
    	height=0.30\textwidth, 
     	width=0.25\textwidth,
        title style={
            yshift=-3
        },
    	ylabel style={yshift=-0.8ex},
    	xlabel style={yshift=0.5ex},
        xticklabel style={
            text width=2.5em,
            align=center,
            /pgf/number format/fixed,
            /pgf/number format/precision=3
        },
        scaled x ticks=false,
    	legend pos = north east,
    	legend style={
    	    fill=white, fill opacity=0.9, draw opacity=1,text opacity=1,
    	    font=\small,  %\tiny, \small
    	},
    ]
    
    \fill[blue!40, opacity=0.6] (axis cs: 0.001264, -0.1) rectangle (axis cs: 0.002282, 0.6);
    \draw [dotted, line width=2.0pt, black] (1.649e-03,-1.0) -- (1.649e-03,8.0);
    \addplot [blue, dashed, mark=o, mark options={solid}, line width=1.2pt] table [y=Y, x=X]{./data/ae_gauss_mu0.250_ny6.dat};
    
    \node[black] at (rel axis cs: 0.1, 0.90) {(b)};
    \end{axis}
\end{tikzpicture}
% ---------------------------------------------------------------------------------------
% ---------------------------------------------------------------------------------------
\caption{\label{fig:two-gaussians} 
(a) Two Gaussians encoded in the intermediate state $\ket{\psi_G}$ produced in the circuit shown in Fig.~\ref{circ:two-gaussians} 
(red dashed line) and computed classically (solid blue line). 
Here, $\mu = 0.250$ is used in Eq.~\eqref{eq:gaussian}, and the resulting width of each Gaussian is $\mu = 0.125$.
b: Probability distribution of measurement results $\tilde{S}_G$ from the circuit shown in Fig.~\ref{circ:two-gaussians}, where $n_y = 6$.
The shaded area marks the interval where the analytical error is bounded as described in Eq.~\eqref{eq:ae-delta}.
The vertical black dotted line corresponds to $S_G$ computed classically using Eq.~\eqref{eq:SG}.
}
\end{figure}

To demonstrate this technique, we construct two Gaussians by using the QSVT and integrate them in space by using AE. 
This is an illustration of the `parallel' computation in quantum circuits: by applying the QSVT circuit only once, we construct two Gauss functions at the same time.
The corresponding circuit is shown in Fig.~\ref{circ:two-gaussians}, where the $x$-grid is prepared by creating a uniform superposition of states in the register $r_x$ and using the parameters from Eqs.~\eqref{sys:alphas-gauss}.
The QSVT circuit is applied only to the first $n_x -1$ qubits of the register $r_x$.
Since the last qubit (the most significant one) is in the superposition $(\ket{0}+\ket{1})/\sqrt{2}$, the QSVT entangles one Gaussian with the zero state of this qubit and another Gaussian with the unit state:
\begin{equation}\label{eq:sup-two-gaussians}
\begin{split}
\ket{\psi_G} = &2^{-n_x/2}\bigg(\ket{0}_{r_x, n_x-1}\ket{G(x)}_{r_x, [0,n_x-2]}\\
     &+ \ket{1}_{r_x, n_x-1}\ket{G(x)}_{r_x, [0,n_x-2]}\bigg),
\end{split}
\end{equation}
where the state $\ket{\psi_G}$ is indicated in Fig.~\ref{circ:two-gaussians}.
The most significant qubit in $r_x$ encodes the left spatial domain, $x < 0$, if the qubit is in the zero state, and the right spatial domain, $x > 0$, if the qubit is in the unit state.
Therefore, the superposition \eqref{eq:sup-two-gaussians} corresponds to one Gaussian in the domain $x=[-1.0,0.0)$ and another Gaussian in $x=(0.0,1.0]$ as shown in Fig.~\ref{fig:two-gaussians}.
Apart from that, if the original Gaussian is defined according to Eq.~\eqref{eq:gaussian}, then the resulting Gaussians have widths $\mu/2$.
Also, due to the initialization circuit (green box in Fig.~\ref{circ:two-gaussians}), the Gaussians are multiplied by the factor $\beta_{\rm init} = 2^{-n_x/2}$. 

The circuit shown in Fig.~\ref{circ:two-gaussians} integrates the Gaussians in the second and third quarters of the spatial domain:
\begin{subequations}\label{eq:SG}
\begin{eqnarray}
    &&S_G = p_{m,1}/N_x,\\
    &&p_{m,1} = \beta_{\rm init}^2 \sum_{x = (-0.5, 0.5)} |G(x)|^2.
\end{eqnarray}
\end{subequations}
The quantum state encoding the desired spatial interval is entangled with $\ket{1}_m$, and then $p_{m,1}$ is computed by using the AE.
As explained in Sec.~\ref{sec:ae}, the AE consists of several queries to the AA operator.
In our case, the latter is defined as 
\begin{equation}
    \text{AA} = U_{\rm prep} \text{REF}_0 U_{\rm prep}^\dagger \text{REF}_G,
\end{equation}
where $U_{\rm prep}$ is marked in Fig.~\ref{circ:two-gaussians} with the black dashed box, $\text{REF}_0$ is the reflector around the initial (in our case, zero) state, $\text{REF}_G$ is the reflector around the state of interest produced by $U_{\rm prep}$.
Here, the state of interest is $\ket{1}_m$, thus, $\text{REF}_G$ is represented by a single Pauli $Z$ gate at the qubit $m$.
The reflector $\text{REF}_0$ is constructed as the sequence $XZX$ at the qubit $m$ controlled by the zero states of the registers $r_x$, $a$, $q$ and `sum'.

The whole circuit shown in Fig.~\ref{circ:two-gaussians} is computed on an emulator of quantum computers.\cite{QSP-code}
The AE of $S_G$ is shown in Fig.~\ref{fig:two-gaussians} and compared with the value computed classically.

% ******************************************
% *** Gaussians: results ***
% ******************************************

% --------------------------------------------------------------------------------
% --- Gaussian as a window function ---
% --------------------------------------------------------------------------------
\subsection{Gaussian as a filter}\label{app:filter-gaussian}
\begin{figure}[t!]
\centering
\begin{quantikz}[transparent, row sep={0.4cm,between origins},column sep={0.2cm}]
\lstick{$\ket{0}_{q_G}$}         &\qw         &\qw                  &\qw                &\gate[7,label style={yshift=-0.4cm}]{\text{Gauss}}  
                                                                                                                 &\qw\\
\lstick{$\ket{0}_{q}$}           &\qw         &\qw                  &\gate[6]{\text{MI}}&\linethrough            &\qw\\
\lstick{$\ket{0}_{a_v}$}         &\qw         &\qw                  &\qw                &\linethrough            &\qw\\
\lstick{$\ket{0}_{a_j}$}         &\qw         &\qw                  &\qw                &\linethrough            &\qw\\
\lstick{$\ket{0}_{a_d}$}         &\qw         &\qw                  &\qw                &\qw                     &\qw\\
\lstick{$\ket{0}_{r_d}$}         &\qw         &\gate[2]{\text{INIT}}&\qw                &\linethrough            &\qw\\
\lstick{$\ket{0}_{r_j}$}         &\qwbundle{} &\qw                  &\qw                &\qw                     &\qw          
\end{quantikz}
\caption{\label{circ:gaussian-field}
The circuit for the computation of the product $G(x)F(x)$, where $G(x)$ is the Gaussian modeled by the QSVT subcircuit `Gauss', $F(x)$ is the field, electric or magnetic, simulated by the QSVT subcircuit `MI'.
The qubits that pass above the circuit `Gauss' are not used by it. 
The initialization `INIT' is the same as discussed in Sec.~\ref{sec:init}. 
}
\end{figure}
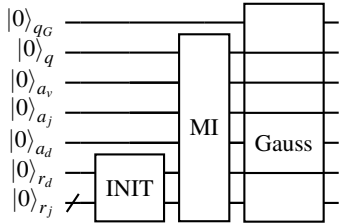

The Gaussian constructed by the QSVT can be used as a filter in real or Fourier space.
In particular, to calculate the product of the EM fields $F(x)$ with the Gaussian, $G(x)F(x)$, one can use the circuit shown in Fig.~\ref{circ:gaussian-field}.
Here, the main question is whether it is possible to use the same ancillae in the QSVT for the computation of $F(x)$ (the `MI' subcircuit) and in the QSVT for the Gaussian calculation (the `Gauss' subcircuit).
The computation shows that one does not need to control the `Gauss' by the zero states of the ancillae used in the `MI' (i.e., ancilla registers $q$, $a_v$, $a_j$), and the same ancilla $a_d$ can be used in both QSVT subcircuits.
However, for the Gaussian QSVT, it is necessary to introduce another ancilla $q_G$ equivalent to the ancilla $q$ in the `MI'. 

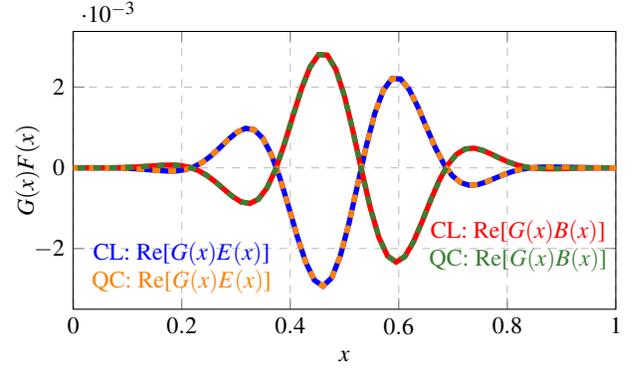
\begin{figure}
\centering
% ---------------------------------------------------------------------------------------
% ---------------------------------------------------------------------------------------
\begin{tikzpicture}
    \begin{axis}[
    	xlabel={$x$},
    	ylabel={$G(x)F(x)$},
    	legend pos=north west,
      	xmin = 0.0,
      	xmax =  1.0,
    % 	ymin = -1,
    %   ymax = 6e-3,
    % 	xtick = {8,16,32,64,128},
    	extra x ticks={},
    	extra y ticks={},
    	xmajorgrids=true, ymajorgrids=true, grid style=dashed,
    	height=0.3\textwidth, 
     	width=0.50\textwidth,
        title style={
            yshift=-3
        },
    	ylabel style={yshift=-2.0ex},
        xticklabel style={
            text width=2.5em,
            align=center
        },
    	legend pos = north east,
    	legend style={
    	    fill=white, fill opacity=0.9, draw opacity=1,text opacity=1,
    	    font=\small,  %\tiny, \small
    	},
    ]
    \addplot [blue,  solid, line width=2.0pt] table [y=Y, x=X]{./data/gauss_E_cl.dat};
    \addplot [red,   solid, line width=2.0pt] table [y=Y, x=X]{./data/gauss_B_cl.dat};
    \addplot [orange, dashed, line width=2.0pt] table [y=Y, x=X]{./data/gauss_E_qc.dat};
    \addplot [OliveGreen,   dashed, line width=2.0pt] table [y=Y, x=X]{./data/gauss_B_qc.dat};
    
    \node[blue]   at (rel axis cs: 0.20, 0.20) {CL: Re[$G(x)E(x)$]};
    \node[orange] at (rel axis cs: 0.20, 0.10) {QC: Re[$G(x)E(x)$]};
    
    \node[red]   at (rel axis cs: 0.82, 0.28)      {CL: Re[$G(x)B(x)$]};
    \node[OliveGreen] at (rel axis cs: 0.82, 0.18) {QC: Re[$G(x)B(x)$]};

    \end{axis}
\end{tikzpicture}
% ---------------------------------------------------------------------------------------
% ---------------------------------------------------------------------------------------
\caption{\label{fig:gaussian-field} 
The product of the Gaussian $G$ with electromagnetic fields simulated classically (CL, solid lines) and by using the circuit shown in Fig.~\ref{circ:gaussian-field} (QC, dashed lines).
}
\end{figure}

Since the `Gauss' is not controlled by the qubit $r_d$, which is responsible for the choice between the electric and magnetic fields, the circuit constructs the product $G(x)F(x)$ for both fields in parallel.
The circuit shown in Fig.~\ref{circ:gaussian-field} outputs the spatial distribution $G(x)E(x)$ (entangled with $\ket{0}_{r_d}$) and $G(x)B(x)$ (entangled with $\ket{1}_{r_d}$) if all ancillae are in the zero state. 
% Here, we do not discuss the measurements.
The signals $G(x)F(x)$ obtained from the emulation of this circuit are shown in Fig.~\ref{fig:gaussian-field} and compared with classical simulations.
% Since the fields are defined over the spatial grid $[0, 1.0]$, and the Gaussian in the QSVT circuit is calculated over $x\in[-1.0, 1.0]$, in classical simulations, one needs to shift the Gaussian by $0.5$ and rescale the Gaussian width, $\mu/2.0$.
% Finally, to compare with QC results, the amplitudes of the classical profiles $G(x)F(x)$ need to be downscaled by the appropriate value of the condition $\kappa$ and $\beta_{\rm sc}$ (as discussed in Eq.~\eqref{eq:qsvt-rescaling}).
% The rescaling coefficient $\beta_{\rm sc}$ in the Gaussian QSVT simulation is $0.98$. (For the Gaussian, it is possible to take $\beta_{\rm sc}$ very close to $1.0$, but this usually increases the computation time of the QSVT angles).

% ******************************************
% *** Circuit: Gaussian(x) * E(x) ***
% ******************************************

% ******************************************
% *** Figure: Gaussian(x) * E(x) ***
% ******************************************

\bibliography{main}

\begin{thebibliography}{10}
\expandafter\ifx\csname url\endcsname\relax
  \def\url#1{\texttt{#1}}\fi
\expandafter\ifx\csname urlprefix\endcsname\relax\def\urlprefix{URL }\fi
\expandafter\ifx\csname href\endcsname\relax
  \def\href#1#2{#2} \def\path#1{#1}\fi

\bibitem{Fisch87}
N.~J. Fisch, \href{https://link.aps.org/doi/10.1103/RevModPhys.59.175}{Theory
  of current drive in plasmas}, Reviews of Modern Physics 59 (1987) 175--234.
\newblock \href {https://doi.org/10.1103/RevModPhys.59.175}
  {\path{doi:10.1103/RevModPhys.59.175}}.
\newline\urlprefix\url{https://link.aps.org/doi/10.1103/RevModPhys.59.175}

\bibitem{Pinsker01}
R.~I. Pinsker,
  \href{https://aip.scitation.org/doi/abs/10.1063/1.1343512}{Introduction to
  wave heating and current drive in magnetized plasmas}, Physics of Plasmas
  8~(4) (2001) 1219--1228.
\newblock \href
  {http://arxiv.org/abs/https://aip.scitation.org/doi/pdf/10.1063/1.1343512}
  {\path{arXiv:https://aip.scitation.org/doi/pdf/10.1063/1.1343512}}, \href
  {https://doi.org/10.1063/1.1343512} {\path{doi:10.1063/1.1343512}}.
\newline\urlprefix\url{https://aip.scitation.org/doi/abs/10.1063/1.1343512}

\bibitem{Prater04}
R.~Prater, \href{https://doi.org/10.1063/1.1690762}{Heating and current drive
  by electron cyclotron waves}, Physics of Plasmas 11~(5) (2004) 2349--2376.
\newblock \href {http://arxiv.org/abs/https://doi.org/10.1063/1.1690762}
  {\path{arXiv:https://doi.org/10.1063/1.1690762}}, \href
  {https://doi.org/10.1063/1.1690762} {\path{doi:10.1063/1.1690762}}.
\newline\urlprefix\url{https://doi.org/10.1063/1.1690762}

\bibitem{Novikau22}
I.~Novikau, E.~A. Startsev, I.~Y. Dodin,
  \href{https://link.aps.org/doi/10.1103/PhysRevA.105.062444}{Quantum signal
  processing for simulating cold plasma waves}, Physical Review A 105 (2022)
  062444.
\newblock \href {https://doi.org/10.1103/PhysRevA.105.062444}
  {\path{doi:10.1103/PhysRevA.105.062444}}.
\newline\urlprefix\url{https://link.aps.org/doi/10.1103/PhysRevA.105.062444}

\bibitem{Engel19}
A.~Engel, G.~Smith, S.~E. Parker,
  \href{https://link.aps.org/doi/10.1103/PhysRevA.100.062315}{Quantum algorithm
  for the vlasov equation}, Physical Review A 100 (2019) 062315.
\newblock \href {https://doi.org/10.1103/PhysRevA.100.062315}
  {\path{doi:10.1103/PhysRevA.100.062315}}.
\newline\urlprefix\url{https://link.aps.org/doi/10.1103/PhysRevA.100.062315}

\bibitem{Lancellotti06}
V.~Lancellotti, D.~Milanesio, R.~Maggiora, G.~Vecchi, V.~Kyrytsya,
  \href{https://doi.org/10.1088/0029-5515/46/7/s10}{{TOPICA}: an accurate and
  efficient numerical tool for analysis and design of {ICRF} antennas}, Nuclear
  Fusion 46~(7) (2006) S476--S499.
\newblock \href {https://doi.org/10.1088/0029-5515/46/7/s10}
  {\path{doi:10.1088/0029-5515/46/7/s10}}.
\newline\urlprefix\url{https://doi.org/10.1088/0029-5515/46/7/s10}

\bibitem{Shiraiwa17}
S.~Shiraiwa, J.~C. Wright, P.~T. Bonoli, T.~Kolev, M.~Stowell,
  \href{https://doi.org/10.1051/epjconf/201715703048}{{RF} wave simulation for
  cold edge plasmas using the {MFEM} library}, EPJ Web Conf. 157 (2017) 03048.
\newblock \href {https://doi.org/10.1051/epjconf/201715703048}
  {\path{doi:10.1051/epjconf/201715703048}}.
\newline\urlprefix\url{https://doi.org/10.1051/epjconf/201715703048}

\bibitem{Svidzinski18}
V.~A. Svidzinski, J.~S. Kim, L.~Zhao, S.~A. Galkin, J.~A. Spencer,
  \href{https://doi.org/10.1063/1.5037110}{Hybrid iterative approach for
  simulation of radio-frequency fields in plasma}, Physics of Plasmas 25~(8)
  (2018) 082509.
\newblock \href {http://arxiv.org/abs/https://doi.org/10.1063/1.5037110}
  {\path{arXiv:https://doi.org/10.1063/1.5037110}}, \href
  {https://doi.org/10.1063/1.5037110} {\path{doi:10.1063/1.5037110}}.
\newline\urlprefix\url{https://doi.org/10.1063/1.5037110}

\bibitem{Hillairet21}
J.~Hillairet, P.~Mollard, L.~Colas, W.~Helou, G.~Urbanczyk, J.-M. Bernard,
  J.-M. Delaplanche, F.~Durand, N.~Faure, P.~Garibaldi, G.~Lombard,
  C.~Bourdelle, C.~Desgranges, E.~Delmas, R.~Dumont, A.~Ekedahl, F.~Ferlay,
  M.~Goniche, C.~Guillemaut, G.~Hoang, P.~Maget, R.~Volpe, Y.~Song, Q.~Yang,
  Z.~Chen, Y.~Wang, H.~Xu, S.~Yuan, Y.~Zhao, F.~Durodie, E.~Lerche, R.~Ragona,
  N.~Bertelli, M.~Ono, S.~Shiraiwa, V.~Bobkov, C.~Klepper, C.~Lau, E.~Martin,
  B.~Lu, R.~Maggiora, D.~Milanesio, K.~Vulliez, G.~Wallace, W.~Team,
  \href{https://doi.org/10.1088/1741-4326/ac1759}{{WEST} actively cooled load
  resilient ion cyclotron resonance heating system results}, Nuclear Fusion
  61~(9) (2021) 096030.
\newblock \href {https://doi.org/10.1088/1741-4326/ac1759}
  {\path{doi:10.1088/1741-4326/ac1759}}.
\newline\urlprefix\url{https://doi.org/10.1088/1741-4326/ac1759}

\bibitem{Zhang22}
W.~Zhang, R.~Bilato, V.~Bobkov, A.~Cathey, A.~D. Siena, M.~Hoelzl, A.~Messiaen,
  J.~Myra, G.~S. L{\'{o}}pez, W.~Tierens, M.~Usoltceva, J.~Wright, the ASDEX
  Upgrade~Team, the EUROfusion MST1~Team,
  \href{https://doi.org/10.1088/1741-4326/ac38c8}{Recent progress in modeling
  {ICRF}-edge plasma interactions with application to {ASDEX} upgrade}, Nuclear
  Fusion 62~(7) (2022) 075001.
\newblock \href {https://doi.org/10.1088/1741-4326/ac38c8}
  {\path{doi:10.1088/1741-4326/ac38c8}}.
\newline\urlprefix\url{https://doi.org/10.1088/1741-4326/ac38c8}

\bibitem{Guttenfelder22}
W.~Guttenfelder, D.~Battaglia, E.~Belova, N.~Bertelli, M.~Boyer, C.~Chang,
  A.~Diallo, V.~Duarte, F.~Ebrahimi, E.~Emdee, N.~Ferraro, E.~Fredrickson,
  N.~Gorelenkov, W.~Heidbrink, Z.~Ilhan, S.~Kaye, E.-H. Kim, A.~Kleiner,
  F.~Laggner, M.~Lampert, J.~Lestz, C.~Liu, D.~Liu, T.~Looby, N.~Mandell,
  R.~Maingi, J.~Myra, S.~Munaretto, M.~Podest{\`{a}}, T.~Rafiq, R.~Raman,
  M.~Reinke, Y.~Ren, J.~R. Ruiz, F.~Scotti, S.~Shiraiwa, V.~Soukhanovskii,
  P.~Vail, Z.~Wang, W.~Wehner, A.~White, R.~White, B.~Woods, J.~Yang,
  S.~Zweben, S.~Banerjee, R.~Barchfeld, R.~Bell, J.~Berkery, A.~Bhattacharjee,
  A.~Bierwage, G.~Canal, X.~Chen, C.~Clauser, N.~Crocker, C.~Domier, T.~Evans,
  M.~Francisquez, K.~Gan, S.~Gerhardt, R.~Goldston, T.~Gray, A.~Hakim,
  G.~Hammett, S.~Jardin, R.~Kaita, B.~Koel, E.~Kolemen, S.-H. Ku, S.~Kubota,
  B.~LeBlanc, F.~Levinton, J.~Lore, N.~Luhmann, R.~Lunsford, R.~Maqueda,
  J.~Menard, J.~Nichols, M.~Ono, J.-K. Park, F.~Poli, T.~Rhodes, J.~Riquezes,
  D.~Russell, S.~Sabbagh, E.~Schuster, D.~Smith, D.~Stotler, B.~Stratton,
  K.~Tritz, W.~Wang, B.~Wirth,
  \href{https://doi.org/10.1088/1741-4326/ac5448}{{NSTX}-u theory, modeling and
  analysis results}, Nuclear Fusion 62~(4) (2022) 042023.
\newblock \href {https://doi.org/10.1088/1741-4326/ac5448}
  {\path{doi:10.1088/1741-4326/ac5448}}.
\newline\urlprefix\url{https://doi.org/10.1088/1741-4326/ac5448}

\bibitem{Dodin20}
I.~Y. Dodin, E.~A. Startsev, \href{https://doi.org/10.1063/5.0056974}{On
  applications of quantum computing to plasma simulations}, Physics of Plasmas
  28~(9) (2021) 092101.
\newblock \href {http://arxiv.org/abs/https://doi.org/10.1063/5.0056974}
  {\path{arXiv:https://doi.org/10.1063/5.0056974}}, \href
  {https://doi.org/10.1063/5.0056974} {\path{doi:10.1063/5.0056974}}.
\newline\urlprefix\url{https://doi.org/10.1063/5.0056974}

\bibitem{HHL09}
A.~W. Harrow, A.~Hassidim, S.~Lloyd,
  \href{https://link.aps.org/doi/10.1103/PhysRevLett.103.150502}{Quantum
  algorithm for linear systems of equations}, Phys. Rev. Lett. 103 (2009)
  150502.
\newblock \href {https://doi.org/10.1103/PhysRevLett.103.150502}
  {\path{doi:10.1103/PhysRevLett.103.150502}}.
\newline\urlprefix\url{https://link.aps.org/doi/10.1103/PhysRevLett.103.150502}

\bibitem{Ambainis12}
A.~Ambainis, \href{http://drops.dagstuhl.de/opus/volltexte/2012/3426}{{Variable
  time amplitude amplification and quantum algorithms for linear algebra
  problems}}, in: C.~D{\"u}rr, T.~Wilke (Eds.), 29th International Symposium on
  Theoretical Aspects of Computer Science (STACS 2012), Vol.~14 of Leibniz
  International Proceedings in Informatics (LIPIcs), Schloss
  Dagstuhl--Leibniz-Zentrum fuer Informatik, Dagstuhl, Germany, 2012, pp.
  636--647.
\newblock \href {https://doi.org/10.4230/LIPIcs.STACS.2012.636}
  {\path{doi:10.4230/LIPIcs.STACS.2012.636}}.
\newline\urlprefix\url{http://drops.dagstuhl.de/opus/volltexte/2012/3426}

\bibitem{Childs17}
A.~M. Childs, R.~Kothari, R.~D. Somma,
  \href{https://doi.org/10.1137/16M1087072}{Quantum algorithm for systems of
  linear equations with exponentially improved dependence on precision}, SIAM
  Journal on Computing 46~(6) (2017) 1920--1950.
\newblock \href {http://arxiv.org/abs/https://doi.org/10.1137/16M1087072}
  {\path{arXiv:https://doi.org/10.1137/16M1087072}}, \href
  {https://doi.org/10.1137/16M1087072} {\path{doi:10.1137/16M1087072}}.
\newline\urlprefix\url{https://doi.org/10.1137/16M1087072}

\bibitem{Scherer17}
A.~Scherer, B.~Valiron, S.-C. Mau, S.~Alexander, E.~van~den Berg, T.~E.
  Chapuran, \href{https://doi.org/10.1007%2Fs11128-016-1495-5}{Concrete
  resource analysis of the quantum linear-system algorithm used to compute the
  electromagnetic scattering cross section of a 2d target}, Quantum Information
  Processing 16~(3) (jan 2017).
\newblock \href {https://doi.org/10.1007/s11128-016-1495-5}
  {\path{doi:10.1007/s11128-016-1495-5}}.
\newline\urlprefix\url{https://doi.org/10.1007%2Fs11128-016-1495-5}

\bibitem{Low17}
G.~H. Low, I.~L. Chuang,
  \href{https://link.aps.org/doi/10.1103/PhysRevLett.118.010501}{Optimal
  {Hamiltonian} simulation by quantum signal processing}, Physical Review
  Letters 118 (2017) 010501.
\newblock \href {https://doi.org/10.1103/PhysRevLett.118.010501}
  {\path{doi:10.1103/PhysRevLett.118.010501}}.
\newline\urlprefix\url{https://link.aps.org/doi/10.1103/PhysRevLett.118.010501}

\bibitem{Low19}
G.~H. Low, I.~L. Chuang,
  \href{https://doi.org/10.22331/q-2019-07-12-163}{Hamiltonian simulation by
  qubitization}, {Quantum} 3 (2019) 163.
\newblock \href {https://doi.org/10.22331/q-2019-07-12-163}
  {\path{doi:10.22331/q-2019-07-12-163}}.
\newline\urlprefix\url{https://doi.org/10.22331/q-2019-07-12-163}

\bibitem{Gilyen19}
A.~Gily\'{e}n, Y.~Su, G.~H. Low, N.~Wiebe,
  \href{https://doi.org/10.1145/3313276.3316366}{Quantum singular value
  transformation and beyond: Exponential improvements for quantum matrix
  arithmetics}, in: Proceedings of the 51st Annual ACM SIGACT Symposium on
  Theory of Computing, STOC 2019, Association for Computing Machinery, New
  York, NY, USA, 2019, p. 193–204.
\newblock \href {https://doi.org/10.1145/3313276.3316366}
  {\path{doi:10.1145/3313276.3316366}}.
\newline\urlprefix\url{https://doi.org/10.1145/3313276.3316366}

\bibitem{Martyn21}
J.~M. Martyn, Z.~M. Rossi, A.~K. Tan, I.~L. Chuang,
  \href{https://link.aps.org/doi/10.1103/PRXQuantum.2.040203}{Grand unification
  of quantum algorithms}, PRX Quantum 2 (2021) 040203.
\newblock \href {https://doi.org/10.1103/PRXQuantum.2.040203}
  {\path{doi:10.1103/PRXQuantum.2.040203}}.
\newline\urlprefix\url{https://link.aps.org/doi/10.1103/PRXQuantum.2.040203}

\bibitem{Montanaro16}
A.~Montanaro, S.~Pallister,
  \href{https://link.aps.org/doi/10.1103/PhysRevA.93.032324}{Quantum algorithms
  and the finite element method}, Physical Review A 93 (2016) 032324.
\newblock \href {https://doi.org/10.1103/PhysRevA.93.032324}
  {\path{doi:10.1103/PhysRevA.93.032324}}.
\newline\urlprefix\url{https://link.aps.org/doi/10.1103/PhysRevA.93.032324}

\bibitem{Lin22}
L.~Lin, \href{https://arxiv.org/abs/2201.08309}{Lecture notes on quantum
  algorithms for scientific computation} (2022).
\newblock \href {https://doi.org/10.48550/ARXIV.2201.08309}
  {\path{doi:10.48550/ARXIV.2201.08309}}.
\newline\urlprefix\url{https://arxiv.org/abs/2201.08309}

\bibitem{William07}
W.~H. Press, S.~A. Teukolsky, W.~T. Vetterling, B.~P. Flannery, Numerical
  Recipes 3rd Edition: The Art of Scientific Computing, 3rd Edition, Cambridge
  University Press, USA, 2007.

\bibitem{Remez34}
E.~Remez, Sur la d\'etermination des polyn\^{o}mes d'approximation de degr\'e
  donn\'ee, Comm. Soc. Math. Kharkov 10 (1934) 41.

\bibitem{Dong21code}
Efficient phase-factor evaluation in quantum signal processing: code,
  \url{https://github.com/qsppack/QSPPACK} (2022).

\bibitem{QSP-code}
Numerical \relax{QSP} framework implemented in {C++} with the application of
  \relax{QuEST} library, \url{https://github.com/ivanNovikau/QSVT\_framework}
  (2022).

\bibitem{Haah20}
J.~Haah, \href{http://dx.doi.org/10.22331/q-2019-10-07-190}{Product
  decomposition of periodic functions in quantum signal processing}, Quantum 3
  (2019) 190.
\newblock \href {https://doi.org/10.22331/q-2019-10-07-190}
  {\path{doi:10.22331/q-2019-10-07-190}}.
\newline\urlprefix\url{http://dx.doi.org/10.22331/q-2019-10-07-190}

\bibitem{Haah20code}
Computation of angles for quantum signal processing in {F}\#,
  \url{https://github.com/microsoft/Quantum-NC/tree/main/src/simulation/qsp}
  (2020).

\bibitem{Dong21}
Y.~Dong, X.~Meng, K.~B. Whaley, L.~Lin,
  \href{http://dx.doi.org/10.1103/PhysRevA.103.042419}{Efficient phase-factor
  evaluation in quantum signal processing}, Physical Review A 103~(4) (Apr
  2021).
\newblock \href {https://doi.org/10.1103/physreva.103.042419}
  {\path{doi:10.1103/physreva.103.042419}}.
\newline\urlprefix\url{http://dx.doi.org/10.1103/PhysRevA.103.042419}

\bibitem{Ying22}
L.~Ying, \href{https://arxiv.org/abs/2202.02671}{Stable factorization for phase
  factors of quantum signal processing} (2022).
\newblock \href {https://doi.org/10.48550/ARXIV.2202.02671}
  {\path{doi:10.48550/ARXIV.2202.02671}}.
\newline\urlprefix\url{https://arxiv.org/abs/2202.02671}

\bibitem{Bank89}
R.~E. Bank, L.~R. Scott, \href{http://www.jstor.org/stable/2157745}{On the
  conditioning of finite element equations with highly refined meshes}, SIAM
  Journal on Numerical Analysis 26~(6) (1989) 1383--1394.
\newline\urlprefix\url{http://www.jstor.org/stable/2157745}

\bibitem{Saad03}
Y.~Saad, Iterative methods for sparse linear systems, Society for Industrial
  and Applied Mathematics, 2003.

\bibitem{Vorst92}
H.~A. van~der Vorst, \href{https://doi.org/10.1137/0913035}{{Bi-CGSTAB}: A fast
  and smoothly converging variant of bi-cg for the solution of nonsymmetric
  linear systems}, SIAM Journal on Scientific and Statistical Computing 13~(2)
  (1992) 631--644.
\newblock \href {http://arxiv.org/abs/https://doi.org/10.1137/0913035}
  {\path{arXiv:https://doi.org/10.1137/0913035}}, \href
  {https://doi.org/10.1137/0913035} {\path{doi:10.1137/0913035}}.
\newline\urlprefix\url{https://doi.org/10.1137/0913035}

\bibitem{Saad86}
Y.~Saad, M.~H. Schultz, \href{https://doi.org/10.1137/0907058}{{GMRES}: A
  generalized minimal residual algorithm for solving nonsymmetric linear
  systems}, SIAM Journal on Scientific and Statistical Computing 7~(3) (1986)
  856--869.
\newblock \href {http://arxiv.org/abs/https://doi.org/10.1137/0907058}
  {\path{arXiv:https://doi.org/10.1137/0907058}}, \href
  {https://doi.org/10.1137/0907058} {\path{doi:10.1137/0907058}}.
\newline\urlprefix\url{https://doi.org/10.1137/0907058}

\bibitem{Freund93}
R.~W. Freund, \href{https://doi.org/10.1137/0914029}{A transpose-free
  quasi-minimal residual algorithm for non-hermitian linear systems}, SIAM
  Journal on Scientific Computing 14~(2) (1993) 470--482.
\newblock \href {http://arxiv.org/abs/https://doi.org/10.1137/0914029}
  {\path{arXiv:https://doi.org/10.1137/0914029}}, \href
  {https://doi.org/10.1137/0914029} {\path{doi:10.1137/0914029}}.
\newline\urlprefix\url{https://doi.org/10.1137/0914029}

\bibitem{Shewchuk94}
J.~R. Shewchuk, \href{https://dl.acm.org/doi/10.5555/865018}{Technical report
  no. cmu-cs-94-125, an introduction to the conjugate gradient method without
  the agonizing pain}, Tech. rep., Carnegie Mellon University, Pittsburgh,
  Pennsylvania, USA (1994).
\newline\urlprefix\url{https://dl.acm.org/doi/10.5555/865018}

\bibitem{Clader13}
B.~D. Clader, B.~C. Jacobs, C.~R. Sprouse,
  \href{https://link.aps.org/doi/10.1103/PhysRevLett.110.250504}{Preconditioned
  quantum linear system algorithm}, Phys. Rev. Lett. 110 (2013) 250504.
\newblock \href {https://doi.org/10.1103/PhysRevLett.110.250504}
  {\path{doi:10.1103/PhysRevLett.110.250504}}.
\newline\urlprefix\url{https://link.aps.org/doi/10.1103/PhysRevLett.110.250504}

\bibitem{Vahala20}
G.~Vahala, L.~Vahala, M.~Soe, A.~K. Ram, Unitary quantum lattice simulations
  for {Maxwell} equations in vacuum and in dielectric media, Journal of Plasma
  Physics 86~(5) (2020) 905860518.
\newblock \href {https://doi.org/10.1017/S0022377820001166}
  {\path{doi:10.1017/S0022377820001166}}.

\bibitem{Ram21}
A.~K. Ram, G.~Vahala, L.~Vahala, M.~Soe,
  \href{https://doi.org/10.1063/5.0067204}{Reflection and transmission of
  electromagnetic pulses at a planar dielectric interface: Theory and quantum
  lattice simulations}, AIP Advances 11~(10) (2021) 105116.
\newblock \href {http://arxiv.org/abs/https://doi.org/10.1063/5.0067204}
  {\path{arXiv:https://doi.org/10.1063/5.0067204}}, \href
  {https://doi.org/10.1063/5.0067204} {\path{doi:10.1063/5.0067204}}.
\newline\urlprefix\url{https://doi.org/10.1063/5.0067204}

\bibitem{Vahala21}
G.~Vahala, M.~Soe, L.~Vahala, A.~K. Ram,
  \href{https://doi.org/10.1080/10420150.2021.1891059}{One- and two-dimensional
  quantum lattice algorithms for {Maxwell} equations in inhomogeneous scalar
  dielectric media. ii: Simulations}, Radiation Effects and Defects in Solids
  176~(1-2) (2021) 64--72.
\newblock \href
  {http://arxiv.org/abs/https://doi.org/10.1080/10420150.2021.1891059}
  {\path{arXiv:https://doi.org/10.1080/10420150.2021.1891059}}, \href
  {https://doi.org/10.1080/10420150.2021.1891059}
  {\path{doi:10.1080/10420150.2021.1891059}}.
\newline\urlprefix\url{https://doi.org/10.1080/10420150.2021.1891059}

\bibitem{Koukoutsis22}
E.~Koukoutsis, K.~Hizanidis, A.~K. Ram, G.~Vahala,
  \href{https://arxiv.org/abs/2209.08523}{Dyson maps and unitary evolution for
  {Maxwell} equations in tensor dielectric media} (2022).
\newblock \href {https://doi.org/10.48550/ARXIV.2209.08523}
  {\path{doi:10.48550/ARXIV.2209.08523}}.
\newline\urlprefix\url{https://arxiv.org/abs/2209.08523}

\bibitem{Jin22}
S.~Jin, N.~Liu, Y.~Yu, \href{https://doi.org/10.1016\%2Fj.jcp.2022.111641}{Time
  complexity analysis of quantum difference methods for linear high dimensional
  and multiscale partial differential equations}, Journal of Computational
  Physics 471 (2022) 111641.
\newblock \href {https://doi.org/10.1016/j.jcp.2022.111641}
  {\path{doi:10.1016/j.jcp.2022.111641}}.
\newline\urlprefix\url{https://doi.org/10.1016\%2Fj.jcp.2022.111641}

\bibitem{Berry12}
D.~W. Berry, A.~M. Childs, Black-box hamiltonian simulation and unitary
  implementation, Quantum Info. Comput. 12~(1–2) (2012) 29–62.

\bibitem{Barenco95}
A.~Barenco, C.~H. Bennett, R.~Cleve, D.~P. DiVincenzo, N.~Margolus, P.~Shor,
  T.~Sleator, J.~A. Smolin, H.~Weinfurter,
  \href{https://link.aps.org/doi/10.1103/PhysRevA.52.3457}{Elementary gates for
  quantum computation}, Physical Review A 52 (1995) 3457--3467.
\newblock \href {https://doi.org/10.1103/PhysRevA.52.3457}
  {\path{doi:10.1103/PhysRevA.52.3457}}.
\newline\urlprefix\url{https://link.aps.org/doi/10.1103/PhysRevA.52.3457}

\bibitem{Jones19}
T.~Jones, A.~Brown, I.~Bush, S.~C. Benjamin,
  \href{https://doi.org/10.1038/s41598-019-47174-9}{Quest and high performance
  simulation of quantum computers}, Scientific Reports 9~(1) (2019) 10736.
\newblock \href {https://doi.org/10.1038/s41598-019-47174-9}
  {\path{doi:10.1038/s41598-019-47174-9}}.
\newline\urlprefix\url{https://doi.org/10.1038/s41598-019-47174-9}

\bibitem{Nielsen10}
M.~A. Nielsen, I.~L. Chuang, Quantum Computation and Quantum Information,
  Cambridge University Press; 10th Anniversary edition, 2010.

\bibitem{Brassard02}
G.~Brassard, P.~H{\o}yer, M.~Mosca, A.~Tapp,
  \href{http://dx.doi.org/10.1090/conm/305/05215}{Quantum amplitude
  amplification and estimation}, Quantum Computation and Information 305 (2002)
  53–74.
\newblock \href {https://doi.org/10.1090/conm/305/05215}
  {\path{doi:10.1090/conm/305/05215}}.
\newline\urlprefix\url{http://dx.doi.org/10.1090/conm/305/05215}

\bibitem{Stix92}
T.~H. Stix, Waves in Plasmas, AIP-Press, 1992.

\bibitem{Tracy14}
E.~R. Tracy, A.~J. Brizard, A.~S. Richardson, A.~N. Kaufman, Ray Tracing and
  Beyond: Phase Space Methods in Plasma Wave Theory, Cambridge University
  Press, New York, 2014.

\bibitem{Svidzinski16}
V.~A. Svidzinski, J.~S. Kim, J.~A. Spencer, L.~Zhao, S.~A. Galkin, E.~G.
  Evstatiev, \href{https://doi.org/10.1063/1.4966638}{Hot plasma dielectric
  response to radio-frequency fields in inhomogeneous magnetic field}, Physics
  of Plasmas 23~(11) (2016) 112101.
\newblock \href {http://arxiv.org/abs/https://doi.org/10.1063/1.4966638}
  {\path{arXiv:https://doi.org/10.1063/1.4966638}}, \href
  {https://doi.org/10.1063/1.4966638} {\path{doi:10.1063/1.4966638}}.
\newline\urlprefix\url{https://doi.org/10.1063/1.4966638}

\bibitem{Ortiz01}
G.~Ortiz, J.~E. Gubernatis, E.~Knill, R.~Laflamme,
  \href{https://link.aps.org/doi/10.1103/PhysRevA.64.022319}{Quantum algorithms
  for fermionic simulations}, Physical Review A 64 (2001) 022319.
\newblock \href {https://doi.org/10.1103/PhysRevA.64.022319}
  {\path{doi:10.1103/PhysRevA.64.022319}}.
\newline\urlprefix\url{https://link.aps.org/doi/10.1103/PhysRevA.64.022319}

\bibitem{Mitarai19}
K.~Mitarai, K.~Fujii,
  \href{https://link.aps.org/doi/10.1103/PhysRevResearch.1.013006}{Methodology
  for replacing indirect measurements with direct measurements}, Physical
  Review Research 1 (2019) 013006.
\newblock \href {https://doi.org/10.1103/PhysRevResearch.1.013006}
  {\path{doi:10.1103/PhysRevResearch.1.013006}}.
\newline\urlprefix\url{https://link.aps.org/doi/10.1103/PhysRevResearch.1.013006}

\bibitem{Buhrman01}
H.~Buhrman, R.~Cleve, J.~Watrous, R.~de~Wolf,
  \href{https://link.aps.org/doi/10.1103/PhysRevLett.87.167902}{Quantum
  fingerprinting}, Phys. Rev. Lett. 87 (2001) 167902.
\newblock \href {https://doi.org/10.1103/PhysRevLett.87.167902}
  {\path{doi:10.1103/PhysRevLett.87.167902}}.
\newline\urlprefix\url{https://link.aps.org/doi/10.1103/PhysRevLett.87.167902}

\bibitem{Green20}
D.~L. Green, E.~D’Azevedo, D.~B. Batchelor, N.~Bertelli, C.~Lau, R.~L.
  Barnett, J.~F.~C. Marin,
  \href{https://aip.scitation.org/doi/abs/10.1063/5.0018579}{A {WKB} based
  preconditioner for the {1D Helmholtz} wave equation}, AIP Conference
  Proceedings 2254~(1) (2020) 060008.
\newblock \href
  {http://arxiv.org/abs/https://aip.scitation.org/doi/pdf/10.1063/5.0018579}
  {\path{arXiv:https://aip.scitation.org/doi/pdf/10.1063/5.0018579}}, \href
  {https://doi.org/10.1063/5.0018579} {\path{doi:10.1063/5.0018579}}.
\newline\urlprefix\url{https://aip.scitation.org/doi/abs/10.1063/5.0018579}

\bibitem{Suau21}
A.~Suau, G.~Staffelbach, H.~Calandra,
  \href{http://dx.doi.org/10.1145/3430030}{Practical quantum computing}, ACM
  Transactions on Quantum Computing 2~(1) (2021) 1–35.
\newblock \href {https://doi.org/10.1145/3430030} {\path{doi:10.1145/3430030}}.
\newline\urlprefix\url{http://dx.doi.org/10.1145/3430030}

\bibitem{Draper00}
T.~G. Draper, \href{https://arxiv.org/abs/quant-ph/0008033}{Addition on a
  quantum computer} (2000).
\newblock \href {https://doi.org/10.48550/ARXIV.QUANT-PH/0008033}
  {\path{doi:10.48550/ARXIV.QUANT-PH/0008033}}.
\newline\urlprefix\url{https://arxiv.org/abs/quant-ph/0008033}

\end{thebibliography}

\end{document}